\documentclass[aps,nofootinbib,longbibliography,preprint]{revtex4}%
\usepackage[dvipdfmx]{graphicx}
\usepackage{graphicx,bm,color}
\usepackage{subfigure}
\usepackage{amsmath}
\usepackage{amssymb}
\usepackage{amsfonts}
\usepackage{cases}
\usepackage{cancel}
\usepackage{hyperref}
\usepackage{ulem}
\usepackage{amsmath,amssymb,bm,color,longtable,mathrsfs,amsfonts,slashed}

\usepackage{here}
\usepackage{tikz}
\usetikzlibrary{matrix}
\newcommand{\Slash}[1]{{\ooalign{\hfil/\hfil\crcr$#1$}}}

\begin{document}
\preprint{ }
\title{Chiral effective model of cold and dense two-color QCD: \\
The linear sigma model approach}

\author{Daiki~Suenaga}
\email[]{suenaga.daiki.j1@f.mail.nagoya-u.ac.jp}
\affiliation{Kobayashi-Maskawa Institute for the Origin of Particles and the Universe, Nagoya University, Nagoya, 464-8602, Japan}

\date{\today}

\begin{abstract}
This review is devoted to summarizing recent developments of the linear sigma model (LSM) in cold and dense two-color QCD (QC$_2$D), in which lattice simulations are straightforwardly applicable thanks to the disappearance of the sign problem. In QC$_2$D, both theoretical and numerical studies derive the presence of the so-called baryon superfluid phase at sufficiently large chemical potential ($\mu_q$), where diquark condensates govern the ground state. The hadron mass spectrum simulated in this phase shows that the mass of an iso-singlet ($I=0$) and $0^-$ state is remarkably reduced, but such a mode cannot be described by the chiral perturbation theory. Motivated by this fact, I invent the LSM constructed upon the linear representation of chiral symmetry, or more precisely the Pauli-G\"ursey symmetry. Then, it is shown that my LSM successfully reproduces the low-lying hadron mass spectrum in a broad range of $\mu_q$ simulated on the lattice. As applications of the LSM, topological susceptibility and sound velocity in cold and dense QC$_2$D are evaluated to compare with lattice results. Besides, generalized Gell-Mann-Oakes-Renner relation and hardon mass spectrum in the presence of a diquark source are analyzed. I also introduce an extended version of the LSM incorporating spin-$1$ hadrons.
\end{abstract}


\maketitle
\tableofcontents

\section{Introduction}
\label{sec:Introduction}

In recent years, elucidation of quantum chromodynamics (QCD) in cold and dense system has been gathering much attention motivated by the progress of neutron star observations~\cite{Baym:2017whm}. In such dense system, quarks confined inside hadrons begin to overlap as the density increases, and finally quark degrees of freedom govern the matter. Due to the complexity stemming from the strong coupling and nonperturbative nature of QCD, however, it is not easy to unveil this transition in detail.

One of the most powerful tools to shed light on the QCD problem is the first-principles {\it lattice QCD simulation}. But lattice simulations with a chemical potential at lower temperature are not straightforward, due to the so-called {\it sign problem} of the Monte-Carlo computation~\cite{Aarts:2015tyj,Nagata:2021ugx}. Besides, considering the current difficulty of accelerator experiments,  cold and dense QCD can be regarded as a frontier of quark-hadron physics.

The sign problem of lattice simulations occurs when the path integral yields complex values. Hence, when we focus on two-color QCD (QC$_2$D) where the gluodynamics is governed by $SU(2)_c$ gauge theory possessing the pseudoreality, the troublesome sign problem disappears. This advantage enables us to apply the lattice simulation straightforwardly even at the cold and dense system. Indeed, thus far, lattice simulations in QC$_2$D with a baryon-number (or a quark-number) chemical potential have been conducted by several groups to explore the phase diagram with order parameters, hadron masses, thermodynamic properties, gluondynamics, transport coefficients, and so on~\cite{Hands:1999md,Kogut:2001na,Hands:2001ee,Muroya:2002ry,Muroya:2003qs,Chandrasekharan:2006tz,Hands:2006ve,Hands:2007uc,Hands:2010gd,Cotter:2012mb,Hands:2012yy,Boz:2013rca,Braguta:2016cpw,Puhr:2016kzp,Boz:2018crd,Astrakhantsev:2018uzd,Iida:2019rah,Wilhelm:2019fvp,Boz:2019enj,Buividovich:2020gnl,Iida:2020emi,Astrakhantsev:2020tdl,Bornyakov:2020kyz,Buividovich:2020dks,Buividovich:2021fsa,Begun:2021nbf,Iida:2022hyy,Begun:2022bxj,Murakami:2023ejc,Braguta:2023yhd,Iida:2024irv}.

In QC$_2$D world, the pseudoreal nature of $SU(2)_c$ gauge theory allows us to treat a quark and an antiquark belonging to ${\bm 2}$ and ${\bm 2}^*$ representations on an equal footing. As a result, in a hadronic level, for instance certain mesons and diquarks share the same properties. In terms of the flavor representation, this is reflected by the enlargement of chiral symmetry; $SU(N_f)_L\times SU(N_f)_R$ chiral symmetry is extended to the so-called {\it Pauli-G\"ursey $SU(2N_f)$ symmetry} in QC$_2$D~\cite{Pauli:1957voo,Gursey:1958fzy}.

Since (anti)diquarks are bosonic in QC$_2$D obeying the Bose-Einstein statistics similarly to mesons, they start to exhibit the Bose-Einstein condensations (BECs) at adequately large chemical potential $\mu_q$. This condensed phase is referred to as the {\it diquark condensed phase} simply, or the {\it baryon superfluid phase} so as to stress the $U(1)$ baryon-number violation with no breakdown of color symmetry. Meanwhile, the calm phase connected to the vacuum (zero temperature and zero chemical potential) is called the {\it hadronic phase}.

\begin{figure}[H]
\centering
\includegraphics[width=9.5cm]{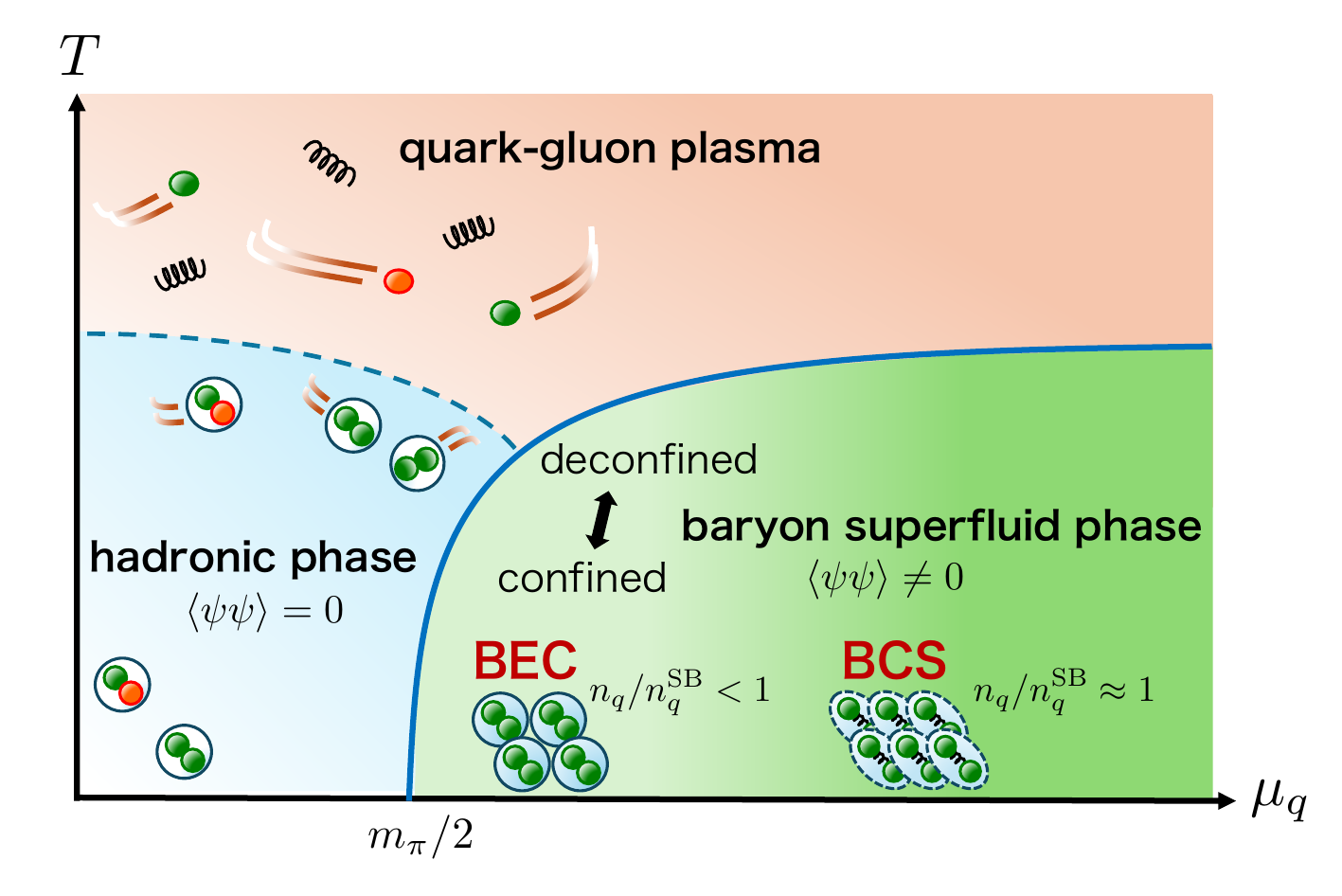}
\caption{A schematic phase diagram of QC$_2$D.}
\label{fig:PhaseDiagram}
\end{figure}

A schematic picture of QC$_2$D phase diagram is depicted in Fig.~\ref{fig:PhaseDiagram}. In this figure, the Bardeen-Cooper-Schrieffer (BCS) regime in the baryon superfluid phase is defined by which the quark density $n_q$ is consistent with the Stefan-Boltzmann-limit value of free quarks $n_q^{\rm SB}$: $n_q/n_q^{\rm SB}\approx1$. Accordingly, in the BEC regime $n_q/n_q^{\rm SB}<1$.

In order to gain qualitative and predictive insights into the {\it numerical experiments} of cold and dense QC$_2$D performed on the lattice, it is inevitable to translate the numerical results in terms of appropriate effective degrees of freedom. In the low-energy regime of QC$_2$D where the system is governed by the highly nonperturbative dynamics, such excitations are played by light hadrons. Hence, hadron effective models can be regarded as a useful tool there. The lattice results predict sufficiently suppressed Polyakov loops even in the dense regime~\cite{Boz:2019enj,Iida:2024irv}, indicating that hadronic and quark matters are connected by crossover. Therefore, hadron effective models would be able to explore the deeper regime of dense QC$_2$D.

The spirit of hadron effective models is expressed by the following matching equality~\cite{Gasser:1983yg,Gasser:1984gg}:
\begin{eqnarray}
Z_{\rm QC_2D} = Z_{\rm eff.\ model}\ , \label{MatchingZ}
\end{eqnarray}
where the left-hand side (LHS) and right-hand side (RHS) stand for the generating functionals of underlying QC$_2$D and of an effective model, respectively. That is, (maybe concise) quantum theory developed in hadron effective models must match that of the nonperturbative QC$_2$D one at low-energy. More practically we make use of
\begin{eqnarray}
\Gamma_{\rm QC_2D} = \Gamma_{\rm eff.\ model} \label{MatchingQCD}
\end{eqnarray}
as the matching condition, with the corresponding effective action $\Gamma = -i{\rm ln}Z$. This $\Gamma$ can be regarded as an action incorporating quantum corrections, so that symmetry properties inhabit QC$_2$D at quantum level must be mimicked by the effective model ones. Those matching properties are the essential points when adopting hadron effective models.

The (approximate) Pauli-G\"ursey $SU(2N_f)$ symmetry is spontaneously broken owing to the emergence of chiral condensates $\langle\bar{\psi}\psi\rangle$ in the vacuum, the breaking pattern of which is $SU(2N_f)\to Sp(2N_f)$ for identical quark masses~\cite{Kogut:1999iv,Kogut:2000ek}. Accordingly, the Nambu-Goldstone (NG) bosons dominate over the low-energy dynamics of QC$_2$D. Due to the equal treatment of certain mesons and diquarks, those NG bosons are played by $N_f^2-1$ pions and $N_f^2-N_f$ flavor-singlet scalar (anti)diquarks.

\begin{figure}[H]
\centering
\includegraphics[width=13cm]{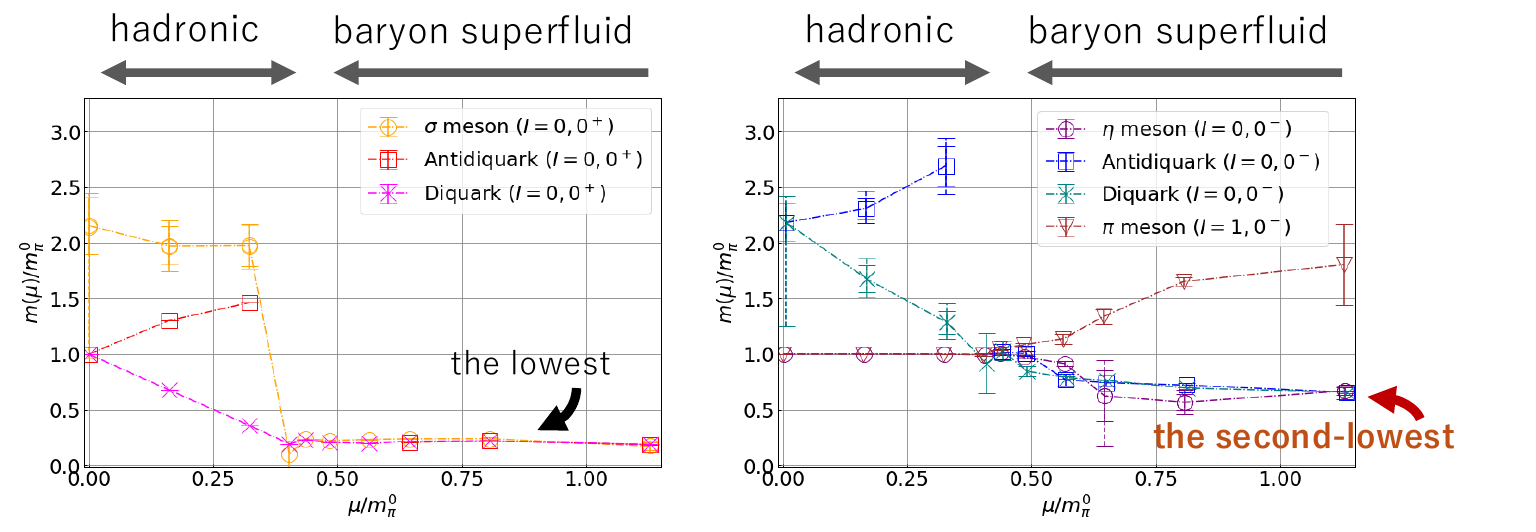}
\caption{$\mu_q$ dependences of the spin-$0$ hadron masses at low temperature, which were computed in Ref.~\cite{Murakami:2022lmq}. The second-lowest state ($I=0$, $0^-$) in the superfluid phase cannot be describe in the ChPT framework.}
\label{fig:IntroMass}
\end{figure}

When describing those NG boson dynamics, the chiral perturbation theory (ChPT) framework is useful thanks to its systematic low-energy expansion, and was developed in Ref~\cite{Kogut:1999iv,Kogut:2000ek}. Indeed, this effective model successfully reproduces, e.g., hadron masses~\cite{Hands:2007uc,Murakami:2022lmq} and sound velocity~\cite{Iida:2022hyy,Iida:2024irv} measured on the lattice for $N_f=2$, in the vicinity of the phase transition to the baryon superfluid phase. However, since the ChPT is based on the low-energy expansion for only the NG bosons where other excitations are integrated out~\cite{Coleman:1969sm,Callan:1969sn}, it is unclear whether the ChPT framework still works at larger $\mu_q$. Moreover, the recent lattice results in Ref.~\cite{Murakami:2022lmq} indicate that the next-lightest excitation in the superfluid phase is an iso-singlet mode carrying a negative parity ($I=0$, $0^-$), which cannot be handled by the ChPT, as depicted in Fig.~\ref{fig:IntroMass}. Those facts require us to extend the ChPT to describe other hadrons including the $I=0$, $0^-$ mode for which the low-energy spectrum of dense QC$_2$D is appropriately delineated.

Motivated by this fact, I invented a linear sigma model (LSM) as an extension of the ChPT based on the linear representation of the Pauli-G\"ursey symmetry for $N_f=2$~\cite{Suenaga:2022uqn}. This effective model allows us to describe not only the NG bosons such as the pions and $0^+$ diquarks but also the scalar mesons and $0^-$ diquarks collectively, although the systematics is rather obscure. The latter hadrons are referred to as the {\it parity partners} or {\it chiral partners} to the NG bosons, which is predicted to degenerate with the NG bosons at the chiral restoration point. Besides, the linear representation of the LSM implies its validity at rather high $\mu_q$ where the nonlinearly realized ChPT framework cannot touch, since chiral symmetry is restored at sufficiently large $\mu_q$.

In addition to the ChPT and LSM approaches, other effective models containing quark-gluon degrees of freedom such as the Nambu-Jona-Lasinio (NJL) and the massive gluon model were employed to theoretically examine cold and dense QC$_2$D properties~\cite{Kogut:1999iv,Kogut:2000ek,Lenaghan:2001sd,Splittorff:2001fy,Ratti:2004ra,Sun:2007fc,Fukushima:2007bj,Brauner:2009gu,Kanazawa:2009ks,Harada:2010vy,Andersen:2010vu,Zhang:2010kn,He:2010nb,Strodthoff:2011tz,Imai:2012hr,Strodthoff:2013cua,Khan:2015puu,Duarte:2015ppa,Chao:2018czo,Adhikari:2018kzh,Contant:2019lwf,Suenaga:2019jjv,Khunjua:2020xws,Kojo:2021knn,Suenaga:2021bjz,Kojo:2021hqh,Khunjua:2021oxf,Suenaga:2022uqn,Kawaguchi:2023olk,Suenaga:2023xwa,Kawaguchi:2024iaw,Acharyya:2024pqj,Khunjua:2024kdc}. 

In this review, I summarize the main points of Refs.~\cite{Suenaga:2022uqn,Kawaguchi:2023olk,Suenaga:2023xwa,Kawaguchi:2024iaw} achieved within the LSM framework. Since effective models such as the LSM must obey QC$_2$D-inspired symmetry properties owing to the matching condition of Eq.~(\ref{MatchingZ}) or Eq.~(\ref{MatchingQCD}), first in Sec.~\ref{sec:QC2DLagrangian}, I explain the Ward-Takahashi identities (WTIs) related to the spontaneous breakdown of the Pauli-G\"ursey symmetry from the underlying QC$_2$D theory. Next, in Sec.~\ref{sec:ChPT} I show a derivation of the ChPT based on the Maurer-Cartan $1$-form and summarize its predictions. Then, in Sec.~\ref{sec:LSM} I construct the LSM and review our works accomplished in Refs.~\cite{Suenaga:2022uqn,Kawaguchi:2023olk,Kawaguchi:2024iaw}  comparing with the ChPT results. Section~\ref{sec:ELSM} is devoted to presenting an extended version of the LSM where additionally spin-$1$ hadrons are incorporated, which is referred to as the extended linear sigma model (eLSM)~\cite{Suenaga:2023xwa}. Finally, in Sec.~\ref{sec:Conclusion} the present article is concluded.

\section{QC$_2 $D Lagrangian for quarks}
\label{sec:QC2DLagrangian}

\subsection{Pauli-G\"ursey $SU(2N_f)$ symmetry}
\label{sec:PGSymmetry}

 The flavor structure, i.e., chiral symmetry of QC$_2$D is extended owing to the pseudoreality of $SU(2)_c$ gauge group, which plays an essential role in describing hadronic excitations in low-energy regime of QC$_2$D. In this subsection we explain how such enlarged symmetry emerges by rewriting the QC$_2$D Lagrangian for quarks.

The Lagrangian for $N_f$ massless quarks interacting with the $SU(2)_c$ gluons is of the form
\begin{eqnarray}
{\cal L}^{\rm kin}_{\rm QC_2D} = \bar{\psi}i\Slash{D}\psi\ , \label{QC2DQuarkL}
\end{eqnarray}
where $\psi = (u,d,\cdots)$ is an $N_f$-components quark field and the covariant derivative reads $D_\mu\psi = (\partial_\mu-ig_s A_\mu)\psi$ with the $SU(2)_c$ gauge field $A_\mu=A_\mu^a \tau_c^a/2$ ($\tau^a_c$ is the Pauli matrix) and the gauge coupling $g_s$. To see chiral structures of the quarks, it is useful to introduce the left-handed  and right-handed quark fields, $\psi_R$ and $\psi_L$, which are eigenstates of the chirality operator $\gamma_5$. When employing the Weyl representation, those fields are expressed to be $\psi_R=\frac{1+\gamma_5}{2}\psi= (\hat{\psi}_R,{\bm 0})^T$ and $\psi_L = \frac{1-\gamma_5}{2}\psi = ({\bm 0},\hat{\psi}_L)^T$. Hence, the Lagrangian~(\ref{QC2DQuarkL}) is rewritten in terms of the two-component spinors $\hat{\psi}_R$ and $\hat{\psi}_L$ as
\begin{eqnarray}
{\cal L}^{\rm kin}_{\rm QC_2D} = \hat{\psi}^\dagger_Ri\sigma^\mu(\partial_\mu -ig_sA_\mu)\hat\psi_R + \hat{\psi}^\dagger_Li\bar{\sigma}^\mu(\partial_\mu-ig_sA_\mu)\hat{\psi}_L\ ,
\label{QC2DQuarkL2}
\end{eqnarray}
with $\sigma^\mu=({\bm 1}, \sigma^i$) and $\bar{\sigma}^{\mu} = ({\bm 1},-\sigma^i$) ($\sigma^i$ is the Pauli matrix inhabiting the spinor space). 

The form of Eq.~(\ref{QC2DQuarkL2}) is universal for any number of colors. The characteristic feature of $SU(2)_c$ gauge theory appears when making use of the pseudoreality of the Pauli matrix, $\tau_c^a = -\tau_c^2(\tau_c^a)^T\tau_c^2$ (and $\sigma^i = -\sigma^2(\sigma^i)^T\sigma^2$). In fact, these relations enable us to reduce Eq.~(\ref{QC2DQuarkL2}) to the following simple form:
\begin{eqnarray}
{\cal L}^{\rm kin}_{\rm QC_2D} = {\Psi}^\dagger i\sigma^\mu\partial_\mu\Psi+g_s\Psi^\dagger \sigma^\mu A_\mu\Psi \ .
\label{QC2DQuarkRed}
\end{eqnarray}
In this equation the extended $2N_f$-components quark filed $\Psi$ is defined by 
\begin{eqnarray}
\Psi \equiv (\hat{\psi}_R,\tilde{\hat{\psi}}_L)^T = (\hat{u}_R,\hat{d}_L,\cdots, \tilde{\hat{u}}_L, \tilde{\hat{d}}_L,\cdots)^T\ , \label{PsiDef}
\end{eqnarray}
with $\tilde{\hat{\psi}}_R = \sigma^2\tau_c^2\hat{\psi}^*_R$ and $\tilde{\hat{\psi}}_L = \sigma^2\tau_c^2\hat{\psi}_L^* $ being the ``conjugate fields'' played by the complex conjugate of $\hat{\psi}_L$. The QC$_2$D Lagrangian expressed in terms of $\Psi$ in Eq.~(\ref{QC2DQuarkRed}) clearly shows a global symmetry under an $SU(2N_f)$ transformation generated by
\begin{eqnarray}
\Psi \to g\Psi\ \ \  {\rm with}\ \ \ g\in G=SU(2N_f)\ . \label{SU4TransPsi}
\end{eqnarray}
Since $\Psi$ is a $2N_f$-component column vector in the (enlarged) flavor space from Eq.~(\ref{PsiDef}), the symmetry by Eq.~(\ref{SU4TransPsi}) is regarded as an extended version of $SU(N_f)_L\times SU(N_f)_R$ chiral symmetry. This is sometimes referred to as the Pauli-G\"{u}rsey $SU(2N_f)$ symmetry~\cite{Pauli:1957voo,Gursey:1958fzy}. Intuitively speaking, the extension of chiral symmetry reflects the blindness of $SU(2)_c$ gluons; Quarks and antiquarks belong to ${\bm 2}$ and ${\bm 2}^*$ representations of $SU(2)_c$, however, due to the pseudoreality ${\bm 2}\simeq{\bm 2}^*$ gluons cannot discriminate quarks and antiquarks. As a result, these states can be treated on an equal footing in a single multiplet, and the flavor structure, i.e., chiral symmetry, is enlarged.

From the above argument, QC$_2$D with massless quarks has been found to possess an $SU(2N_f)$ symmetry generated by Eq.~(\ref{SU4TransPsi}). Meanwhile, the quark mass term reads
 \begin{eqnarray}
{\cal L}_{\rm QC_2D}^{\rm mass} = -\frac{1}{2}\left(\Psi^T\sigma^2\tau_c^2{\cal M}_q\Psi + {\rm H.c.} \right) = -\bar{\psi}M_q\psi \ , \label{QC2DMass}
\end{eqnarray}
 where the quark mass matrix in terms of the extended quark multiplet~(\ref{PsiDef}) takes the form of the following $2N_f\times 2N_f$ matrix:
\begin{eqnarray}
{\cal M}_q = \left(
\begin{array}{cc}
0 & -M_q \\
M_q & 0 \\
\end{array}
\right) 
\end{eqnarray} 
with ${M}_q = {\rm diag.}(m_u,m_d,\cdots)$. This mass term obviously breaks the $SU(2N_f)$ symmetry, similarly to the (explicit) chiral-symmetry breaking in three-color QCD. In particular, when all the quark masses are identical, $m_q\equiv m_u=m_d=\cdots$, the mass term~(\ref{QC2DMass}) is reduced to
\begin{eqnarray}
{\cal L}_{\rm QC_2D}^{\rm mass} = -\frac{m_q}{2}\left(\Psi^T\sigma^2\tau_c^2E^T\Psi + {\rm H.c.} \right) = -m_q\bar{\psi}\psi \ , \label{QC2DMassIsospin}
\end{eqnarray}
where $E$ is a $2N_f\times 2N_f$ symplectic matrix
\begin{eqnarray}
E  = \begin{pmatrix}
{\bm 0} & {\bm 1} \\
-{\bm 1} & {\bm 0} \\
\end{pmatrix} \ .
\end{eqnarray}
The operator $\Psi^T\sigma^2\tau_c^2 E^T\Psi$ is not generally invariant under $g\in G$ ($=SU(2N_f)$) but only invariant under $h\in H$ belonging to a subgroup of $G$, which satisfies 
\begin{eqnarray}
h^T E h = E\ . \label{SymplecticCond}
\end{eqnarray}
This relation is nothing but the definition of $Sp(2N_f)$ group. Therefore, $H=Sp(2N_f)$ and the symmetry breaking pattern reads $SU(2N_f)\to Sp(2N_f)$ in this particular case. The number of generators of $SU(2N_f)$ and $Sp(2N_f)$ are $4N_f^2-1$ and $N_f(2N_f+1)$, respectively, hence the number of NG bosons associated with the breaking of $SU(2N_f)\to Sp(2N_f)$ is
\begin{eqnarray}
4N_f^2-1-N_f(2N_f+1) = 2N_f^2-N_f-1\ . \label{NumberNG}
\end{eqnarray}
As in three-color QCD, $N_f^2-1$ pseudoscalar mesons are responsible for the NG bosons, which cannot cover the whole number of Eq.~(\ref{NumberNG}). That is, in QC$_2$D, $N_f^2-N_f=2\,_{N_f} C_2$ NG bosons additionally emerge other than the pseudoscalar mesons. These additional NG bosons are played by the flavor-antisymmetric and scalar (anti)diquarks which are the lightest (anti)baryonic modes, as can be indeed understood by the combination factor $2\,_{N_f} C_2$. The simultaneous emergence of mesonic and (anti)baryonic NG modes also stems from the ``blindness'' of $SU(2)_c$ gluons.

The $U(1)$ baryon-number and $U(1)$ axial transformations of $\Psi$ are easily understood from the definition~(\ref{PsiDef}). That is, the upper and lower $N_f$ components of $\Psi$ carry opposite baryon charges and the identical axial charges. Thus, when we assign quark-number $+1$ (baryon-number $+1/2$) for $\hat{\psi}$, the resultant $U(1)_B$ transformation law of $\Psi$ reads
\begin{eqnarray}
\Psi \overset{U((1)_B}{\to} {\rm e}^{-i\theta_B J}\Psi\  \ \ {\rm with}\ \ \ J = 
\left(
\begin{array}{cc}
{\bm 1} & {\bm 0} \\
{\bm 0} & -{\bm 1} \\
\end{array}
\right) \ . \label{BaryonTrans}
\end{eqnarray}
Similarly, 
\begin{eqnarray}
\Psi \overset{U((1)_A}{\to} {\rm e}^{-i\theta_A}\Psi\ ,
\end{eqnarray}
under the $U(1)$ axial transformation. It should be noted that the $U(1)$ baryon-number transformation~(\ref{BaryonTrans}) belongs to a subgroup of $SU(2N_f)$. Meanwhile, the $U(1)_A$ rotation simply changes the overall phase of $\Psi$ which is not generated by any of the $SU(2N_f)$.

In what follows, we restrict ourselves into two-flavor ($N_f=2$) with an exact isospin symmetry: $m_q \equiv m_u= m_d$, which corresponds to the often used lattice-simulation setup~\cite{Braguta:2023yhd}, otherwise stated. In this particular case, the symmetry breaking pattern is $SU(4)\to Sp(4)$.

\subsection{
Algebra of $SU(4)$ and $Sp(4)$}
\label{sec:Algebra}

For $N_f=2$ with isospin symmetry, the Pauri-G\"{u}rsey symmetry turns to be $G=SU(4)$ which contains $15$ generators. Since the symmetry breaking pattern is $SU(4)\to Sp(4)$, it is convenient to separate the $15$ generators into those belonging to the algebra of $H=Sp(4)$ and of $G/H=SU(4)/Sp(4)$: $S^i \in {\cal H}$ ($i=1-10$) and $X^a\in {\cal G}-{\cal H}$ ($a=1-5$). In this paper we employ
\begin{eqnarray}
S^{i=1-4} = \frac{1}{2\sqrt{2}}\left(
\begin{array}{cc}
\tau_f^i & 0 \\
0 & -(\tau_f^i)^T \\
\end{array}
\right)\ \ \ ,  \ \ \ S^{i=5-10} = \frac{1}{2\sqrt{2}}\left(
\begin{array}{cc}
0& B_f^i \\
(B_f^i)^\dagger &0 \\
\end{array}
\right)\  , \label{Generators}
\end{eqnarray}
and
\begin{eqnarray}
X^{a=1-3} = \frac{1}{2\sqrt{2}}\left(
\begin{array}{cc}
\tau_f^a & 0 \\
0 &(\tau_f^a)^T \\
\end{array}
\right)\  \ \ , \  \ \ X^{a=4,5} = \frac{1}{2\sqrt{2}}\left(
\begin{array}{cc}
0& D_f^a \\
(D_f^a)^\dagger & 0 \\
\end{array}
\right) \ , \label{GeneratorX}
\end{eqnarray}
to parametrize them, in which $\tau_f^4 = {\bm 1}$, $B_f^5={\bm 1}$, $B_f^6=i{\bm 1}$, $B_f^7=\tau_f^3$, $B_f^8=i\tau_f^3$, $B^9=\tau_f^1$, $B^{10}=i\tau_f^1$, $D^4=\tau_f^2$, $D^5=i\tau_f^2$, with $\tau_f^{1,2,3}$ being the Pauli matrices acting on the flavor space. The generators belonging to the algebra of the unbroken $Sp(4)$ satisfy
\begin{eqnarray}
S^i E = -E (S^i)^T\ \ \ \ \ \ ( S^i \in {\cal H}) \label{SAlgebra}
\end{eqnarray}
from Eq.~(\ref{SymplecticCond}). Accordingly the broken generators $X^a$ obey
\begin{eqnarray}
X^a E= E(X^a)^T \ \ \ \ \ \  (X^a \in {\cal G} - {\cal H})\ . \label{XAlgebra}
\end{eqnarray}
The above basis are convenient since mesonic and baryonic modes can be properly separated once an effective Lagrangian is constructed. 

For later convenience we further define
\begin{eqnarray}
X^{a=0} = \frac{1}{2\sqrt{2}}\left(
\begin{array}{cc}
{\bm 1} & 0 \\
0 & {\bm 1} \\
\end{array}
\right)
\end{eqnarray}
to parametrize the trivial algebra.

\subsection{Spurion fields}
\label{sec:Spurion}

Since QC$_2$D with massless quarks preserves the Pauli-G\"usey $SU(4)$ symmetry when $N_f=2$, it is convenient to regard, e.g., quark masses, as an external-source contribution breaking the $SU(4)$ symmetry properly. In this subsection, we introduce the so-called {\it spurion fields} so as to formulate such a systematic inclusion of the breaking effects~\cite{Gasser:1983yg,Gasser:1984gg}.

The source term of QC$_2$D takes the form of
\begin{eqnarray}
{\cal L}^{\rm source}_{\rm QC_2D} = -\Psi^T\sigma^2\tau_c^2\zeta^\dagger\Psi - \Psi^\dagger\sigma^2\tau_c^2 \zeta \Psi^*  + \Psi^\dagger\sigma_\mu\zeta^\mu\Psi \ , \label{QC2DSourceRed}
\end{eqnarray}
where the spurion fields $\zeta$ ($=-\zeta^T$) and $\zeta_\mu$ ($=\zeta^\dagger_\mu$) transform under local $SU(4)$ transformation as
\begin{eqnarray}
\zeta \to g \zeta g^T\ , \ \ \zeta_\mu \to g \zeta g^\dagger -i\partial_\mu gg^\dagger\ . \label{SpurionTransQC2D}
\end{eqnarray}
In this way, the whole QC$_2$D Lagrangian
\begin{eqnarray}
{\cal L}^{q}_{\rm QC_2D} = {\cal L}^{\rm kin}_{\rm QC_2D} + {\cal L}^{\rm source}_{\rm QC_2D} \label{LQuarkQC2D}
\end{eqnarray}
preserves the local $SU(4)$ symmetry. We note that the spin-$1$ spurion $\zeta_\mu$ is introduced as if to be a gauge field with respect to $G=SU(4)$ symmetry. The spurions can be decomposed into real fields $s^a$, $p^a$, $V_\mu^i$ and $V_\mu'^a$ as
\begin{eqnarray}
\zeta = \sqrt{2}\sum_{a=0}^5(s^a-ip^a)X^aE\  \ , \ \ \ \ \zeta_\mu = 2\sqrt{2}\left(\sum_{i=1}^{10}v^i_\mu S^i - \sum_{a=0}^5v'^a_\mu X^a\right)\ . \label{SpurionDec}
\end{eqnarray}

In the source contribution~(\ref{QC2DSourceRed}), by replacing the scalar field $s^{a=0}$ with its vacuum expectation value (VEV) as $\langle s^{a=0}\rangle=m_q$ and setting all other fields to be vanishing, one can obtain
\begin{eqnarray}
{\cal L}_{\rm QC_2D}^{\rm source}\Big|_{\langle s^{0}\rangle=m_q} = -\frac{m_q}{2}\left(\Psi^T\sigma^2\tau_c^2E^T\Psi + {\rm H.c.} \right) = -m_q\bar{\psi}\psi \ .
\end{eqnarray}
This form is, indeed, identical to the mass term in Eq.~(\ref{QC2DMassIsospin}). Similarly, a quark chemical potential $\mu_q$ can be introduced by choosing the VEV of spin-$1$ spurions as $\langle v_{\mu=0}^{a=4}\rangle=\mu_q$. Moreover, a diquark source term which leads to condensations of the isospin-singlet and color-singlet scalar diquarks breaking $U(1)_B$ symmetry can also be realized by $\langle p^{a=5}\rangle=j$.

To summarize, within the basis~(\ref{SpurionDec}), when taking the following VEVs:
\begin{eqnarray}
\langle s^{a=0}\rangle=m_q\ , \ \ \langle p^{a=5}\rangle = j\ , \ \  \langle v_{\mu=0}^{i=4}\rangle = \mu_q\ , \label{SpurionReplace}
\end{eqnarray} \
the source term~(\ref{QC2DSourceRed}) is reduced to
\begin{eqnarray}
{\cal L}_{\rm QC_2D}^{\rm source}\Big|_{\langle \zeta\rangle,\langle\zeta_\mu\rangle} &=&- \sqrt{2}m_q\left(\Psi^T\sigma^2\tau_c^2 E^TX^0 \Psi + {\rm h.c.}\right) - \sqrt{2}j\left(i\Psi^T\sigma^2\tau_c^2E^T X^5 \Psi + {\rm h.c.}\right) \nonumber\\
&& + 2\sqrt{2}\mu_q \Psi^\dagger S^4\Psi \nonumber\\
&=& -m_q\bar{\psi}\psi-j\left( -\frac{i}{2}\psi^TC\gamma_5\tau_c^2\tau_f^2\psi + {\rm h.c.}\right) + \mu_q\bar{\psi}\gamma^0\psi\ , \label{SourceSystematic}
\end{eqnarray}
which, indeed, correctly reproduce the quark mass, diquark source and chemical potential terms.

In addition to the systematic inclusion of the physical parameters as in Eq.~(\ref{SourceSystematic}), our spurion-field treatment plays a key role in matching QC$_2$D and low-energy effective models. For instance, taking a functional derivative of $\Gamma_{\rm QC_2D}$ with respect to $s^0$ and $p^5$ and imposing the VEVs~(\ref{SpurionReplace}), one can easily obtain formulas
\begin{eqnarray}
\langle\bar{\psi}\psi\rangle = \frac{\delta\Gamma_{\rm QC_2D}}{-\delta s^0}\Big|_{\langle\zeta\rangle,\langle\zeta_\mu\rangle}\ , \ \ \langle{\psi}\psi\rangle = \frac{\delta\Gamma_{\rm QC_2D}}{-\delta p^0}\Big|_{\langle\zeta\rangle,\langle\zeta_\mu\rangle}\ ,
\end{eqnarray}
respectively, with a shorthand notation
\begin{eqnarray}
\psi\psi \equiv  -\frac{i}{2}\psi^TC\gamma_5\tau_c^2\tau_f^2\psi + {\rm h.c.}\ .
\end{eqnarray}
Therefore, making use of the matching condition~(\ref{MatchingQCD}), the condensates are found to be evaluated within effective models as
\begin{eqnarray}
\langle\bar{\psi}\psi\rangle = \frac{\delta\Gamma_{\rm eff.\ model}}{-\delta s^0}\Big|_{\langle\zeta\rangle,\langle\zeta_\mu\rangle}\ , \ \ \langle{\psi}\psi\rangle = \frac{\delta\Gamma_{\rm eff.\ model}}{-\delta p^0}\Big|_{\langle\zeta\rangle,\langle\zeta_\mu\rangle}\ . \label{MatchingCondensate}
\end{eqnarray}
In the same way any $n$-point functions of underlying QC$_2$D can be matched with those of the low-energy effective models.

\subsection{Ward-Takahashi identities}
\label{sec:WTIQC2D}

Here, we derive the WTIs from symmetry properties of the $SU(4)$ which also play a central role when matching QC$_2$D and the effective models. Here we only focus on the WTIs connecting spin-$0$ operators.

Let us define the following spin-$0$ composite operators:
\begin{eqnarray}
{\cal O}_X^a \equiv \Psi^T\sigma^2\tau_c^2 E X^a\Psi\ \ \ \ \ (a=0-5)\ . \label{CompositeOperator}
\end{eqnarray}
Under the infinitesimal transformation driven by the broken generators $X^a$, $\Psi\to {\rm e}^{-i\theta^aX^a}\Psi$, those operators are transformed as
\begin{eqnarray}
{\cal O}_X^a &\overset{G/H}{\to}& {\cal O}_X^a - i\theta^b \Psi^T\sigma^2\tau_c^2\left((X^b)^TEX^a  + EX^a X^b\right)\Psi \nonumber\\
&=& {\cal O}_X^a - \frac{i}{\sqrt{2}}\theta^a {\cal O}_X^0  \ \ \ \ (a=1-5) \ , \label{XaG/H}
\end{eqnarray}
and
\begin{eqnarray}
{\cal O}_X^0 &\overset{G/H}{\to}& {\cal O}_X^0 - i\theta^a \Psi^T\sigma^2\tau_c^2\left[(X^a)^TE + EX^a\right]X^0\Psi \nonumber\\
&=&{\cal O}_X^0 - \frac{i}{\sqrt{2}} \theta^a{\cal O}_X^a  \ \ \ \ (a=1-5) \ , \label{X0G/H}
\end{eqnarray}
where the algebras in Eq.~(\ref{XAlgebra}) and $\{X^a,X^b\} = \delta^{ab}E/4$ ($a,b=1-5$) are used. Similarly under $U(1)_B$ transformation $\Psi\to{\rm e}^{-i\theta_BJ}\Psi$, one can see
\begin{eqnarray}
{\cal O}_X^a \overset{U(1)_B}{\to} {\cal O}_X^a - i\theta_B\Psi^T\sigma^2\tau_c^2\{J,EX^a\}\Psi = \left\{
\begin{array}{cc}
{\cal O}_X^a & (a=0-3) \\
{\cal O}_X^4 + 2\theta_B{\cal O}_X^5 & (a=4) \\
{\cal O}_X^5 - 2\theta_B{\cal O}_X^4 & (a=5) \\
\end{array}
\right.\  . \label{XaU1B}
\end{eqnarray}
Although the composite operators~(\ref{CompositeOperator}) exhibit concise transformation properties, they include positive-parity and negative-parity states collectively and are not useful as building blocks of the physical states. For this reason we also define
\begin{eqnarray}
{\cal O}_{\sigma} &\equiv& \bar{\psi}\psi = -\sqrt{2}{\cal O}_X^0-\sqrt{2}{\cal O}_X^{\dagger 0} \ , \nonumber\\
{\cal O}_{a_0}^a &\equiv&  \bar{\psi}\tau_f^a\psi = -\sqrt{2}{\cal O}_X^a-\sqrt{2}{\cal O}_X^{\dagger a}   \ \ \ \ (a=1-3)\ , \nonumber\\ 
 {\cal O}_{\eta} &\equiv& \bar{\psi}i\gamma_5\psi = -\sqrt{2}i{\cal O}_X^0+\sqrt{2}i{\cal O}_X^{\dagger 0}\ , \nonumber\\
 {\cal O}_{\pi}^a &\equiv& \bar{\psi}i\gamma_5\tau_f^a\psi   = -\sqrt{2}i{\cal O}_X^a+\sqrt{2}i{\cal O}_X^{\dagger a}  \ \ \ \ (a=1-3)\ , \label{OperatorMeson}
\end{eqnarray}
for mesonic operators, while
\begin{eqnarray}
{\cal O}_{B_4} &\equiv& \frac{1}{2}\psi^TC\gamma_5\tau_c^2\tau_f^2\psi + {\rm h.c.} = -\sqrt{2}i{\cal O}_X^4 +\sqrt{2}i{\cal O}_X^{\dagger4}\ , \nonumber\\
{\cal O}_{B_5} &\equiv& -\frac{i}{2}\psi^TC\gamma_5\tau_c^2\tau_f^2\psi + {\rm h.c.} = -\sqrt{2}i{\cal O}_X^5 + \sqrt{2}i{\cal O}_X^{\dagger5}\ , \nonumber\\
{\cal O}_{B'_4} &\equiv& -\frac{i}{2}\psi^TC\tau_c^2\tau_f^2\psi + {\rm h.c.} = -\sqrt{2}{\cal O}_X^4 -\sqrt{2}{\cal O}_X^{\dagger4}\ , \nonumber\\ 
{\cal O}_{B_5'} &\equiv& -\frac{1}{2}\psi^TC\tau_c^2\tau_f^2\psi + {\rm h.c.} = -\sqrt{2}{\cal O}_X^5 -\sqrt{2}{\cal O}_X^{\dagger5} \ , \label{OperatorBaryon}
\end{eqnarray}
for baryonic ones. Their transformation laws are easily read off from Eqs.~(\ref{XaG/H}) -~(\ref{XaU1B}).

Inserting the VEVs~(\ref{SpurionReplace}) of the spurion fields, the QC$_2$D Lagrangian~(\ref{LQuarkQC2D}) now takes the form of
\begin{eqnarray}
{\cal L}_{\rm QC_2D}^q = \Psi^\dagger i\sigma_\mu {\cal D}^\mu\Psi + \sqrt{2}m_q\left({\cal O}_X^0 + {\cal O}_X^{0\dagger}\right) + \sqrt{2}j\left(i{\cal O}_X^5 - i{\cal O}_X^{5\dagger}\right) \ , \label{LQC2DWTI}
\end{eqnarray}
with ${\cal D}_\mu\Psi = (\partial_\mu-ig_sA_\mu-i\mu_q J)\Psi$ and the composite operators are defined by Eq.~(\ref{CompositeOperator}). Under the local transformation generated by $X^a\in {\cal G}-{\cal H}$ ($a=1-5$) this Lagrangian transforms as
\begin{eqnarray}
{\cal L}^q_{\rm QC_2D} &\overset{G/H}{\to}& {\cal L}^q_{\rm QC_2D} -\theta^a\left[D_\mu j_X^{\mu a}-m_q\left(i{\cal O}_X^a-i{\cal O}_X^{a\dagger}\right) + j^a\delta^{5a}\left({\cal O}_X^0+{\cal O}_X^{ 0\dagger}\right) \right]\ ,\label{LQC2DG/HTrans}
\end{eqnarray}
in which we have made use of the integration by parts to collectively treat the corrections. The broken current $j_X^{\mu a}$ is given by 
\begin{eqnarray}
j_X^{\mu a} = \Psi^\dagger \sigma^\mu X^a\Psi\ , \label{JXDef}
\end{eqnarray}
with the covariant derivative of the form
\begin{eqnarray}
D_\mu j_X^{\mu a} = \partial_\mu j_X^a-i\mu_q\delta_{\mu0}\Psi^\dagger[X^a,J] = 
\left\{
\begin{array}{cc}
\partial_\mu j_X^{\mu a} & (a=1-3) \\
\partial_\mu j_X^{\mu 4} +2 \mu_q\delta_{\mu0}j_X^{\mu 5} & (a=4)\\
\partial_\mu j_X^{\mu 5}-2\mu_q\delta_{\mu0}j_X^{\mu 4}  & (a=5) \\
\end{array}
\right.\ .
\end{eqnarray}
Here, let us focus on an arbitrary functional which takes the form of
\begin{eqnarray}
{\cal I}[{\hat{\cal O}}(y)] \equiv \int[d\bar{\psi} d\psi][dA]\hat{\cal O}(y){\rm e}^{i\int d^4x{\cal L}^q_{\rm QC_2D}}\ .\label{IHatO}
\end{eqnarray}
Assuming that the $G/H$ transformation law of the operator $\hat{\cal O}$ reads $\hat{\cal O} \overset{G/H}{\to} \hat{\cal O} + \theta^a\delta^a\hat{\cal O}$, invariance of the functional ${\cal I}[{\hat{\cal O}}(y)]$ yields the following identity for Green's functions:
\begin{eqnarray}
 \left\langle\delta^a\hat{\cal O}\right\rangle \delta^{ab}\delta(x)  = i\Big\langle{\rm T}^*\Big[\partial_\mu j_X^{\mu b}+m_q\left(i{\cal O}_X^b - i{\cal O}_X^{b\dagger}\right) + \frac{j}{\sqrt{2}}\delta^{b5}{\cal O}_{\sigma}\Big](x)\hat{\cal O}(0)\Big\rangle\ . \label{WTIG/H}
\end{eqnarray}
In this identity, we have set $y=0$ and the symbol ${\rm T}^*$ stands for the time-ordering operator but commutes with any derivatives that maintains the explicit Lorentz covariance of the path-integral formulation. Therefore, choosing ${\cal O}_\pi^a$, ${\cal O}_{B_4}$, ${\cal O}_{B_5}$, and ${\cal O}_\sigma$ for $\hat{\cal O}$, from Eq.~(\ref{WTIG/H}) we arrive at the following WTIs:
\begin{eqnarray}
 \left\langle{\cal O}_{\sigma}\right\rangle\delta^{ab} \delta(x) = \sqrt{2} \langle{\rm T}^*i\partial_\mu^xj_X^{\mu b}(x){\cal O}^a_\pi(0)\rangle - i m_q\langle {\rm T}{\cal O}^b_\pi(x){\cal O}^a_\pi(0)\rangle\ \ \ \ (a,b=1-3)\ ,  \label{WTIPi}
\end{eqnarray}
\begin{eqnarray} 
 \left\langle{\cal O}_{\sigma}\right\rangle \delta(x) = \sqrt{2}\langle{\rm T}^* iD_\mu^xj_X^{\mu 4}(x){\cal O}_{B_4}(0)\rangle - i m_q\langle {\rm T}{\cal O}_{B_4}(x){\cal O}_{B_4}(0)\rangle\ ,  \label{WTIB4}
\end{eqnarray}
\begin{eqnarray}
 \left\langle{\cal O}_{\sigma}\right\rangle \delta(x) = \sqrt{2}\langle{\rm T}^*iD^x_\mu j_X^{\mu 5}(x){\cal O}_{B_5}(0)\rangle - i m_q\langle {\rm T}{\cal O}_{B_5}(x){\cal O}_{B_5}(0)\rangle + ij\langle{\rm T}{\cal O}_\sigma(x){\cal O}_{B_5}(0)\rangle\ , \label{WTIB5}
\end{eqnarray}
and
\begin{eqnarray}
-\langle{\cal O}_{B_5}\rangle\delta(x) = \sqrt{2}\langle{\rm T}^*iD_\mu^xj_X^{\mu 5}(x){\cal O}_\sigma(0)\rangle - im_q\langle{\rm T}{\cal O}_{B_5}(x){\cal O}_\sigma(0)\rangle+ ij\langle{\rm T}{\cal O}_\sigma(x){\cal O}_\sigma(0)\rangle\ , \label{WTIB52}
\end{eqnarray}
respectively, with the help of the transformation laws presented in Eqs.~(\ref{XaG/H}) and~(\ref{X0G/H}).

Likewise, the local  $U(1)_B$ transformation law of the Lagrangian~(\ref{LQC2DWTI}) reads
\begin{eqnarray}
{\cal L}^q_{\rm QC_2D} &\overset{U(1)_B}{\to}&  {\cal L}^q_{\rm QC_2D} -\theta_B\left[ \partial_\mu j_B^\mu + 2\sqrt{2} j(i{\cal O}_X^4-i{\cal O}_X^{\dagger4})\right] 
\end{eqnarray}
from $\Psi\to {\rm e}^{-i\theta_B J}\Psi$, with $j_B^\mu = \Psi^\dagger\sigma^\mu J\Psi$ being the baryon-number current. Hence, tracing the same procedure below Eq.~(\ref{IHatO}), 
\begin{eqnarray}
 \left\langle\delta_B\hat{\cal O}\right\rangle \delta(x)  = i\Big\langle{\rm T}^*\Big[\partial_\mu j_B^{\mu} - 2j{\cal O}_{B_4}\Big](x)\hat{\cal O}(0)\Big\rangle
\end{eqnarray}
is derived, where the transformed part $\delta_B\hat{\cal O}$ has been defined through $\hat{\cal O} \overset{U(1)_B}{\to} \hat{\cal O} + \theta_B\delta_B\hat{O}$. Taking ${\cal O}_{B_4}$ for $\hat{\cal O}$ in this identity, one can find the following WTI in terms of $U(1)_B$ symmetry:
\begin{eqnarray}
2\langle{\cal O}_{B_5}\rangle\delta(x) = i\partial_\mu^x\langle {\rm T} j_B^\mu(x){\cal O}_{B_4}(0)\rangle - 2ij\langle{\rm T}{\cal O}_{B_4}(x){\cal O}_{B_4}(0)\rangle \ . \label{WTIBaryon}
\end{eqnarray}

It should be noted that all the WTIs derived in this subsection are valid at any temperature and density since only the symmetry properties of the functional~(\ref{IHatO}) is utilized in the derivations.

\subsection{Gell-Mann-Oakes-Renner relations with the diquark source}
\label{sec:GORRelation}

In Sec.~\ref{sec:WTIQC2D} the WTIs connecting the chiral and diquark condensates to the particular two-point functions have been derived from the appropriate invariance of the path-integral formalism. Here, based on the identities we present the so-called Gell-Mann-Oakes-Renner (GOR) relations~\cite{Cheng:1985bj} in the presence of the diquark source $j$, that is valid as long as we stick to low-energy QC$_2$D.

First, we focus on the pion sector~(\ref{WTIPi}) that is separated from the baryonic sectors and explicit chemical potential effects. By inserting only a pion one-particle state
\begin{eqnarray}
\int\frac{d^3p}{(2\pi)^32E_\pi}|\pi^c({\bm p})\rangle\langle\pi^c({\bm p})| \label{CompetePi}
\end{eqnarray} 
as a part of the complete set ($E_\pi$ is a pion dispersion relation), the first term in the RHS of Eq.~(\ref{WTIPi}) can be simplified to be (${\rm T}$ is the ordinary time-ordering operator)
\begin{eqnarray}
 \langle{\rm T}^*i\partial_\mu^xj_X^{\mu b}(x){\cal O}^a_\pi(0)\rangle \sim {\rm T}\int\frac{d^3p}{(2\pi)^32E_\pi} if_\pi p^2{\cal A}_{\pi\pi}{\rm e}^{-ip_\pi\cdot x}  \ , \label{PiWTIJPi}
\end{eqnarray}
where we have assumed that the amplitude can be evaluated as
\begin{eqnarray}
 \langle0|j_X^{\mu b}(x)|\pi^a({\bm p})\rangle &=& if_\pi p^\mu\delta^{ab}{\rm e}^{-ip_\pi\cdot x} \ , \nonumber\\
\langle0|{\cal O}_\pi^b(x)|\pi^a({\bm p})\rangle &=& {\cal A}_{\pi\pi} \delta^{ab} {\rm e}^{-ip_\pi\cdot x} \ , \label{Amplitudes}
\end{eqnarray}
with a pion decay constant $f_\pi$ and a ${\bm p}$-independent matrix element ${\cal A}_{\pi\pi}$. The momentum $p_\pi^\mu$ is defined by $p_\pi^\mu=(E_\pi,{\bm p})$. Thus, the matrix element~(\ref{PiWTIJPi}) is reduced to 
\begin{eqnarray}
\langle{\rm T}^*i\partial_\mu^xj_X^{\mu b}(x){\cal O}^a_\pi(0)\rangle \sim -\int\frac{d^4p}{(2\pi)^4}\frac{f_\pi p^2{\cal A}_{\pi\pi}}{p_0^2-E_\pi^2} {\rm e}^{-ip\cdot x}\ .
\end{eqnarray}
In a similar way the second term of the RHS of Eq.~(\ref{WTIPi}) reads
\begin{eqnarray}
\langle{\rm T}{\cal O}_\pi^b(x){\cal O}^a_\pi(0)\rangle \sim  {\rm T}\int\frac{d^3p}{(2\pi)^32E_\pi}\left|{\cal A}_{\pi\pi}\right|^2{\rm e}^{-ip_\pi\cdot x} = \int\frac{d^4p}{(2\pi)^4}\frac{i\left|{\cal A}_{\pi\pi}\right|^2}{p_0^2-E_\pi^2} {\rm e}^{-ip\cdot x}\ ,
\end{eqnarray}
and hence one can arrive at (${\cal A}_{\pi\pi}^*={\cal A}_{\pi\pi}$)
\begin{eqnarray}
\langle{\cal O}_\sigma\rangle = -\frac{\sqrt{2}f_\pi p^2{\cal A}_{\pi\pi}}{p_0^2-E_\pi^2}+\frac{m_q{\cal A}_{\pi\pi}^2}{p_0^2-E_\pi^2}\ . \label{PiPiGORWTI}
\end{eqnarray}
When taking $p_0\to 0$ and $p_0\to E_\pi$ after choosing the rest frame ${\bm p}={\bm 0}$, two equations of
\begin{eqnarray}
\langle{\cal O}_\sigma\rangle = -\frac{m_q}{m_\pi^2}{\cal A}_{\pi\pi}^2\ \ , \ \ \ \ 0=-\sqrt{2}f_\pi m_\pi^2{\cal A}_{\pi\pi} +m_q{\cal A}_{\pi\pi}^2\ , \label{ChiralWTI}
\end{eqnarray}
are obtained, with $m_\pi \equiv E_\pi|_{{\bm p}={\bm 0}}$ being the pion mass. Therefore, eliminating the matrix element ${\cal A}_{\pi\pi}$ we finally find
\begin{eqnarray}
f_\pi^2m_\pi^2 = -\frac{m_q\langle{\cal O}_\sigma\rangle}{2}\ , \label{GORPion}
\end{eqnarray}
which is nothing but the familiar GOR relation. The factor $1/2$ in the RHS is due to the normalization of $f_\pi$ in QC$_2$D as will be explained in Sec.~\ref{sec:DecayConstant}. It should be noted that this relation holds at any density as long as the one-pion saturation of the complete set and momentum independence of ${\cal A}_{\pi\pi}$ are reasonably satisfied.

Next, we move on to the baryonic sector. The $\mu_q$ independence of the GOR relation for the pion sector is due to decouplings from the baryonic sector, meanwhile, the baryonic WTIs are easily affected by $\mu_q$ and contaminations from ${\cal O}_\sigma$ owing to the $U(1)_B$ violation too. In order to reduce these difficulties, here we restrict ourselves to $\mu_q=0$. In this case the WTI for ${\cal O}_{B_4}$, Eq.~(\ref{WTIB4}), coincides with the pion one. Besides, due to the charge-conjugation symmetry, ${\cal O}_{B_4}$ is always separated from ${\cal O}_{B_5}$ and ${\cal O}_{\sigma}$.\footnote{Only ${\cal O}_{B_4}$ carries $C=-1$ while ${\cal O}_{B_5}$ and ${\cal O}_\sigma$ carry $C=+1$.} Thus, the GOR relation from Eq.~(\ref{WTIB4}) coincides with Eq.~(\ref{GORPion}), from which the mass of $B_4$ is equal to the pion mass. On the other hand, the WTI for ${\cal O}_{B_5}$ is still complicated due to mixings from ${\cal O}_\sigma$, stemming from the $U(1)_B$ violation. 

The WTIs for ${\cal O}_{B_5}$ and ${\cal O}_{\sigma}$ are combined into a single relation, as shown below. By defining the mass eigenoperators ${\cal O}_{\tilde{B}_5}$ and ${\cal O}_{\tilde{\sigma}}$ through
\begin{eqnarray}
\left(
\begin{array}{c}
{\cal O}_{\tilde{B}_5} \\
{\cal O}_{\tilde{\sigma}} \\
\end{array}
\right)
=
\left(
\begin{array}{cc}
\cos\theta & -\sin\theta \\
\sin\theta & \cos\theta \\
\end{array}
\right)
\left(
\begin{array}{c}
{\cal O}_{{B}_5} \\
{\cal O}_{\sigma} \\
\end{array}
\right)\ ,
\end{eqnarray}
overlaps of the operators ${\cal O}_{B_5}$ and ${\cal O}_{\sigma}$ between $\tilde{B}_5$ state and vacuum can be evaluated to be
\begin{eqnarray}
&& \langle0|{\cal O}_{B_5}|\tilde{B}_5({\bm p})\rangle = \cos\theta \langle0|{\cal O}_{\tilde{B}_5}|\tilde{B}_5({\bm p})\rangle=   {\cal A}_{\tilde{B}_5\tilde{B}_5}\cos\theta \ , \nonumber\\
&& \langle0|{\cal O}_{\sigma}|\tilde{B}_5({\bm p})\rangle = -\sin\theta \langle0|{\cal O}_{\tilde{B}_5}|\tilde{B}_5({\bm p})\rangle=  -  {\cal A}_{\tilde{B}_5\tilde{B}_5}\sin\theta  \ ,
\end{eqnarray}
where ${\cal A}_{\tilde{B}_5\tilde{B}_5}$ has been defined similarly to Eq.~(\ref{Amplitudes}). Then, introducing the decay constant $f_5$ by
\begin{eqnarray}
\langle0| j_X^{\mu 5}(0)|\tilde{B}_5({\bm p})\rangle = if_5 p^\mu \ , \label{f5Def}
\end{eqnarray}
from Eqs.~(\ref{WTIB5}) and~(\ref{WTIB52}) one can derive
\begin{eqnarray}
\langle{\cal O}_\sigma\rangle &=& -\frac{\sqrt{2}f_5p^2{\cal A}_{\tilde{B}_5\tilde{B}_5}\cos\theta}{p_0^2-{E}_{\tilde{B}_5}^2} + \frac{m_q{\cal A}_{\tilde{B}_5\tilde{B}_5}^2\cos^2\theta}{p_0^2-{E}_{\tilde{B}_5}^2}+ \frac{j{\cal A}_{\tilde{B}_5\tilde{B}_5}^2\sin\theta\cos\theta}{p_0^2-{E}_{\tilde{B}_5}^2}\ , \nonumber\\
\langle{\cal O}_{B_5}\rangle &=& -\frac{\sqrt{2}{\cal A}_{\tilde{B}_5\tilde{B}_5}f_5p^2\sin\theta}{p_0^2-{E}_{\tilde{B}_5}^2} + \frac{m_q{\cal A}_{\tilde{B}_5\tilde{B}_5}^2\sin\theta\cos\theta}{p_0^2-{E}_{\tilde{B}_5}^2}  + \frac{j{\cal A}_{\tilde{B}_5\tilde{B}_5}^2\sin^2\theta}{p_0^2-{E}_{\tilde{B}_5}^2} \ , \label{SigmaB5WTI}
\end{eqnarray}
as siblings of Eq.~(\ref{PiPiGORWTI}). We note that all transitions to $|\tilde{\sigma}({\bm p})\rangle$ have been omitted in the derivation since only the (approximate) NG bosons are assumed to saturate the low-energy physics of QC$_2$D. Taking $p_0\to 0$ and $p_0\to m_{\pi}$ at the rest frame ${\bm p}={\bm 0}$ (the mass of $\tilde{B}_5$ is identical to the pion mass) in Eq.~(\ref{SigmaB5WTI}),
\begin{eqnarray}
{\cal A}_{\tilde{B}_5\tilde{B}_5} = -\frac{\langle{\cal O}_\sigma\rangle}{\sqrt{2}f_5\cos\theta} =  -\frac{\langle{\cal O}_{B_5}\rangle}{\sqrt{2}f_5\sin\theta}
\end{eqnarray}
is found, so inserting this relation into either of Eq.~(\ref{SigmaB5WTI}) at $p\to0$ yields
\begin{eqnarray}
f_5^2m_\pi^2 = -\frac{m_q\langle{\cal O}_\sigma\rangle}{2} - \frac{j\langle{\cal O}_{B_5}\rangle}{2}\ .
\end{eqnarray}

Finally, the WTI~(\ref{WTIBaryon}) related to $U(1)_B$ symmetry is easily derived to be 
\begin{eqnarray}
\left(\frac{f_B}{2\sqrt{2}}\right)^2m_{\pi}^2 = -\frac{j\langle{\cal O}_{B_5}\rangle}{2} \ ,
\end{eqnarray}
where the decay constant associated with the baryon-number current $f_B$ has been defined through
\begin{eqnarray}
\langle0| j_B^{\mu}(0)|{B}_4({\bm p})\rangle = if_B p^\mu\ . \label{fBDef}
\end{eqnarray}

In summary, the GOR relation with respect to the broken current $j_X^{\mu a}$ and the $U(1)$ baryon-number current $j_B^\mu$ are obtained as
\begin{eqnarray}
f_\pi^2m_\pi^2 &=& -\frac{m_q\langle\bar{\psi}\psi\rangle}{2} \ \ \ \ \ \ ({\rm at\ any}\ \mu_q)\  , \label{GORPi} \\
f_5^2m_\pi^2 &=& -\frac{m_q\langle\bar{\psi}\psi\rangle}{2} - \frac{j\langle\psi\psi\rangle}{2} \ \ \ \ \ \ ({\rm only\ at}\ \mu_q=0)\ , \label{GORB5} \\
\left(\frac{f_B}{2\sqrt{2}}\right)^2m_{\pi}^2 &=& -\frac{j\langle\psi\psi\rangle}{2} \ \ \ \ \ \ ({\rm only\ at}\ \mu_q=0) \ , \label{GORBaryon}
\end{eqnarray}
where the decay constants are introduced from Eqs.~(\ref{Amplitudes}),~(\ref{f5Def}) and~(\ref{fBDef}). Besides, we have used ${\cal O}_\sigma=\bar{\psi}\psi$ and shorthand notation for the diquark operator ${\cal O}_{B_5}=\psi\psi$. 

\subsection{Comment on the decay constant $f_\pi$}
\label{sec:DecayConstant}

In Sec.~\ref{sec:GORRelation}, the pion decay constant $f_\pi$ has been introduced through the matrix element~(\ref{Amplitudes}) associated with the broken current $j_X^{\mu a}$ and one-pion state $|\pi^a({\bm p})\rangle$. The broken current for the pion sector can be expressed in terms of the ordinary quark field $\psi$ as
\begin{eqnarray}
j_X^{\mu a} = \Psi^\dagger\sigma^\mu X^a\Psi = \frac{1}{\sqrt{2}}\bar{\psi}\gamma^\mu\gamma_5 T_f^a\psi = j_5^{\mu a}\ \ \ (a=1-3)\ ,
\end{eqnarray}
where $T_f^a=\tau_f^a/2$ and the familiar axial current: $j_5^{\mu a} \equiv \bar{\psi}\gamma^\mu\gamma_5 T_f^a\psi$ has been defined. Then, the matrix element is rewritten into
\begin{eqnarray}
\langle0|j_X^{\mu a}(x)|\pi^b({\bm p})\rangle = \frac{1}{\sqrt{2}}\langle0|j_5^{\mu a}(x)|\pi^b({\bm p})\rangle\ . \label{JXJ5Amplitude}
\end{eqnarray}

The decay constant in three-color QCD, $f_\pi^{N_c=3}=93\, {\rm MeV}$, is introduce with respect to the familiar broken current $j_5^{\mu a}$ though
\begin{eqnarray}
\langle0|j_5^{\mu a}(x)|\pi^b({\bm p})\rangle = if^{N_c=3}_\pi p^\mu\delta^{ab}{\rm e}^{-ip_\pi\cdot x} \ .
\end{eqnarray}
Hence, Eq.~(\ref{JXJ5Amplitude}) can be expressed in terms of $f_\pi^{N_c=3}$ as
\begin{eqnarray}
\langle0|j_X^{\mu a}(x)|\pi^b({\bm p})\rangle =  \frac{i}{\sqrt{2}}f^{N_c=3}_\pi p^\mu\delta^{ab}{\rm e}^{-ip_\pi\cdot x} \ ,
\end{eqnarray}
and comparing this equation with the QC$_2$D definition in Eq.~(\ref{Amplitudes}), one can find
\begin{eqnarray}
f_\pi = \frac{1}{\sqrt{2}}f_\pi^{N_c=3}\ . \label{FPiNormalization}
\end{eqnarray}

Equation~(\ref{FPiNormalization}) implies that the decay constant in QC$_2$D is different from the three-color QCD one by a factor $1/\sqrt{2}$. Within chiral effective models such as the ChPT, LSM and NJL model in QC$_2$D, the decay constant is, of course, defined through the broken current associated with the generator $X^a$, which corresponds to not $f_\pi^{N_c=3}$ but $f_\pi$.

\section{Chiral perturbation theory}
\label{sec:ChPT}

\subsection{Model construction based on the Maurer-Cartan $1$-form}
\label{sec:ModelChPT}

Among hadron effective models, the ChPT which treats NG bosons in association with a certain symmetry breaking is one of the powerful and standard models due to its systematic low-energy expansion. Then, in this subsection we explain derivation of the ChPT in QC$_2$D in terms of the so-called Maurer-Cartan $1$-form~\cite{Bando:1987br}.

Let us introduce the following representative which parametrizes the coset space $G/H=SU(4)/Sp(4)$:
\begin{eqnarray}
\xi = {\rm exp}\left(i\pi^aX^a/f_0\right)\ , \label{XiDef}
\end{eqnarray}
where $\pi^a$'s can be regarded as the NG bosons: three pions, diquark and antidiquark. Besides, $f_0$ is a parameter having mass dimension $+1$ which corresponds to the pion decay constant in the lowest-order of ChPT, at vanishing $\mu_q$.

From properties of the coset and representative, one can choose such that $\xi$ defined in Eq.~(\ref{XiDef}) transforms under the global $SU(4)$ transformation as~\cite{Bando:1987br}
\begin{eqnarray}
\xi \to g\xi h^\dagger(g,\pi)\ . \label{XiTrans}
\end{eqnarray}
We note that $h(g,\pi)$ must be a function of $g$ and $\pi$ for which the representative $\xi$ correctly transforms. Here, for later convenience we introduce the Maurer-Cartan 1-form
\begin{eqnarray}
\alpha_\mu \equiv i^{-1}\partial_\mu\xi^\dagger\xi\ . \label{MC1Form}
\end{eqnarray}
This 1-form is indeed useful to construct a Lagrangian from the viewpoint of low-energy expansion since it includes one derivative and the $G$-transformation law is simply generated by $h(g,\pi)$:
\begin{eqnarray}
\alpha_\mu \to h(g,\pi)\alpha_\mu h^\dagger(g,\pi)-i\partial_\mu h(g,\pi)h^\dagger(g,\pi)\ . \label{AlphaTrans}
\end{eqnarray}
The 1-form~(\ref{MC1Form}) generally belongs to the algebra of both ${\cal H}$ and ${\cal G}-{\cal H}$, so that we try to separate them. The decomposition is performed by introducing a sibling of $\xi$ as
\begin{eqnarray}
\tilde{\xi} \equiv E^T{\xi}^* E\ . \label{XiTildeDef}
\end{eqnarray}
In fact, when defining 
\begin{eqnarray}
\alpha_{\perp,\mu} &=& \frac{1}{2i}(\partial_\mu\xi^\dagger\xi-\partial_\mu\tilde{\xi}^\dagger\tilde{\xi}) \ ,\nonumber\\
\alpha_{\parallel,\mu} &=& \frac{1}{2i}(\partial_\mu\xi^\dagger\xi+\partial_\mu\tilde{\xi}^\dagger\tilde{\xi}) \ , \label{AlphaDef}
\end{eqnarray}
so as to satisfy $a_\mu = a_{\perp,\mu} + a_{\parallel,\mu}$, those are expanded as
\begin{eqnarray}
\alpha_{\perp,\mu} = -\frac{\partial_\mu\pi^a}{f_0}X^a + \cdots\ \ , \ \ \ \ \alpha_{\parallel,\mu} = \frac{\partial_\mu\pi^a\pi^b}{2if_0^2}[X^a,X^b] + \cdots\ .
\end{eqnarray}
Here, $[X^a,X^b]E=-E [X^a,X^b]^T$ follows from Eq.~(\ref{XAlgebra}), then, the commutator $[X^a,X^b]$ is understood to belong to the algebra ${\cal H}$ from the definition~(\ref{SAlgebra}). In this way, we can conclude that 
\begin{eqnarray}
\alpha_{\perp,\mu}\in {\cal G}-{\cal H}\ \ \ \ \ {\rm while}\ \ \  \ \ \alpha_{\parallel,\mu}\in {\cal H}\ .
\end{eqnarray}
Besides, the transformation laws of $\alpha_{\perp,\mu}$ and $\alpha_{\parallel,\mu}$ under $G=SU(4)$ read
\begin{eqnarray}
\alpha_{\perp,\mu} &\to& h(g,\pi)\alpha_{\perp,\mu} h^\dagger(g,\pi)\ , \nonumber\\
\alpha_{\parallel,\mu} &\to& h(g,\pi)\alpha_{\parallel,\mu} h^\dagger(g,\pi)-i\partial_\mu h(g,\pi)h^\dagger(g,\pi)\ , \label{AlphaSepTrans}
\end{eqnarray}
respectively, with the help of the following property:
\begin{eqnarray}
\tilde{\xi} \to E^Tg^*\xi^*h^T(g,\pi)E =E^T g^*E\tilde{\xi}h^\dagger(g,\pi) \label{TildeXiTrans}
\end{eqnarray}
following from the algebras in Eqs.~(\ref{SAlgebra}) and~(\ref{XAlgebra}).

Based on the above building blocks, the $SU(4)$-invariant ChPT Lagrangian of ${\cal O}(p^2)$ is constructed as
\begin{eqnarray}
{\cal L}^{{\cal O}(p^2)}_{\rm ChPT} = f_0^2{\rm tr}[\alpha_{\perp,\mu}\alpha^\mu_\perp] + f_0^2{\rm tr}[\hat{\zeta} + \hat{\zeta}^\dagger]\ , \label{ChPTLowest}
\end{eqnarray}
where we have defined
\begin{eqnarray}
\hat{\zeta} = B_0\xi^\dagger \zeta E^T \tilde{\xi}  \label{ZetaHat}
\end{eqnarray}
with $B_0$ being a constant having the mass dimension $+1$, the $SU(4)$ transformation law of which is
\begin{eqnarray}
\hat{\zeta} \to h(g,\pi)\hat{\zeta} h^\dagger(g,\pi)\ .
\end{eqnarray}
The field $\zeta$ in Eq.~(\ref{ZetaHat}) is the spin-$0$ spurion field, which is replaced by its VEV to incorporate, e.g., the quark mass $m_q$ effect in the end, as explained in Sec.~\ref{sec:Spurion}. 


Our main aim to employ the ChPT is to explore low-energy physics of cold and dense QC$_2$D, so that we need to incorporate a quark chemical potential. The chemical potential is introduced systematically by gauging Eq.~(\ref{ChPTLowest}) with respect to $SU(4)$ to incorporate the spin-$1$ spurion field $\zeta_\mu$, and replacing it by the VEV with $\langle v_{\mu=0}^{i=4}\rangle =\mu_q$ from Eqs.~(\ref{SpurionDec}) and~(\ref{SpurionReplace}). Then, in the following analysis we will use 
\begin{eqnarray}
\alpha_{\perp,\mu} = \frac{1}{2i}(D_\mu\xi^\dagger\xi-D_\mu\tilde{\xi}^\dagger\tilde{\xi})
\end{eqnarray}
as the 1-form, where the covariant derivative reads
\begin{eqnarray}
D_\mu\xi^\dagger \equiv \partial_\mu\xi^\dagger+i\xi^\dagger \zeta_\mu\ \ , \ \ \ \ 
D_\mu\tilde{\xi}^\dagger \equiv \partial_\mu\tilde{\xi}^\dagger - i\tilde{\xi}^\dagger E^T\zeta_\mu^TE\ ,  \label{CovariantDXi}
\end{eqnarray}
with $\zeta_\mu \to\langle\zeta_\mu\rangle= \mu_q\delta_{\mu0}J$, from the transformation properties~(\ref{XiTrans}) and~(\ref{TildeXiTrans}). On the other hand, for a while we ignore the diquark source $j$.

\subsection{ChPT in the hadronic phase}
\label{sec:HadronicChPT}

In Sec.~\ref{sec:ModelChPT} we have constructed the ChPT Lagrangian of ${\cal O}(p^2)$ in terms of the Maurer-Cartan $1$-form in Eq.~(\ref{ChPTLowest}). Defining $U=\xi E^T\xi^T$, the Lagrangian is rewritten to
\begin{eqnarray}
{\cal L}^{{\cal O}(p^2)}_{\rm ChPT} = \frac{f_0^2}{4}{\rm tr}[D_\mu U^\dagger D^\mu U ] + {\rm tr}[U\zeta^\dagger + U^\dagger\zeta]\ ,  \label{ChPTOp2}
\end{eqnarray}
which is, indeed, identical to the Lagrangian invented by Kogut~et.~al.~\cite{Kogut:1999iv,Kogut:2000ek}. In this equation the covariant derivative reads
\begin{eqnarray}
D_\mu U  = \partial_\mu U-i\zeta_\mu U-iU \zeta^T_\mu\ . \label{CovariantSigma}
\end{eqnarray}

Setting $\langle\pi^a\rangle=0$, or $\langle\xi\rangle=1$, one finds $\langle U\rangle=E^T$ from its definition. This VEV must be associated with the ground-state configuration of low-energy QC$_2$D, i.e., the chiral condensate $\langle\bar{\psi}\psi\rangle$ in the hadronic phase, as long as the diquark source $j$ is switched off and $\mu_q$ is adequately small. In other words, conceptionally the VEV takes the form of $\langle U\rangle \propto \langle\bar{\psi}\psi\rangle E^T$ for which the remaining $Sp(4)$ symmetry of QC$_2$D is properly built-in. To gain more insights into this structure, we introduce a quark bilinear $\Phi_{ij}$ with flavor indices uncontracted as 
\begin{eqnarray}
\Phi_{ij} \equiv \Psi_j^T\sigma^2\tau_c^2\Psi_i\ . \label{PhiIJ}
\end{eqnarray}
Using the definition~(\ref{PsiDef}), one can easily show that the VEV of a scalar operator $\langle\bar{\psi}\psi\rangle$ can be embedded into $\langle\Phi\rangle$ as $\langle\Phi\rangle=-(1/4) \langle\bar{\psi}\psi\rangle E^T$, and hence, the VEV of $\Phi$ corresponds to the correct ground-state configuration of $\langle U\rangle$: $\langle U\rangle \propto \langle\Phi\rangle $. This fact implies that when $U$ is expanded up to ${\cal O}(\pi)$, its quark-bilinear representation is identical to the linear operator $\Phi$. In this linearization $U$ reads
\begin{eqnarray}
U \approx E^T + \frac{i}{\sqrt{2}f_0}\left(
\begin{array}{cccc}
0 & \pi^5-i\pi^4 &  -\pi^3 & -(\pi^1-i\pi^2) \\
-(\pi^5-i\pi^4) & 0 & -(\pi^1+i\pi^2) & \pi^3 \\
\pi^3 & \pi^1+i\pi^2 & 0 & \pi^5+i\pi^4 \\
\pi^1-i\pi^2 & -\pi^3 & -(\pi^5+i\pi^4 )& 0 \\
\end{array}
\right) \ . \label{SigmaExpansion}
\end{eqnarray}
Meanwhile, pionic and baryonic bilinear operators ($\tau_f^\pm = \tau_f^1\pm i\tau_f^2$)
\begin{eqnarray}
&& \pi^\pm \sim\frac{1}{\sqrt{2}} \bar{\psi}i\gamma_5\tau_f^\mp\psi \ , \ \ \pi^0 \sim \bar{\psi}i\gamma_5\tau_f^3\psi\ , \nonumber\\
&& B \sim -\frac{i}{\sqrt{2}} \psi^TC\gamma_5\tau^2\tau_f^2\psi \ , \ \ \bar{B} \sim -\frac{i}{\sqrt{2}}\psi^\dagger C\gamma_5\tau^2\tau_f^2\psi^*\ ,
\end{eqnarray} 
are involved in $\Phi$ as (regardless of the normalization)
\begin{eqnarray}
\Phi \sim \left(
\begin{array}{cccc}
0 & \sqrt{2}iB &-i\pi^0 & -\sqrt{2}i\pi^+ \\
-\sqrt{2}iB & 0 & -\sqrt{2}i\pi^-& i\pi^0 \\
i\pi^0 & \sqrt{2}i\pi^- & 0 &  \sqrt{2}i\bar{B} \\
\sqrt{2}i\pi^+ & -i\pi^0 & -\sqrt{2}i\bar{B} & 0 \\
\end{array}
\right) \ ,\label{PhiChPT}
\end{eqnarray}
from Eq.~(\ref{PhiIJ}). Therefore, comparing the second term of Eq.~(\ref{SigmaExpansion}) and Eq.~(\ref{PhiChPT}) enables us to identify pions and (anti)diquarks as
\begin{eqnarray}
\pi^\pm = \frac{\pi^1\mp i\pi^2}{\sqrt{2}}\ , \ \ \pi^0 = \pi^3 \ ,\ \ B = \frac{\pi^5-i\pi^4}{\sqrt{2}}\ , \ \ \bar{B} = \frac{\pi^5+i\pi^4}{\sqrt{2}}\  ,
\end{eqnarray}
by choosing normalizations appropriately.
We note that the $E^T$ part in Eq.~(\ref{SigmaExpansion}) simply denotes the vacuum configuration in the hadronic phase: $\langle U\rangle =E^T$.

By reading off the quadratic term of $\pi^a$ ($a=1$ - $3$) in the ChPT Lagrangian~(\ref{ChPTOp2}), the pion masses are derived to be
\begin{eqnarray}
\big(m_\pi^{\rm (H)}\big)^2 = 2B_0 m_q\ , \label{MPiHadron}
\end{eqnarray}
meanwhile, the diquark and antidiquark masses read
\begin{eqnarray}
m_B^{\rm (H)} = m_\pi^{\rm (H)}-2\mu_q \ , \ \ m_{\bar{B}}^{\rm (H)} = m_\pi^{\rm (H)}+2\mu_q\ . \label{BMassHadron}
\end{eqnarray}
In these equations the superscript $({\rm H})$ has been attached to emphasize that the mass formulas are valid only in the hadronic phase.

\subsection{ChPT in the baryon superfluid phase}
\label{sec:SuperfluidityChPT}

The ground-state configuration $\langle U\rangle=E^T$ corresponding to the hadronic phase is indeed realized as a stationary point of the effective potential $V^{{\cal O}(p^2)}_{\rm ChPT} \equiv -\left\langle{\cal L}^{{\cal O}(p^2)}_{\rm ChPT}\right\rangle$, unless the chemical potential is sufficiently large (or $j\neq0$). On the other hand, this configuration is altered for $\mu_q> m_\pi^{\rm (H)}/2$ due to the emergence of diquark condensates, resulting in the baryon superfluid phase. This phase transition is also signaled by the diquark mass; When $\mu_q> m_\pi^{\rm (H)}/2$ the diquark mass turns to negative as seen from Eq.~(\ref{BMassHadron}). In this subsection we explain how the ChPT Lagrangian is modified in the superfluid phase.

In the baryon superfluid phase, the VEV of $\langle U\rangle$ is rotated from $E^T$ to
\begin{eqnarray}
U_\alpha \equiv V_\alpha E^T V_\alpha^T = V^2_\alpha E^T\ \ \ \ {\rm with}\ \ \ \ V_\alpha^2 = {\rm e}^{i\alpha\bar{X}}\ .\label{SigmaAC}
\end{eqnarray}
Here, following Refs.~\cite{Kogut:1999iv,Kogut:2000ek} we employ $\bar{X} = -2\sqrt{2}X^5$ as the rotation axis, in such a way that Eq.~(\ref{SigmaAC}) at sufficiently large $\mu_q$ approaches $U_d \equiv {\rm diag}(\tau_f^2,\tau_f^2)$ and the diquark condensates dominate over the ground state. Thus, deviation from $\alpha=0$ denotes the beginning of the chiral restoration and emergence of the superfluidity. In association with the rotation of the ground-state configuration~(\ref{SigmaAC}), it is useful to rotate the subgroup $H$ so as to keep parametrizing $\pi^a$ as the representative of $G/H$ appropriately. When we define the rotated generators
\begin{eqnarray}
S_\alpha^i = V_\alpha S^i V_\alpha^\dagger\ \ \ , \ \  \ \ X_\alpha^a = V_\alpha X^a V_\alpha^\dagger\ ,
\end{eqnarray}
one can easily show that they satisfy the following algebras of ${\cal H}$ and ${\cal G}-{\cal H}$ correctly:
\begin{eqnarray}
S_\alpha^i U_\alpha = -U_\alpha (S^i)^T \ \ \  ,\ \  \ \ X^a U_\alpha =  U_\alpha (X^a)^T\ ,
\end{eqnarray}
similarly to Eqs.~(\ref{SAlgebra}) and~(\ref{XAlgebra}). Thus, adopting these generators the representative of rotated-$G/H$ is provided by
\begin{eqnarray}
\xi_\alpha \equiv {\rm e}^{i\pi^a X_\alpha^a} = V_\alpha\xi V^\dagger_\alpha\ \ \ \ \  \ \ \ \Big( \xi_\alpha \overset{G}{\to} g\xi_ah_\alpha^\dagger\ \ \ {\rm with}\ \ \ h_\alpha\in H\Big)\ ,
\end{eqnarray}
with $\xi$ being Eq.~(\ref{XiDef}). Similarly, the other important building block in constructing the ChPT Lagrangian is provided by
\begin{eqnarray}
\tilde{\xi}_\alpha \equiv U_\alpha\xi_\alpha^* U_\alpha^\dagger\ \ \ \ \  \ \ \ \Big( \tilde{\xi}_\alpha \overset{G}{\to} U_\alpha g^* U_\alpha^\dagger \tilde{\xi}_\alpha h_\alpha^\dagger\Big)\ ,
\end{eqnarray}
like Eq.~(\ref{XiTildeDef}). With these quantities, tracing the same procedure as in Sec.~\ref{sec:ModelChPT}, one can finally arrive at\footnote{The rotated field $\hat{\zeta}_\alpha$ would be given by $\hat{\zeta}_\alpha = B_0\xi_\alpha^\dagger\zeta \Sigma_\alpha\xi_\alpha$, the $G$-transformation low of which is $\zeta_\alpha\to h_\alpha \zeta_\alpha h_\alpha^\dagger$.}
\begin{eqnarray}
{\cal L}^{{\cal O}(p^2)}_{\rm ChPT} = \frac{f_0^2}{4}{\rm tr}[D_\mu U_\alpha^\dagger D^\mu U_\alpha] + {\rm tr}[U_\alpha\zeta_\alpha^\dagger + U_\alpha^\dagger\zeta_\alpha]\ . \label{ChPTBS} 
\end{eqnarray}
We note that the structure of $G$ is not modified even in the superfluid phase whereas the subgroup $H$ is rotated, and thus the covariant derivative associated with the gauge symmetry of $G$ takes the same form as Eq.~(\ref{CovariantSigma}).

The value of angle $\alpha$ fixing the ground-state configuration is determined by a stationary condition of the potential within the mean-field approximation:
\begin{eqnarray}
V^{{\cal O}(p^2)}_{\rm ChPT} \equiv -\left\langle{\cal L}^{{\cal O}(p^2)}_{\rm ChPT}\right\rangle = 2f_0^2\Big[\mu_q^2(1-\cos2\alpha) + \big(m_\pi^{\rm (H)}\big)^2\cos\alpha\Big]\ ,
\end{eqnarray}
namely, 
\begin{eqnarray}
2\mu_q^2\sin2\alpha = \big(m_\pi^{\rm (H)}\big)^2\sin\alpha\ ,
\end{eqnarray}
which yields ($\mu_{\rm cr} \equiv m_\pi^{\rm (H)}/2$)
\begin{eqnarray}
&\bullet& {\rm for}\ \ \ \mu_q< \mu_{\rm cr}: \ \ \alpha = 0 \ ,\nonumber\\
&\bullet& {\rm for}\ \ \ \mu_{\rm cr}\leq \mu_q: \ \  \cos \alpha=\frac{\big(m_\pi^{\rm (H)}\big)^2}{4\mu_q^2} \ . \label{TwoPhaseChPT}
\end{eqnarray}
The former and latter solutions represent the hadronic and baryon superfluid phases, respectively. Upon this ground state, the ChPT Lagrangian is expanded to be
\begin{eqnarray}
{\cal L}^{{\cal O}(p^2)}_{\rm ChPT} &=& \frac{1}{2}\partial_\mu\pi^a\partial^\mu \pi^a + 2\mu_q\cos\alpha\, (\partial_0\pi^4\pi^5-\pi^4\partial_0\pi^5) \nonumber\\
&& -\sum_{a=1,2,3}\frac{{m}_\pi^2}{2}\pi^a\pi^a  -\frac{{m}_4^2}{2}\pi^4\pi^4-\frac{{m}_5^2}{2}\pi^5\pi^5+ \cdots \ , \label{LHighDense}
\end{eqnarray}
in which we have defined
\begin{eqnarray}
{m}_\pi^2 &=& \big(m_\pi^{\rm (H)}\big)^2\cos\alpha-2\mu_q^2(\cos2\alpha-1) = 4\mu_q^2\ , \nonumber\\
{m}_4^2 &=& \big(m_\pi^{\rm (H)}\big)^2\cos\alpha-2\mu_q^2(\cos2\alpha+1) = 0\ ,\nonumber\\
{m}_5^2 &=& \big(m_\pi^{\rm (H)}\big)^2\cos\alpha-4\mu_q^2\cos2\alpha = 4\mu_q^2-\frac{\big(m_\pi^{\rm (H)}\big)^4}{4\mu_q^2}\ . \label{ChPTMassBS}
\end{eqnarray}
Therefore, the pion mass is found to be simply given by $2\mu_q$ in the superfluid phase. In this phase, the $U(1)$ baryon-number violation leads to a rotated kinetic mixing with $\cos\alpha$ accompanied for diquark and antidiquark states, as shown in the second term in Eq.~(\ref{LHighDense}). 

\begin{figure}[H]
  \begin{center}
    \begin{tabular}{cc}

      \begin{minipage}[c]{0.5\hsize}
       \centering
       \hspace*{-1cm} 
         \includegraphics*[scale=0.22]{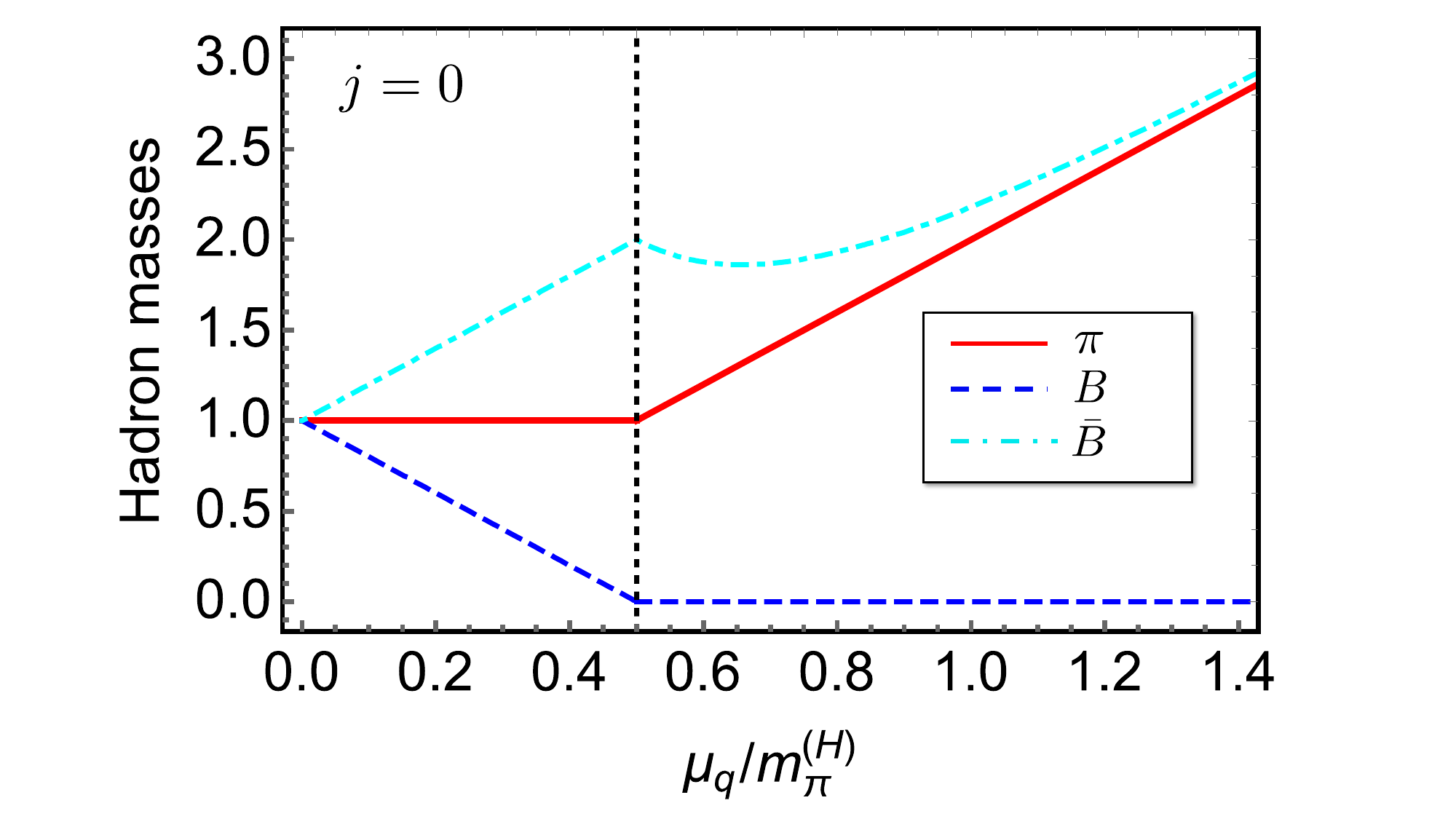}\\
         \end{minipage}

      \begin{minipage}[c]{0.4\hsize}
       \centering
        \hspace*{-1cm} 
          \includegraphics*[scale=0.22]{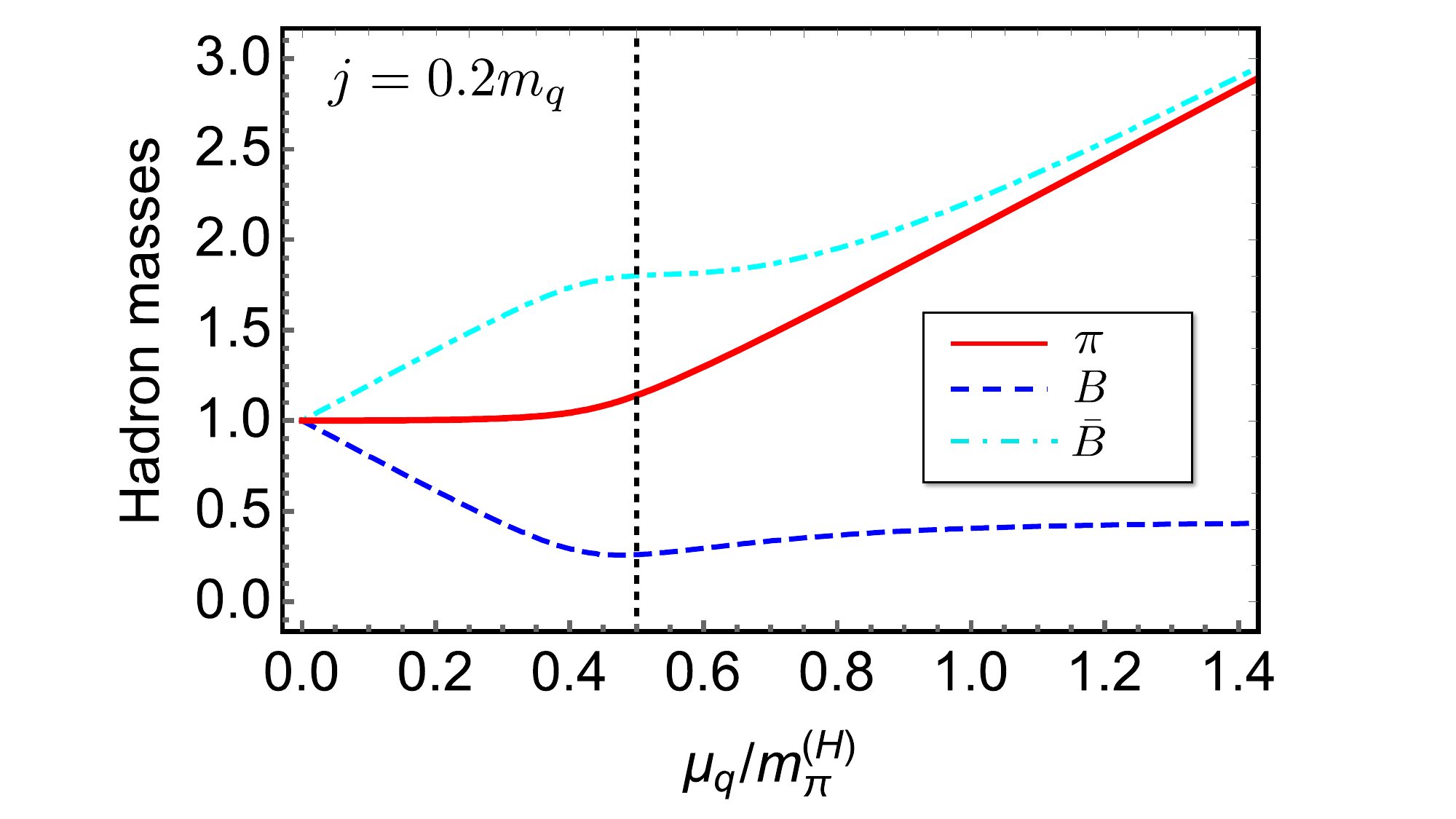}\\
      \end{minipage}

      \end{tabular}
 \caption{$\mu_q$ dependences of the hadron masses evaluated by the ${\cal O}(p^2)$ ChPT with $j=0$ (left) and $j=0.2m_q$ (right).} 
\label{fig:ChPTMass}
  \end{center}
\end{figure}

Based on the derived mass formulas of pions and (anti)diquarks, we can depict $\mu_q$ dependences of the masses predicted by the ${\cal O}(p^2)$ ChPT with vanishing diquark source, in the left panel of Fig.~\ref{fig:ChPTMass}. In the hadronic phase the hadron masses exhibit stable $\mu_q$ dependences as analytically evaluated in Eqs.~(\ref{MPiHadron}) and~(\ref{BMassHadron}). In the baryon superfluid phase, in addition to the monotonic pion mass increment with the formula~(\ref{ChPTMassBS}), the diquark mass is found to be always zero indicating that this state is responsible for the NG boson associated with the breakdown of $U(1)$ baryon-number symmetry~\cite{Kogut:1999iv,Kogut:2000ek}.

\subsection{ChPT with a diquark source $j$}
\label{sec:ChPTWithJ}

In order to achieve the baryon superfluid phase appropriately on the lattice, it is necessary to incorporate a diquark source $j$, then, we take, or extrapolate, $j\to0$ limit in the end. Sometimes this extrapolation is not easily done and the diquark source effects would be remained in the actual lattice simulation. Then, in order to gain insights into the source effect baed on effective models, here we keep $j$ to be finite.

As demonstrated in Sec.~\ref{sec:Spurion} the diquark source $j$ is introduced by replacing the spin-$0$ spurion field as $p^5\to\langle p^5\rangle = j$. Proceeding with this treatment within the ChPT framework, from Eq.~(\ref{ChPTBS}) the phase of the bare mass term is modified as $\cos\alpha \to \cos(\alpha-\phi)$ with $\tan\phi=j/m_q$. Hence, in the presence of $j$ the phase $\alpha$ is not simply fixed by Eq.~(\ref{TwoPhaseChPT}) but is determined by the following modified stationary condition:
\begin{eqnarray}
2\mu_q^2\sin2\alpha = \big(m_\pi^{\rm (H)}\big)^2\sin(\alpha-\phi)\ . \label{GapModifiedChPT}
\end{eqnarray}
This equation implies that $\alpha$ is nonzero when $j\neq0$ even in the vacuum ($\mu_q=0$), namely, the superfluidity always governs the system due to the explicit $U(1)_B$ symmetry breaking.

With finite $j$, the NG boson masses read
\begin{eqnarray}
{m}_\pi^2 &=& \bar{m}_\pi^2\cos(\alpha-\phi)-2\mu_q^2(\cos2\alpha-1) = \frac{\cos\phi}{\cos\alpha}\bar{m}_\pi^2\ ,\nonumber\\
{m}_4^2 &=& \bar{m}_\pi^2\cos(\alpha-\phi)-2\mu_q^2(\cos2\alpha+1) = \frac{\sin\phi}{\sin\alpha}\bar{m}_\pi^2 \ ,\nonumber\\
{m}_5^2 &=& \bar{m}_\pi^2\cos(\alpha-\phi)-4\mu_q^2\cos2\alpha = \left(\frac{\cos^2\alpha}{\sin\alpha}\sin\phi + \frac{\sin^2\alpha}{\cos\alpha}\cos\phi\right)\bar{m}_\pi^2\ , \nonumber\\
\label{NGMassBare}
\end{eqnarray}
where $\bar{m}^2_\pi \equiv 2B_0\sqrt{m_q^2+j^2}$ is the vacuum pion mass. In these equations we have made use of the corrected stationary condition~(\ref{GapModifiedChPT}) to find the right-most expressions. When we take $j=0.2m_q$ as a demonstration, the hadron mass spectra at finite $\mu_q$ is obtained as depicted in the right panel of Fig.~\ref{fig:ChPTMass}, where $\mu_q$ dependences are slightly modified. In particular, the NG mode disappears reflecting the explicit $U(1)_B$ symmetry violation.
 

From the matching~(\ref{MatchingCondensate}), the chiral and diquark condensates can be evaluated within the ChPT as
\begin{eqnarray}
\langle\bar{\psi}\psi\rangle &=& \frac{\partial{\cal L}^{{\cal O}(p^2)}_{\rm ChPT}}{-\partial s^0}\Big|_{\langle\zeta\rangle,\langle\zeta_\mu\rangle} = -4Gf_0^2\cos\alpha\ , \nonumber\\
\langle{\psi}\psi\rangle &=& \frac{\partial{\cal L}^{{\cal O}(p^2)}_{\rm ChPT}}{-\partial p^0}\Big|_{\langle\zeta\rangle,\langle\zeta_\mu\rangle} = -4Gf_0^2\sin\alpha\ ,  \label{CondensateChPT}
\end{eqnarray}
respectively. This expression is universal to any value of $\mu_q$.

The broken current within the ChPT is evaluated by taking a derivative of ${\cal L}^{{\cal O}(p^2)}_{\rm ChPT}$ with respect to $\zeta_X^{\mu a} \equiv -2\sqrt{2}V_\mu'^a $ and setting Eq.~(\ref{SpurionReplace}), resulting in
\begin{eqnarray}
j_{X\mu}^a &=& -f_0\cos\alpha\, \partial_\mu\pi^a + \cdots\ \ \ \ \ \ ({\rm for}\ \ a=1-3 )\ ,\nonumber\\
j_{X\mu}^4 &=& -f_0\cos\alpha\partial_\mu\pi^4 -2f_0\mu_q\delta_{\mu0}\cos2\alpha\,  \pi^5 + \cdots\ \ \ \ \ \ ({\rm for}\ \ a=4 )\ ,\nonumber\\
j_{X\mu}^5 &=& -f_0\partial_\mu\pi^5+2f_0\mu_q\delta_{\mu0}\cos\alpha\,  \pi^4 + \cdots\ \ \ \ \ \ ({\rm for}\ \ a=5 )\ . \label{jPiChPTEach}
\end{eqnarray}
Similarly the $U(1)_B$ current is calculated to be
\begin{eqnarray}
j_{B\mu} = -2\sqrt{2}f_\pi \sin\alpha  \partial_\mu\pi^4-4\sqrt{2}f_\pi\mu_q\delta_{\mu0}\sin2\alpha\, \pi^5 + \cdots\ ,
\end{eqnarray}
by taking a derivative with respect to $\zeta_B^\mu \equiv V^{\mu, i=4}$. From the pionic sector in Eq.~(\ref{jPiChPTEach}) one can easily see
\begin{eqnarray}
f_\pi = f_0\cos\alpha
\end{eqnarray}
at any $\mu_q$, from the argument in Sec.~\ref{sec:GORRelation}. Thus, utilizing the pion mass formula in Eq.~(\ref{NGMassBare}) together with the chiral condensate~(\ref{CondensateChPT}), the GOR relation~(\ref{GORPi}) is readily confirmed.

As for the baryonic sector, when taking $\mu_q=0$ so as to eliminate difficulties due to the mixings, the decay constants associated with the baryonic broken current and $U(1)_B$ current read
\begin{eqnarray}
f_5 &=& f_0 \ \ \ \ \ (a=5) \ , \label{FaChPT}
\end{eqnarray}
and
\begin{eqnarray}
f_B = 2\sqrt{2}f_\pi\sin\alpha \ ,\label{FBChPT}
\end{eqnarray} 
respectively. Thus, using the mass formulas~(\ref{NGMassBare}) and decay constants~(\ref{FaChPT}) and~(\ref{FBChPT}), we can easily verify that the GOR relations~(\ref{GORB5}) and~(\ref{GORBaryon}) certainly hold within the ChPT.\footnote{it seems that $\left(\frac{f_B}{2\sqrt{2}}\right)^2m_4^2=-\frac{\langle\psi\psi\rangle}{2}$ holds at any $\mu_q$.}


\subsection{Thermodynamic properties}
\label{sec:Thermodynamic}

The ChPT Lagrangian in the hadronic and superfluid phases have been derived in Sec~\ref{sec:HadronicChPT} and Sec~\ref{sec:SuperfluidityChPT}, respectively, and thus we are now ready to evaluate thermodynamic properties such as pressure, energy density and sound velocity. Here we exhibit $\mu_q$ dependences of those quantities with vanishing $j$~\cite{Hands:2006ve,Son:2000by}.

From the Lagrangians~(\ref{ChPTOp2}) and~(\ref{ChPTBS}), the pressure $p=\langle{\cal L}\rangle$ in the hadronic and superfluid phases are evaluated to be
\begin{eqnarray}
p_{\rm ChPT}^{{\rm (H)}} &=& 2f_0^2\big(m_\pi^{\rm (H)}\big)^2\ ,  \nonumber\\
p_{\rm ChPT}^{{\rm (BS)}} &=& f_0^2\big(m_\pi^{\rm (H)}\big)^2\left(\bar{\mu}^2 + \frac{1}{\bar{\mu}^2}\right)\ ,
\end{eqnarray}
respectively, where $\bar{\mu} = \mu_q/\mu_{\rm cr} = 2\mu_q/m_\pi^{\rm (H)}$. The stability of the vacuum ($\mu_q=0$) requires that the vacuum pressure must be zero, and thus the correct pressure in the superfluid phase is the following subtracted one:
\begin{eqnarray}
p_{\rm ChPT}^{\rm sub} \equiv p_{\rm ChPT}^{{\rm (BS)}} - p_{\rm ChPT}^{{\rm (H)}} = f_0^2\big(m_\pi^{\rm (H)}\big)^2\left(\bar{\mu} - \frac{1}{\bar{\mu}}\right)^2\ . \label{PressureChPT}
\end{eqnarray}
With this subtracted pressure the baryon-number density and baryon susceptibility are derived to be
\begin{eqnarray}
n_{\rm ChPT} &=& \frac{\partial p_{\rm ChPT}^{\rm sub}}{\partial\mu_q} = \frac{2f_0^2\big(m_\pi^{\rm (H)}\big)^2}{\mu_q}\left(\bar{\mu}^2 - \frac{1}{\bar{\mu}^2}\right)\ , \nonumber\\
\chi_{\rm ChPT} &=& \frac{\partial^2 p_{\rm ChPT}^{\rm sub}}{\partial\mu_q^2} = 8f_0^2\left(1+\frac{3}{\bar{\mu}^4}\right)\ , \label{DensityChPT}
\end{eqnarray}
respectively. Moreover, the (subtracted) energy density is also straightforwardly evaluated as
\begin{eqnarray}
\epsilon_{\rm ChPT}^{\rm sub}  = -p_{\rm ChPT}^{\rm sub}+\mu_q n_{\rm ChPT} = f_0^2\big(m_\pi^{\rm (H)}\big)^2\frac{(\bar{\mu}^2+3)(\bar{\mu}^2-1)}{\bar{\mu}^2}\ .
\end{eqnarray}

Another significant quantity which characterizes a dense matter is the (squared) sound velocity $c_s^2 = \partial p/\partial \epsilon$ defined along the isentropic trajectory. As long as we stick to zero temperature, the trajectory is identical to $T=0$ line and the sound velocity is simply evaluated by 
\begin{eqnarray}
c_s^2 = \frac{n}{\mu_q\chi}\ . \label{SoundVelocityFormula}
\end{eqnarray}
 Hence, from Eq.~(\ref{DensityChPT}) one can find
\begin{eqnarray}
\big(c_s^{\rm ChPT}\big)^2 = \frac{n_{\rm ChPT}}{\mu_q\chi_{\rm ChPT}} = \frac{1-1/\bar{\mu}^4}{1+3/\bar{\mu}^4}\ .
\end{eqnarray}
This formula will be used in Sec.~\ref{sec:SoundVelocity} to see a difference between the ChPT and LSM results focusing on the bulk structure of dense QC$_2$D.

\subsection{Hidden local symmetry}
\label{sec:HLS}

The ChPT is capable of describing NG boson dynamics based on a systematic low-energy expansion, since the theory is constructed upon the Maurer-Cartan $1$-form~(\ref{MC1Form}) including a derivative. However, the expansion cannot converge as the energy scale is increased due to appearances of other hadronic modes. Among them, spin-$1$ hadrons such as $\rho$ mesons and axialvector diquarks can also be treated in the systematic-expansion scheme as an extension of the ChPT, by regarding them as gauge bosons associated with the subgroup $H$. This systematic treatment of the spin-$1$ hadrons is called {\it hidden local symmetry (HLS)} technique~\cite{Harada:2003jx}. In this subsection we briefly review how the HLS extension is done in the ChPT of QC$_2$D. For a detailed argument please see Ref.~\cite{Harada:2010vy}.

In the decomposition of $\Sigma=\xi E^T\xi^T$, one can find redundant degrees of freedom $\sigma^i$ incorporated via
\begin{eqnarray}
\xi= \xi(\pi) \xi(\sigma) \ \  \ \ \ {\rm with}\ \ \ \ \ \xi(\pi) = {\rm e}^{i\pi^aX^a/f_\pi}\ \ \ {\rm and}\ \ \ \xi(\sigma) = {\rm e}^{i\sigma^iS^i/f_\sigma}\ ,
\end{eqnarray}
which is hidden in $\Sigma$ because of Eq.~(\ref{SAlgebra}). These secret fields can be identified as NG bosons of the spontaneous breakdown of $H_{\rm local} = [Sp(4)]_{\rm local}$. In other words, now the whole symmetry is extended from $SU(4)$ to $SU(4)\times[Sp(4)]_{\rm local}$, and $\xi(\pi)$ and $\xi(\sigma)$ transform as 
\begin{eqnarray}
\xi(\pi) \to g \xi(\pi)h^\dagger(x)\ \ , \ \ \ \xi(\sigma) \to h(x)\xi(\sigma)h^\dagger(x)\ , \label{XiHLS}
\end{eqnarray}
respectively. Accordingly, the gauge bosons associated with $H_{\rm local} = [Sp(4)]_{\rm local}$, $V_\mu$, which transform as
\begin{eqnarray}
V_\mu \to h(x)V_\mu h^\dagger(x)-i\partial_\mu h(x)h^\dagger(x)\ ,
\end{eqnarray}
join the low-energy spectrum. This $V_\mu$ belongs to the algebra of ${\cal H}$ containing $10$ degrees of freedom: $V_\mu = V_\mu^i S^i$. They are corresponding to three $\rho$ mesons, one $\omega$ meson, three axialvector diquark baryons, and three axialvector antidiquark baryons.

With the transformation laws~(\ref{XiHLS}), we only need to change $h(g,\pi)$ to $h(x)$ in the formulas~(\ref{XiTrans}) and~(\ref{TildeXiTrans}). Therefore, when we define
\begin{eqnarray}
\hat{\alpha}_{\perp,\mu} &\equiv&  \frac{1}{2i}({\cal D}_\mu\xi^\dagger\xi-{\cal D}_\mu\tilde{\xi}^\dagger\tilde{\xi}) \ ,\nonumber\\
\hat{\alpha}_{\parallel,\mu} &\equiv& \frac{1}{2i}({\cal D}_\mu\xi^\dagger\xi+{\cal D}_\mu\tilde{\xi}^\dagger\tilde{\xi}) \ ,
\end{eqnarray}
as extensions of Eq.~(\ref{CovariantDXi}) with the covariant derivatives
\begin{eqnarray}
{\cal D}_\mu\xi^\dagger &=& \partial_\mu\xi^\dagger-iV_\mu\xi^\dagger+i\xi^\dagger \zeta_\mu\ , \nonumber\\
{\cal D}_\mu\tilde{\xi}^\dagger &\equiv& \partial_\mu\tilde{\xi}^\dagger-iV_\mu\tilde{\xi}^\dagger - i\tilde{\xi}^\dagger E^T\zeta_\mu^TE\ , 
\end{eqnarray}
one can easily check 
\begin{eqnarray}
\hat{\alpha}_{\perp,\mu} \to h(x)\hat{\alpha}_{\perp,\mu} h^\dagger(x) \ \ , \ \ \ \hat{\alpha}_{\parallel,\mu} \to h(x)\hat{\alpha}_{\parallel,\mu} h^\dagger(x)\ ,
\end{eqnarray}
and the HLS Lagrangian is readily constructed as
\begin{eqnarray}
{\cal L}_{\rm HLS}^{{\cal O}(p^2)} = - \frac{1}{2g_\rho^2}{\rm tr}[V_{\mu\nu}V^{\mu\nu}] + f_\pi^2{\rm tr}[\hat{\alpha}_{\perp,\mu}\hat{\alpha}_\perp^\mu] + f_\sigma^2{\rm tr}[\hat{\alpha}_{\parallel,\mu}\hat{\alpha}_\parallel^\mu]  + f_\pi^2{\rm tr}[\hat{\zeta} + \hat{\zeta}^\dagger]\ . \label{HLSLagrangian}
\end{eqnarray}
In this Lagrangian the (dressed) spurion field transforms as $\hat{\zeta} \to h(x)\hat{\zeta} h^\dagger(x)$, and we have incorporated the kinetic term of the vector bosons from their field strength
\begin{eqnarray}
V_{\mu\nu} = \partial_\mu V_\nu-\partial_\nu V_\mu - i[V_\mu,V_\nu]\ ,
\end{eqnarray}
with an HLS-gauge coupling $g_\rho$. Within the unitary gauge the NG bosons are simply absorbed by the longitudinal modes of the vector bosons, leading to $\sigma^i=0$.

The HLS Lagrangian~(\ref{HLSLagrangian}) only includes ${\cal O}(p^2)$ contributions. The ${\cal O}(p^4)$ terms are listed in Ref.~\cite{Harada:2010vy} and their contributions to spin-$1$ hadron masses at finite $\mu_q$ is also explored in this literature.

\section{Linear sigma model}
\label{sec:LSM}

The ChPT which describes five NG bosons: three pions, diquark and antidiquark, has been reviewed in Sec.~\ref{sec:ChPT}, as the low-energy effective model of QC$_2$D. One way to extend the model to incorporate spin-$1$ bosons systematically based on the HLS technique has also been shortly explained in Sec.~\ref{sec:HLS}. Although those frameworks are powerful thanks to their systematic expansion with the power counting, we know that QC$_2$D involves other light excitations, e.g., the scalar mesons and negative-parity diquark baryons that cannot be treated by those models. Lattice simulations have been, indeed, measuring those hadrons. In particular, the recent lattice simulation claims that in the superfluid phase, there exists an iso-singlet $0^-$ mode as the second-lightest hadron, which is lighter than the pions~\cite{Murakami:2022lmq}. This fact implies that the ChPT is no longer a correct low-energy effective model of dense QC$_2$D, so that it is inevitable to construct another effective model which is capable of describing such hadrons as well based on the Pauli-G\"ursey $SU(4)$ symmetry. Then, in this section, we introduce the LSM pursuant to the linear representation of the $SU(4)$ symmetry treating $0^\pm$ mesons and diquark baryons comprehensively.

\subsection{Model construction}
\label{sec:ConstructionLSM}

In Eq.~(\ref{PhiIJ}) the following $4\times4$-matrix bilinear operator made of $\Psi$:
\begin{eqnarray}
\Phi_{ij} = \Psi_j^T\sigma^2\tau_c^2\Psi_i\ , \label{PhiIJ2}
\end{eqnarray}
has been introduced to understand the bilinear representation of NG boson $\pi^a$'s, which are incorporated nonlinearly within the ChPT framework. Meanwhile, the bilinear operator $\Phi$ contains $12$ degrees of freedom as real numbers since $\Phi=-\Phi^T$, implying that we can assign $12$ hadronic states to parametrize $\Phi$ maximumly when employing the linear representation of the Pauli-G\"ursey $SU(4)$ symmetry. Thus, the following $12$ hadron fields:
\begin{eqnarray}
&& \eta \sim \bar{\psi}i\gamma_5\psi\ , \ \ \pi^\pm \sim\frac{1}{\sqrt{2}} \bar{\psi}i\gamma_5\tau_f^\mp\psi \ , \ \ \pi^0 \sim \bar{\psi}i\gamma_5\tau_f^3\psi\ , \ \  \sigma \sim \bar{\psi}\psi\ , \ \ a_0^\pm \sim\frac{1}{\sqrt{2}} \bar{\psi}\tau_f^\mp\psi \ ,  \nonumber\\
&& a_0^0 \sim \bar{\psi}\tau_f^3\psi\ , \ \ B \sim -\frac{i}{\sqrt{2}} \psi^TC\gamma_5\tau^2\tau_f^2\psi \ , \ \ \bar{B} \sim -\frac{i}{\sqrt{2}}\psi^\dagger C\gamma_5\tau^2\tau_f^2\psi^*\ , \nonumber\\
&& B' \sim- \frac{1}{\sqrt{2}} \psi^TC\tau^2\tau_f^2\psi \ , \ \ \bar{B}' \sim \frac{1}{\sqrt{2}}\psi^\dagger C\tau^2\tau_f^2\psi^*\ ,
\end{eqnarray} 
can be embedded into $\Phi$ as
\begin{eqnarray}
\Phi\sim \Sigma \equiv \frac{1}{2}\left(
\begin{array}{cccc}
0 & -B'+iB &\frac{\sigma-i\eta+a_0^0-i\pi^0}{\sqrt{2}} & a_0^+-i\pi^+ \\
 B'-iB & 0 & a_0^--i\pi^- & \frac{\sigma-i\eta-a_0^0+i\pi^0}{\sqrt{2}} \\
-\frac{\sigma-i\eta+a_0^0-i\pi^0}{\sqrt{2}} & -a_0^-+i\pi^- & 0 & - \bar{B}' + i\bar{B}\\
-a_0^++i\pi^+ &-\frac{\sigma-i\eta-a_0^0+i\pi^0}{\sqrt{2}} & \bar{B}'-i\bar{B} & 0 \\
\end{array}
\right) \ . \label{PhiPAssign}
\end{eqnarray}
In this equation $\Sigma$ has been defined as a mass-dimension $+1$ matrix with a normalization factor $1/2$ chosen for later convenience. The matrix~(\ref{PhiPAssign}) is expressed concisely in terms of generators $X^a$ together with the symplectic matrix $E$ as
\begin{eqnarray}
\Sigma = ({\cal S}^a - i{\cal P}^a)X^aE\ ,
\end{eqnarray}
where ${\cal S}^a$'s and ${\cal P}^a$'s ($a=0$ - $5$) are related to the hadron fields by
\begin{eqnarray}
&& \eta = {\cal P}^0\ , \ \  \pi^\pm=\frac{{\cal P}^1\mp i{\cal P}^2}{\sqrt{2}} \ , \ \  \pi^0 = {\cal P}^3\ , \ \  B = \frac{{\cal P}^5-i{\cal P}^4}{\sqrt{2}}\ , \ \ \bar{B} =  \frac{{\cal P}^5+i{\cal P}^4}{\sqrt{2}} \ , \nonumber\\
&&\sigma = {\cal S}^0\ , \ \ a_0^\pm=\frac{{\cal S}^1\mp i{\cal S}^2}{\sqrt{2}} \ , \ \ a_0^0 = {\cal S}^3\ , \ \  B' = \frac{{\cal S}^5-i{\cal S}^4}{\sqrt{2}}\ , \ \ \bar{B}' =  \frac{{\cal S}^5+i{\cal S}^4}{\sqrt{2}} \ . \label{Spin0Hadrons}
\end{eqnarray}
Quantum numbers of these spin-$0$ hadrons are tabulated in Table~\ref{tab:spin0}. 

\begin{table}[htbp]
\caption{Quantum numbers of the hadrons in Eq.~(\ref{Spin0Hadrons}). }
\label{tab:spin0}
\begin{center}
  \begin{tabular}{c||c|c|c} \hline\hline
\textbf{Hadrons} & \textbf{Spin and parity ($J^P$)} & \textbf{Quark number} & \textbf{Isospin}\\ \hline
$\eta$ & $0^-$ & $0$ & $0$\\
$\pi$ & $0^-$ & $0$ & $1$ \\
$\sigma$ & $0^+$ & $0$ & $0$ \\
$a_0$ & $0^+$ & $0$ & $1$ \\
$B$ ($\bar{B}$) & $0^+$ & $+2$ ($-2$) & $0$ \\
$B'$ ($\bar{B}'$) & $0^-$ & $+2$ ($-2$) & $0$ \\ \hline \hline
\end{tabular}
\end{center}
\end{table}

From the interpolating field~(\ref{PhiIJ2}) the $SU(4)$ transformation law of $\Sigma$ is readily understood to be
\begin{eqnarray}
\Sigma \to g\Sigma g^T \ \ \ {\rm with}\ \ \  g\in SU(4)\ . \label{SigmaSU4}
\end{eqnarray}
Thus one can construct an LSM Lagrangian preserving the $SU(4)$ symmetry as~\cite{Suenaga:2022uqn}
\begin{eqnarray}
{\cal L}_{\rm LSM} &=& {\rm tr}[D_\mu \Sigma^\dagger D^\mu\Sigma]-m_0^2{\rm tr}[\Sigma^\dagger\Sigma]-\lambda_1\big({\rm tr}[\Sigma^\dagger\Sigma]\big)^2-\lambda_2{\rm tr}[(\Sigma^\dagger\Sigma)^2] \nonumber\\
&& +\bar{c}\, {\rm tr}[\zeta^\dagger\Sigma+\Sigma^\dagger \zeta] + {\cal L}_{\rm anom.}\ ,
 \label{LSMTwoColor}
\end{eqnarray}
where ${\cal L}_{\rm anom.}$ is responsible for the $U(1)_A$ anomaly of QC$_2$D, which generally takes the form of
\begin{eqnarray}
{\cal L}_{\rm anom.} = \frac{a}{2}{\rm tr}[\tilde{\Sigma}\Sigma + \tilde{\Sigma}^\dagger\Sigma^\dagger] + \frac{c_1}{4}\big({\rm tr}[\tilde{\Sigma}\Sigma + \tilde{\Sigma}^\dagger\Sigma^\dagger] \big)^2 + \frac{c_2}{2}{\rm tr}[\Sigma^\dagger\Sigma]{\rm tr}[\tilde{\Sigma}\Sigma + \tilde{\Sigma}^\dagger\Sigma^\dagger] \ ,
\end{eqnarray}
with $\tilde{\Sigma}_{ij} = \frac{1}{2}\epsilon_{ijkl}\Sigma_{kl}$. These terms indeed break $U(1)_A$ symmetry with which the $U(1)_A$ transformation of $\Sigma$ is simply generated by $\Sigma \to {\rm e}^{-i\theta_A }\Sigma {\rm e}^{-i\theta_A}={\rm e}^{-2i\theta_A }\Sigma$. In Eq.~(\ref{LSMTwoColor}) the covariant derivative is defined by
\begin{eqnarray}
D_\mu\Sigma = \partial_\mu\Sigma-i\zeta_\mu\Sigma-i\Sigma\zeta_\mu^T\ ,
\end{eqnarray}
and the spurion fields $\zeta$ and $\zeta_\mu$ exhibit transformation laws of
\begin{eqnarray}
\zeta \to g\zeta g^T , \ \  \zeta_\mu \to g\zeta_\mu g^\dagger-i\partial_\mu gg^\dagger  \ ,
\end{eqnarray}
which are, of course, the same as the ones introduced in QC$_2$D Lagrangian in Eq.~(\ref{QC2DSourceRed}). In the end we replace them by the VEVs in Eq.~(\ref{SpurionReplace}) with Eq.~(\ref{SpurionDec}) to take into account the quark mass, diquark source and chemical potential effects. We note that ${\rm det}\Sigma + {\rm det}\Sigma^\dagger$ term for the anomaly effects adopted in Ref.~\cite{Suenaga:2022uqn} is obtained from the $c_1$ term with the help of the following identity:
\begin{eqnarray}
\big({\rm tr}[\tilde{\Sigma}\Sigma + \tilde{\Sigma}^\dagger\Sigma^\dagger] \big)^2 = -8{\rm tr}[(\Sigma^\dagger\Sigma)^2] + 4 \big({\rm tr}[\Sigma^\dagger\Sigma]\big)^2 + 16({\rm det}\Sigma + {\rm det}\Sigma^\dagger)\ .
\end{eqnarray}

The latest lattice result where disconnected diagrams are also included seems to imply that $m_\eta^{\rm (H)}/m_\pi^{\rm (H)}$ is close to unity, and the $U(1)_A$ anomaly effects may be suppressed at least in the vacuum~\cite{Murakami2022}. Hence, in the following arguments we will ignore the anomalous contributions: $a=c_1=c_2=0$, otherwise stated.



\subsection{Phase structure from the LSM}
\label{Sec:HadronMassLSM}

As in the ChPT analysis the current LSM undergoes a phase transition to the baryon superfluid phase driven by the emergence of diquark condensates. Unlike the ChPT, within the LSM based on the linear representation such effects can be represented directly by a mean field of the positive-parity diquark baryon. Hence, here we consider 
\begin{eqnarray}
\sigma_0 \equiv \langle \sigma\rangle\ ,\ \ \Delta \equiv \langle {\cal P}^5\rangle\ ,
\end{eqnarray}
for the chiral condensate and diquark condensate. From the formula~(\ref{MatchingCondensate}) matching underlying QC$_2$D, diquark condensates within the LSM are evaluated to be
\begin{eqnarray}
\langle\bar{\psi}\psi\rangle &=& \frac{\partial{\cal L}_{\rm LSM}}{-\partial s^0}\Big|_{\langle\zeta\rangle,\langle\zeta_\mu\rangle} = -\sqrt{2}\bar{c}\sigma_0\ , \nonumber\\
\langle{\psi}\psi\rangle &=& \frac{\partial{\cal L}_{\rm LSM}}{-\partial p^0}\Big|_{\langle\zeta\rangle,\langle\zeta_\mu\rangle} = -\sqrt{2}\bar{c}\Delta\ . \label{CondensateLSM}
\end{eqnarray}

At the mean-field level the effective potential takes the form of
\begin{eqnarray}
V_{\rm LSM}^{\rm eff} = -2\mu_q^2\Delta^2 + \frac{m_0^2}{2}(\Delta^2+\sigma_0^2)+\frac{\tilde{\lambda}}{4}(\sigma_0^2+\Delta^2)^2 -\sqrt{2}\bar{c}(m_q\sigma_0 + j\Delta) \ , \label{VEffLSM}
\end{eqnarray}
where $\tilde{\lambda} = (4\lambda_1+\lambda_2)/4$. The phase structures, i.e., $\mu_q$ dependences of $\sigma_0$ and $\Delta$ are determined by finding stationary points of this potential with respect to these mean fields:
\begin{eqnarray}
\left(m_\pi^2-\frac{\sqrt{2}\bar{c}m_q}{\sigma_0}\right)\sigma_0 = 0 \  \ , \ \  \ \left(m_\pi^2-4\mu_q^2-\frac{\sqrt{2}\bar{c}j}{\Delta}\right)\Delta = 0 \ ,\label{GapEqSLSM}
\end{eqnarray}
from which the pion mass at any $\mu_q$ reads
\begin{eqnarray}
m_\pi^2 = m_0^2 +\tilde{\lambda}(\sigma_0^2+\Delta^2)  \label{MPiLSM}
\end{eqnarray}
by expanding the Lagrangian~(\ref{LSMTwoColor}) upon $\sigma_0$ and $\Delta$. Here, let us take $j=0$ to exclude the diquark condensates in the vacuum. In this case, solving the gap equations yields
\begin{eqnarray}
&\bullet& {\rm for}\ \ \ \mu_q< \mu_{\rm cr}: \ \  \sigma_0=\sigma_0^{\rm (H)} = {\rm (constant)}\ , \ \ \Delta= 0\ . \nonumber\\
&\bullet&{\rm for}\ \ \ \mu_{\rm cr }\leq \mu_q: \ \  \sigma_0 = \frac{m_q\bar{c}}{2\sqrt{2}}\mu_q^{-2}\ , \ \ \Delta = \left[\big(\sigma_0^{\rm (H)}\big)^2 -\sigma_0^2+ \frac{1}{\tilde{\lambda}}\Big(4\mu_q^2-\big(m^{\rm (H)}_\pi\big)^2\Big)\right]^{1/2} \ , \nonumber\\ \label{PhaseLSM}
\end{eqnarray}
where $m_\pi^{\rm (H)}=\sqrt{2}m_q\bar{c}/\sigma_0^{\rm (H)}$ is the pion mass in the hadronic phase. Thus, the critical chemical potential $\mu_{\rm cr} = m_\pi^{\rm (H)}/2$ separating the hadronic and baryon superfluid phases is identical to the one found in the ChPT framework. It should be noted that the NJL model analysis also derives the same $\mu_{\rm cr}$~\cite{Ratti:2004ra}.

\begin{figure}[H]
  \begin{center}
    \begin{tabular}{cc}

      \begin{minipage}[c]{0.5\hsize}
       \centering
       \hspace*{-1cm} 
         \includegraphics*[scale=0.22]{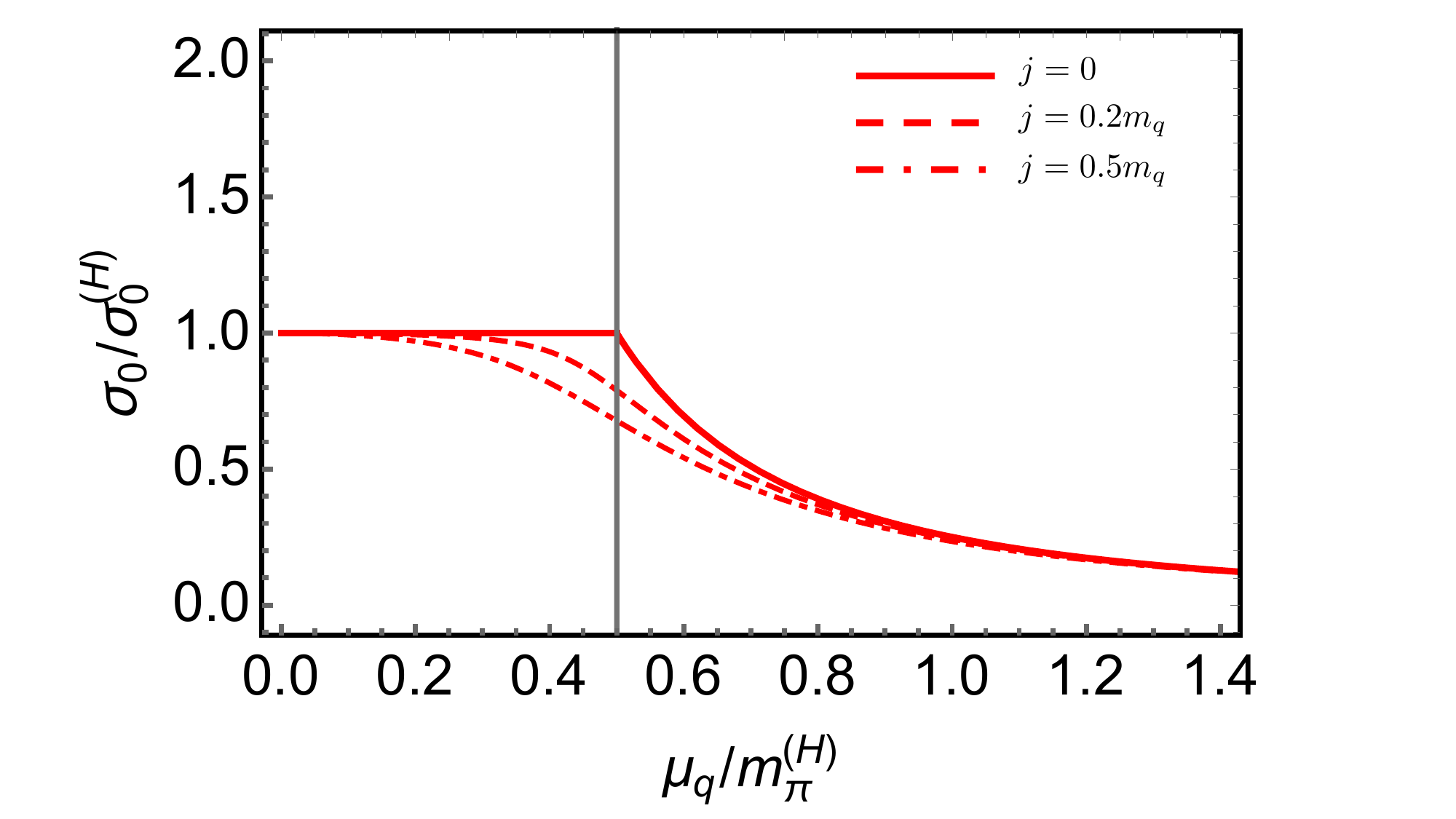}\\
         \end{minipage}

      \begin{minipage}[c]{0.4\hsize}
       \centering
        \hspace*{-1cm} 
          \includegraphics*[scale=0.22]{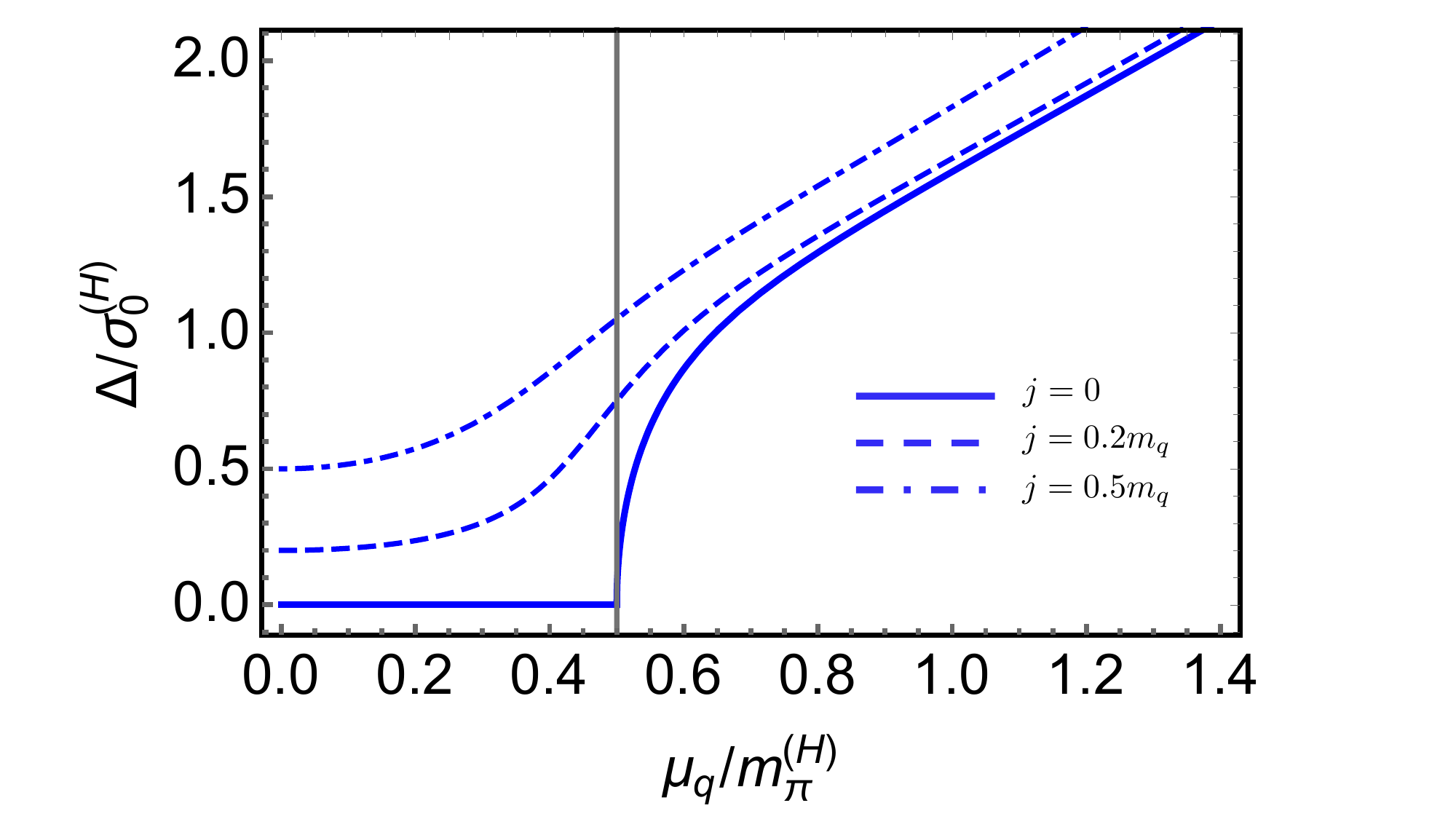}\\
      \end{minipage}

      \end{tabular}
 \caption{$\mu_q$ dependences of the mean fields $\sigma_0$ and $\Delta$ with $j=0$ (solid), $j=0.2m_q$ (dashed) and $j=0.5m_q$ (dot-dashed). The vertical gray line represents $\mu_q/m_\pi^{\rm (H)} = 1/2$.} 
\label{fig:Sigma0AndDelta}
  \end{center}
\end{figure}

The $\mu_q$ dependences of $\sigma_0$ and $\Delta$ with $j=0$, $j=0.2m_q$ and $j=0.5m_q$ are depicted in Fig.~\ref{fig:Sigma0AndDelta}. In plotting this figure we have used the large-$N_c$ suppression~\cite{Witten:1979kh} for the parameters, i.e., $\lambda_1=0$, and adopted
\begin{eqnarray}
m_\pi^{\rm (H)} = 738\, {\rm MeV}\ , \ \ m_{B'(\bar{B}')}^{\rm (H)} = 1611\, {\rm MeV}\ , \label{InputSpin0}
\end{eqnarray}
as inputs from the measured hadron masses on the lattice~\cite{Murakami:2022lmq}. In addition,
\begin{eqnarray}
\sigma_0^{\rm (H)} = 250\, {\rm MeV} \label{TypicalSigma}
\end{eqnarray}
has been employed as a typical value for the chiral condensate, in order to fix the remaining parameter. The figure indicates that when $j$ is finite, $\Delta$ always acquires nonzero values leading to the superfluid phase, whereas the hadronic and superfluid phases are well separated for $j=0$ as analytically found in Eq.~(\ref{PhaseLSM}). As long as $j$ is not sufficiently large, the prominent chiral restoration and evolution of $\Delta$ start at $\mu_q\approx m_\pi^{\rm (H)}/2$.

\subsection{Hadron mass spectrum at finite $\mu_q$}
\label{sec:HadronMassLSM}

In this subsection, we restrict ourselves to vanishing diquark source, $j=0$. In this limit, the parameters are fixed to be
\begin{eqnarray}
 \lambda_1=0\ , \ \ \lambda_2=65.6\ , \ \ m_0^2= -(693\, {\rm MeV})^2\ , \ \ m_q\bar{c} = (456\, {\rm MeV})^3\ , \label{PSet1}
\end{eqnarray}
where $\lambda_1=0$ stems from the large-$N_c$ expansion~\cite{Witten:1979kh}. With these parameters we are ready to numerically explore the hadron masses in cold and dense QC$_2$D matter with the LSM. The pion mass formula has been provided in Eq.~(\ref{MPiLSM}). The other hadron masses are evaluated by reading off the quadratic terms of each fields in Eq.~(\ref{LSMTwoColor}), which reads
\begin{eqnarray}
&& m_{a_0}^2 = m_\pi^2 + \frac{\lambda_2}{2}(\sigma_0^2+\Delta^2) \ , \ \ 
m_{{\cal P}^4}^2 = m_\pi^2 - 4\mu_q^2\ , \ \ 
m_{{\cal P}^5}^2 = m_\pi^2-4\mu_q^2 + 2\tilde{\lambda}\Delta^2\ , \nonumber\\
&& m_\sigma^2 = m_\pi^2 +2\tilde{\lambda}\sigma_0^2\ , \ \ m_{{\cal P}^5\sigma}^2 = 2\tilde{\lambda}\sigma_0\Delta\ ,
\end{eqnarray}
\begin{eqnarray}
&& m_{{\cal S}^4}^2 = m_\pi^2-4\mu_q^2 + \frac{\lambda_2}{2}(\sigma_0^2+\Delta^2)  \ , \ \ 
m_{{\cal S}^5}^2 = m_\pi^2-4\mu_q^2 + \frac{\lambda_2}{2}\sigma_0^2 \ , \nonumber\\
&& m_\eta^2 = m_\pi^2 + \frac{\lambda_2}{2}\Delta^2 \ ,\ \ 
m_{{\cal S}^5\eta}^2 = \frac{\lambda_2}{2}\sigma_0\Delta\ . \label{MassPartner}
\end{eqnarray}
We note that (${\cal P}^4,{\cal P}^5,\sigma$) and (${\cal S}^4,{\cal S}^5,\eta$) exhibit state mixings owing to the baryon-number violation, leading to the following $3\times3$ propagator-inverse matrices in the momentum space:
\begin{eqnarray} 
iD^{-1}_{{\cal P}^4{\cal P}^5\sigma}(p) = \left(
\begin{array}{ccc}
p^2-m_{{\cal P}^4}^2 & 2i\mu_q p_0 & 0 \\
-2i\mu_qp_0 & p^2-m_{{\cal P}^5}^2 & -m_{{\cal P}^5\sigma}^2 \\
0 & -m_{{\cal P}^5\sigma}^2 & p^2-m_\sigma^2 \\
\end{array}
\right)\ ,  \label{MixingPositiveP}
\end{eqnarray}
\begin{eqnarray} 
iD^{-1}_{{\cal S}^4{\cal S}^5\eta}(p) = \left(
\begin{array}{ccc}
p^2-m_{{\cal S}^4}^2 & 2i\mu_q p_0 & 0 \\
-2i\mu_qp_0 & p^2-m_{{\cal S}^5}^2 & -m_{{\cal S}^5\eta}^2 \\
0 & -m_{{\cal S}^5\eta}^2 & p^2-m_\eta^2 \\
\end{array}
\right)\ . \label{MixingNegativeP}
\end{eqnarray}
The former hadrons share $I=0$ and $0^+$ while the latter share $I=0$ and $0^-$.

\begin{figure}[H]
  \begin{center}
    \begin{tabular}{cc}

      \begin{minipage}[c]{0.5\hsize}
       \centering
       \hspace*{-1cm} 
         \includegraphics*[scale=0.22]{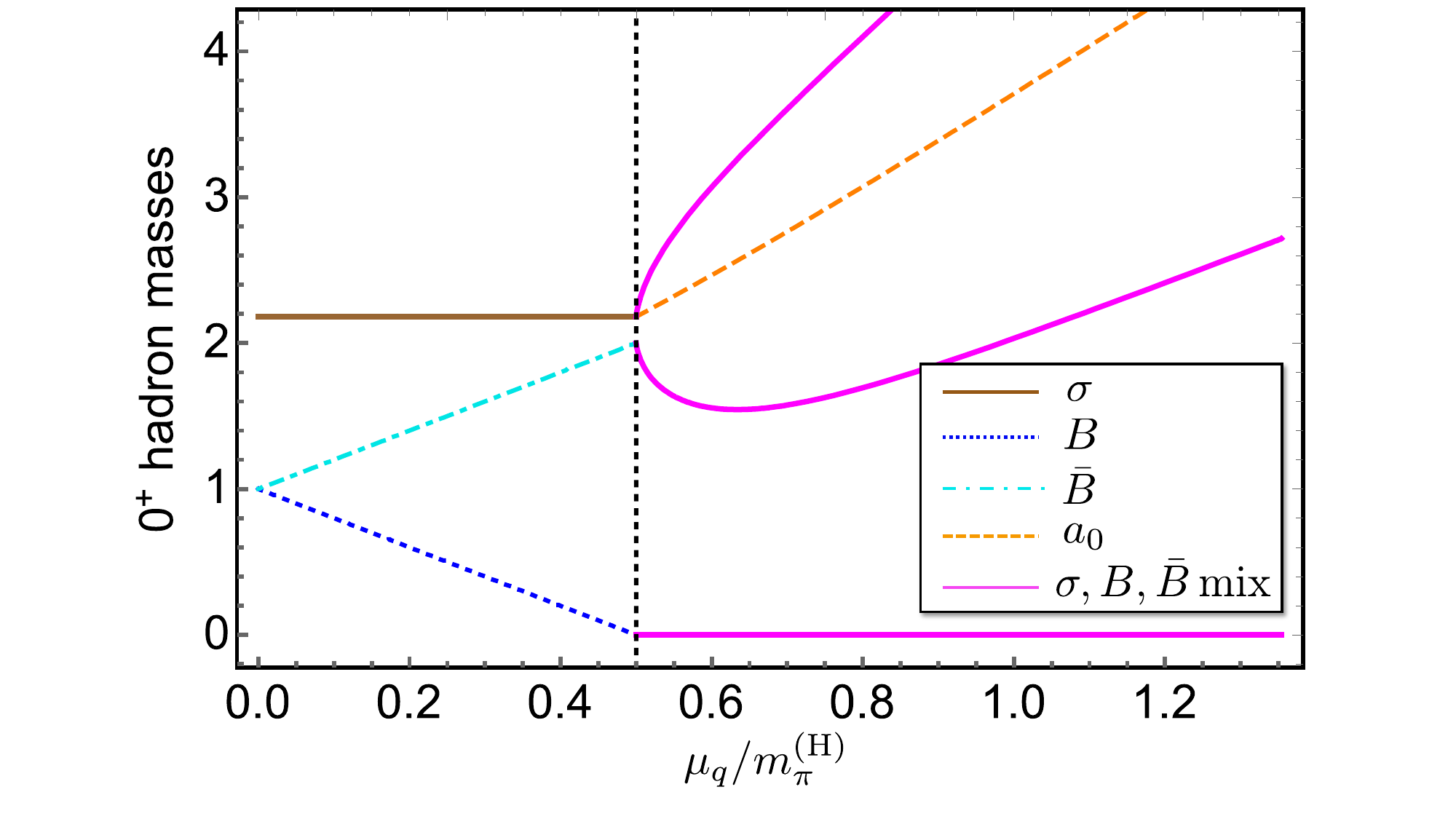}\\
         \end{minipage}

      \begin{minipage}[c]{0.4\hsize}
       \centering
        \hspace*{-1cm} 
          \includegraphics*[scale=0.22]{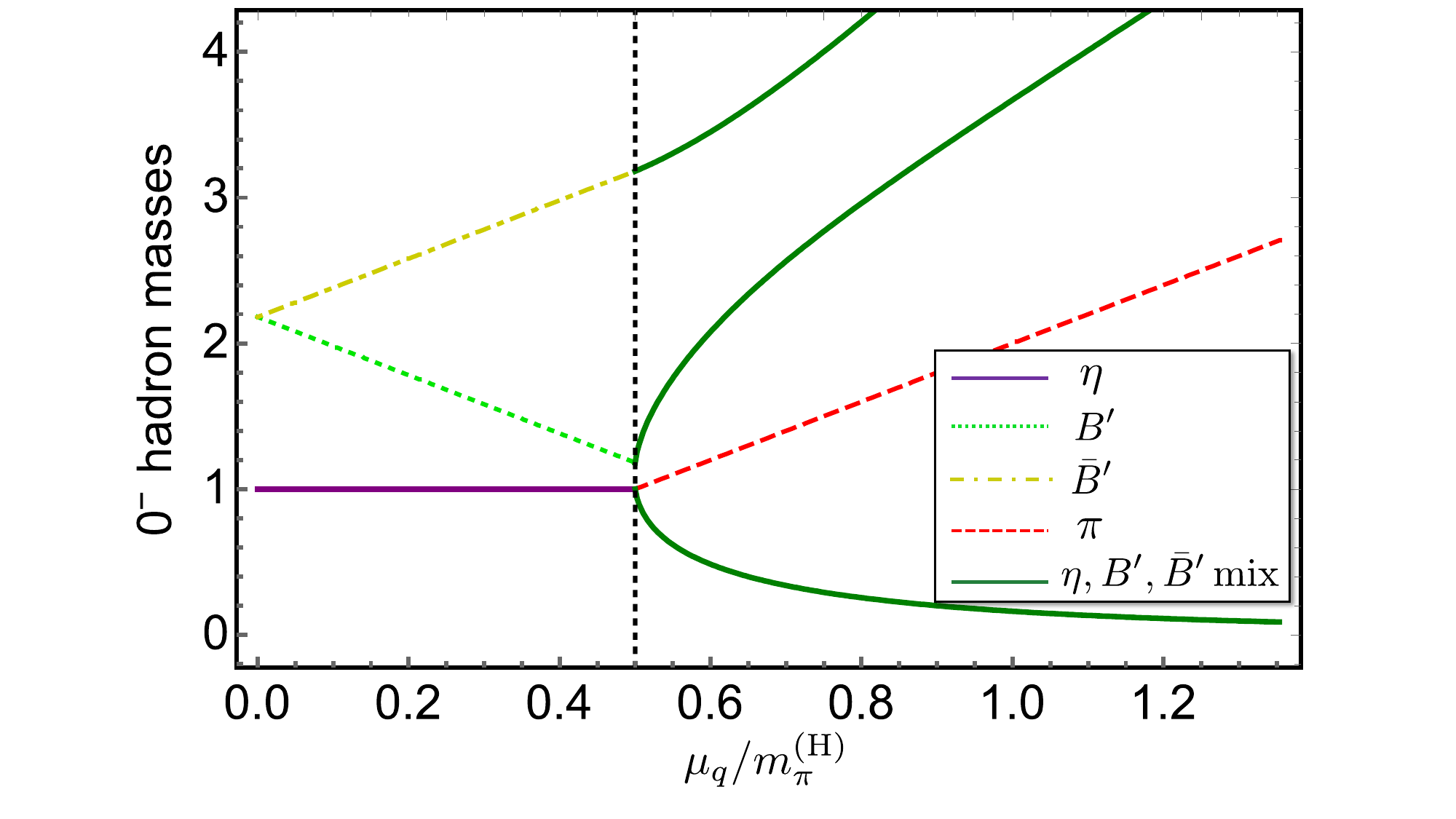}\\
      \end{minipage}

      \end{tabular}
 \caption{Mass spectra of $0^+$ (left) and $0^-$ (right) hadrons evaluated within the present LSM. The masses are scaled by $m_\pi^{\rm (H)}$. The figures are taken from Ref.~\cite{Suenaga:2022uqn} where legends are slightly modified.} 
\label{fig:Spin0Mass}
  \end{center}
\end{figure}

\begin{figure}[H]
  \begin{center}
    \begin{tabular}{cc}

      \begin{minipage}[c]{0.5\hsize}
       \centering
       \hspace*{-1cm} 
         \includegraphics*[scale=0.23]{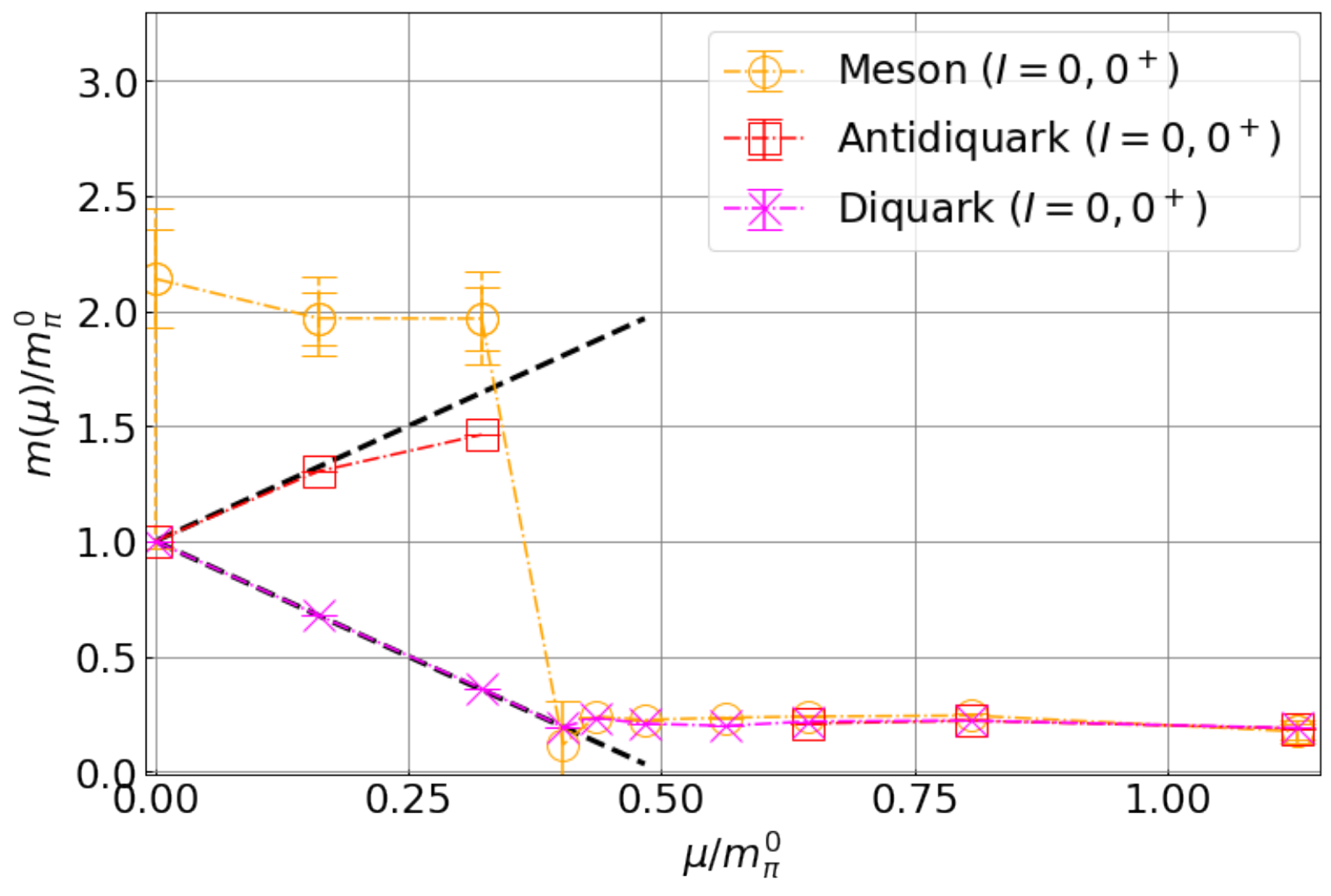}\\
         \end{minipage}

      \begin{minipage}[c]{0.4\hsize}
       \centering
        \hspace*{-1cm} 
          \includegraphics*[scale=0.23]{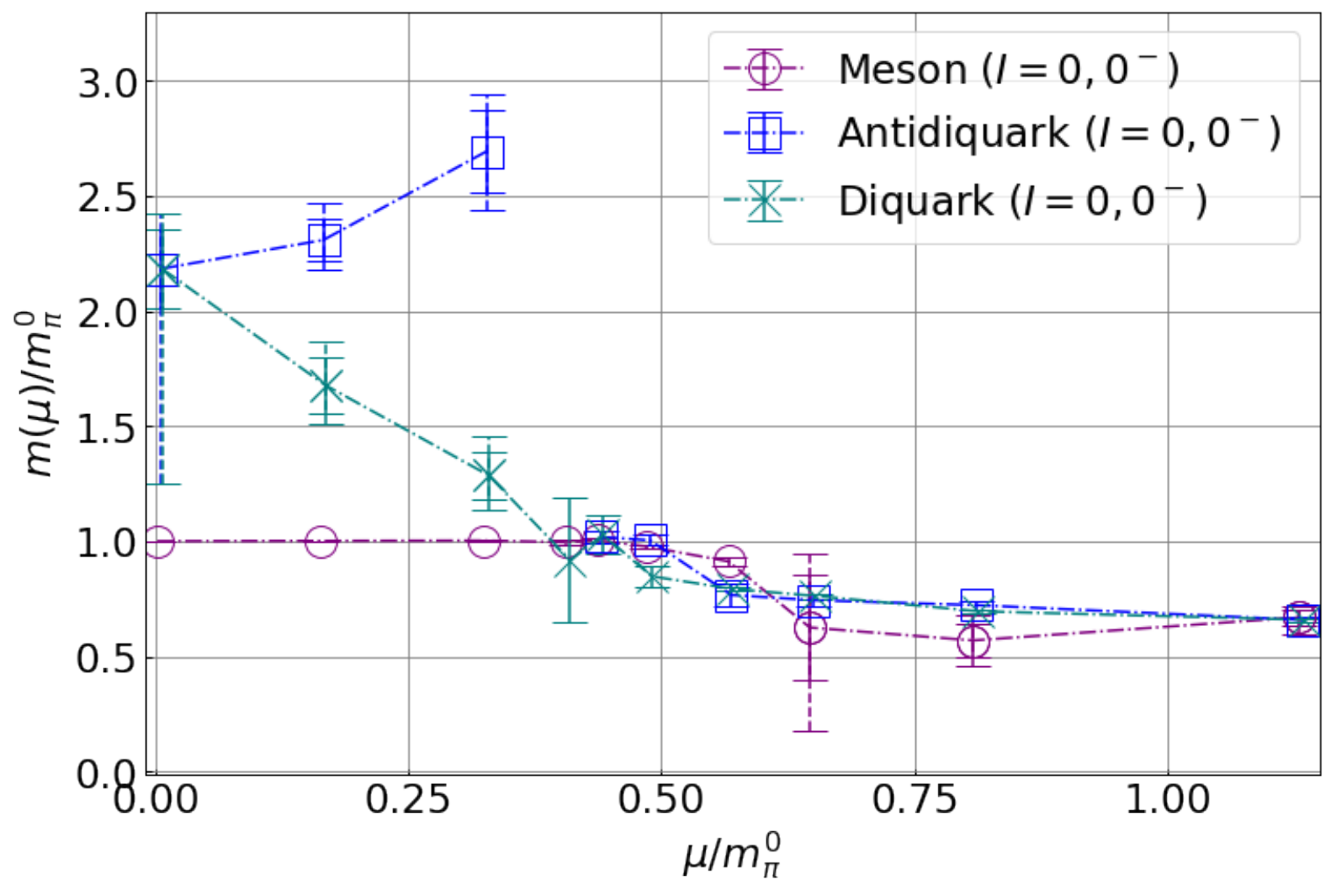}\\
      \end{minipage}

      \end{tabular}
 \caption{Mass spectra of $0^+$ (left) and $0^-$ (right) iso-singlet hadrons measured on the lattice. The figures are taken from Ref.~\cite{Murakami:2022lmq}. } 
\label{fig:HadronMassLattice}
  \end{center}
\end{figure}

Depicted in Fig.~\ref{fig:Spin0Mass} is $\mu_q$ dependences of the mass of $0^+$ hadrons (left) and of $0^-$ hadrons (right) with the parameter set~(\ref{PSet1}). Both figures indicate the stable $\mu_q$ dependences of the hadron masses in the hadronic phase, reflecting the so-called {\it Silver-Braze property}. In the baryon superfluid phase, meanwhile, notable behaviors are found. For instance, $\sigma$, $B$ and $\bar{B}$ mix while $\eta$, $B'$ and $\bar{B}'$ do, due to the $U(1)_B$ violation. Among the  $\sigma$-$B$-$\bar{B}$ mixed states, a massless mode is obtained, which corresponds to the NG boson accompanied by the $U(1)_B$ breaking. Besides, the nonlinear mass suppression of the lightest mode of the $\eta$-$B'$-$\bar{B}'$ mixed state which was observed by the lattice simulation~\cite{Murakami:2022lmq} is successfully reproduced, in contrast to the ChPT framework. From this reproduction, one can conclude that the present LSM is regarded as a plausible effective model which correctly describes the low-energy hadron spectrum in cold and dense QC$_2$D. For comparison we exhibit the simulated mass spectra of iso-singlet $0^\pm$ hadrons at finite $\mu_q$ in Fig.~\ref{fig:HadronMassLattice}, although some artifacts originating from a finite diquark source $j$ contaminate the spectra. On the lattice, the mixings are indicated by the mass degeneracies. We note that the pion mass is analytically evaluated to be $m_\pi = 2\mu_q$, which is consistent with the lattice simulations~\cite{Murakami:2022lmq}.

Quantitatively, the nonlinear suppression of the mass of the lightest $\eta$-$B'$-$\bar{B}'$ mixed state measured on the lattice is rather mild, while the present LSM result in the absence of $U(1)_A$ anomaly effects exhibits a substantial mass reduction, as shown in Figs.~\ref{fig:Spin0Mass} and~\ref{fig:HadronMassLattice}. In Ref.~\cite{Suenaga:2022uqn}, it was demonstrated that as the $U(1)_A$ anomaly effects get enhanced within the LSM analysis, the suppression is weakened so as to approach the correct behavior measured on the lattice. This observation suggests that the anomaly effects for hadrons would be enhanced in the superfluid phase, while in the vacuum the effects seem to be significantly suppressed. A similar anomaly enhancement at finite density was also discussed in three-color QCD by means of the functional renormalization group (FRG) method~\cite{Fejos:2016hbp,Fejos:2017kpq}.

\begin{figure}[H]
\centering
\includegraphics[width=11cm]{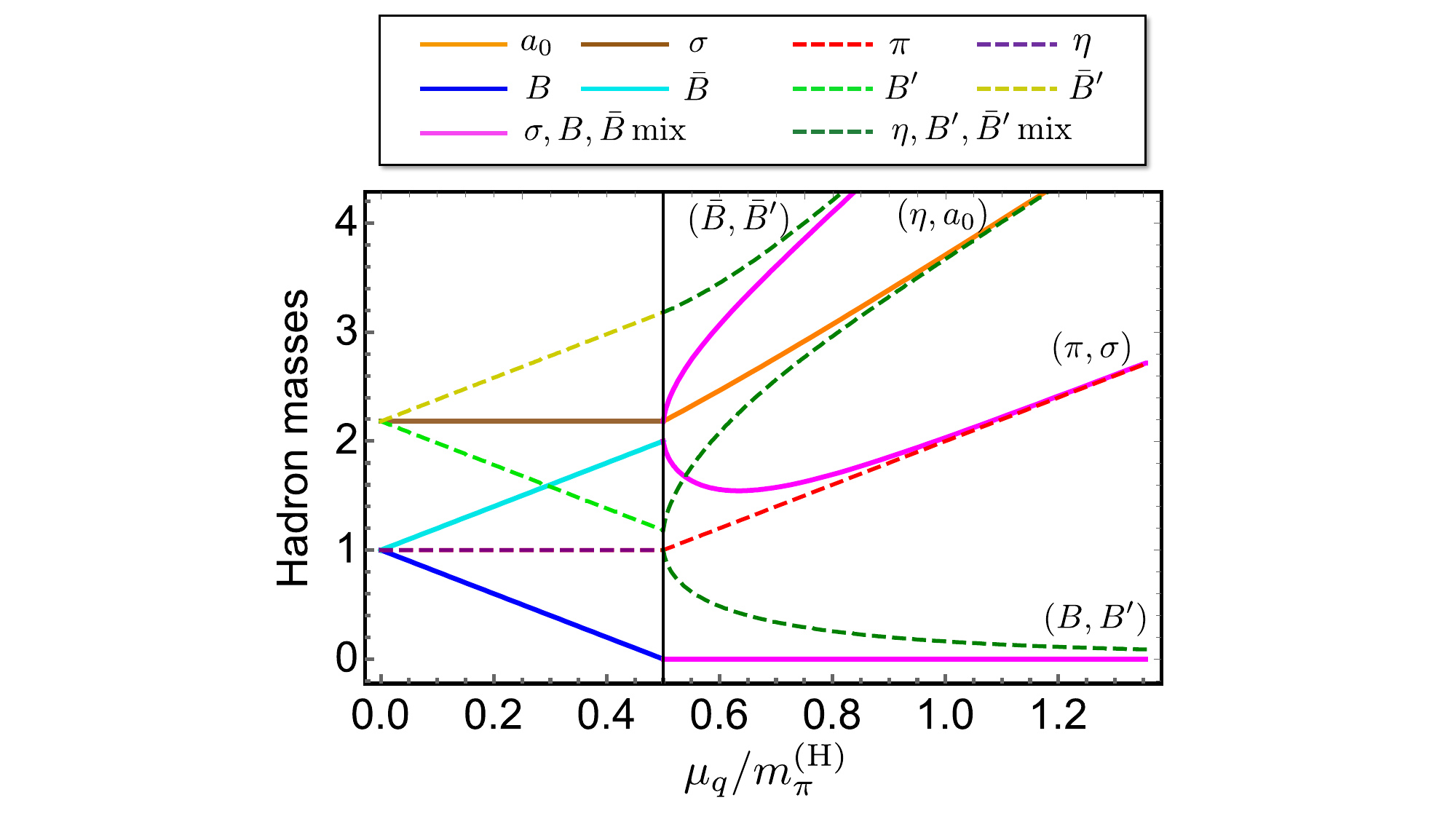}
\caption{$\mu_q$ dependences of $0^\pm$ hadron masses. The figure is taken from Ref.~\cite{Suenaga:2022uqn} where legends are slightly modified.}
\label{fig:ChiralPartner0}
\end{figure}

In Fig.~\ref{fig:ChiralPartner0}, we depict the mass spectrum of $0^+$ and $0^-$ hadrons collectively for which mass degeneracies of the parity partners are clearly seen. At sufficiently large $\mu_q$, the mass degeneracies hold for pairs of $(\pi,\sigma)$, $(\eta,a_0)$, $(B,B')$ and $(\bar{B},\bar{B}')$ where the mixings disappear.

\subsection{LSM with a diquark source $j$}
\label{sec:LSMSource}

The inclusion of the diquark source $j$ has no influence on the hadron mass formulas directly but modifies the effective potential, since $j$ couples to $\Delta$ linearly as in Eq.~(\ref{VEffLSM}) . As a result $\mu_q$ dependences of $\sigma_0$ and $\Delta$ are altered as demonstrated in Fig.~\ref{fig:Sigma0AndDelta}, and accordingly the hadron mass spectrum is changed.

\begin{figure}[H]
\centering
\includegraphics[width=9cm]{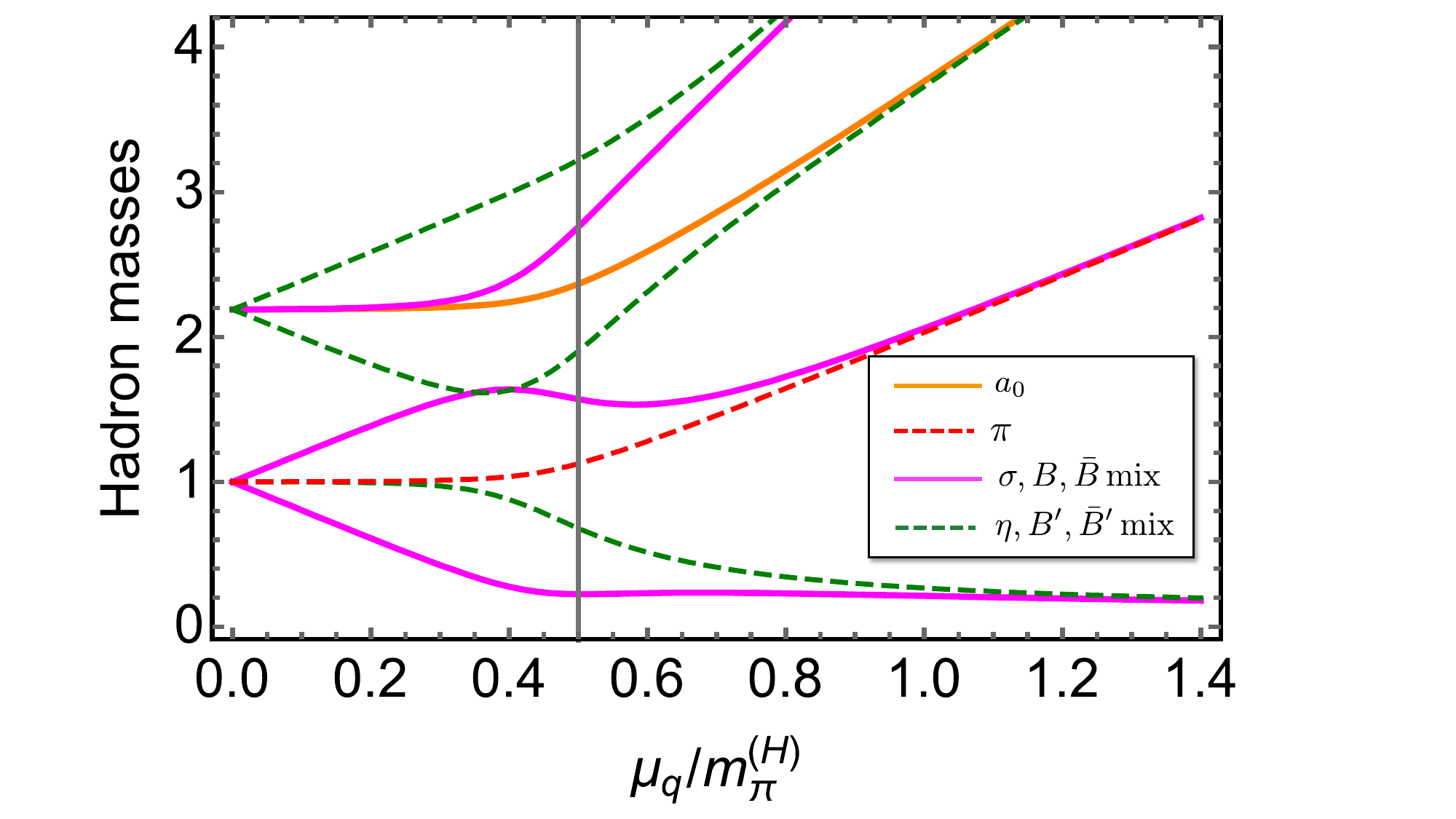}
\caption{$\mu_q$ dependences of $0^\pm$ hadron masses with $j=0.2m_q$.}
\label{fig:0pmMassWithJ}
\end{figure}

Depicted in Fig.~\ref{fig:0pmMassWithJ} exhibits $\mu_q$ dependences of the hadron mass with $j=0.2m_q$. For finite $j$, $\Delta$ is always non-vanishing and $\sigma$-$B$-$\bar{B}$ mixing and $\eta$-$B'$-$\bar{B}'$ mixing occur at any $\mu_q$. Besides, the NG mode does not emerge since $U(1)_B$ symmetry is explicitly broken. The figure shows that the mass degeneracies between the chiral partners are clearly realized for large $\mu_q$.

In what follows, we check the GOR relation analytically presented in Sec.~\ref{sec:GORRelation} within the LSM. The broken current within the LSM is obtained by taking a derivative of ${\cal L}_{\rm LSM}^{\rm eff}$ with respect to $\zeta_{X\mu}^{a} \equiv -2\sqrt{2}V'^a_\mu$ as done in Sec.~\ref{sec:ChPTWithJ}, which yields
\begin{eqnarray}
j_{X\mu}^a &=& \frac{\sigma_0}{\sqrt{2}}\partial_\mu\pi^a  + \cdots\ \ \ \ \ \ ({\rm for}\ a=1-3)\ , \nonumber\\
j_{X\mu}^4 &=&  \frac{\sigma_0}{\sqrt{2}}\partial_\mu{\cal P}^4 + \sqrt{2}\mu_q\sigma_0\delta_{\mu0}{\cal P}^5 + \cdots\ \ \ \ \ \ ({\rm for}\ a=4)\ , \nonumber\\
j_{X\mu}^{5} &=& \frac{\sigma_0}{\sqrt{2}}\partial_\mu{\cal P}^5 -\frac{\Delta}{\sqrt{2}}\partial_\mu\sigma - \sqrt{2}\mu_q\sigma_0\delta_{\mu0}{\cal P}^4 + \cdots\ \ \ \ \ \ ({\rm for}\ a=5)\ . \label{JXLSM}
\end{eqnarray}
Similarly the $U(1)_B$ current is derived to be
\begin{eqnarray}
j_B^\mu = 2\Delta\partial_\mu {\cal P}^4+8\mu_q\Delta\delta_{\mu0} {\cal P}^5  +\cdots\ , \label{JBLSM}
\end{eqnarray}
by taking a derivative of the Lagrangian with respect to $\zeta_B^\mu\equiv V^{\mu,i=4}$. From the pion sector immediately
\begin{eqnarray}
f_\pi = \frac{\sigma_0}{\sqrt{2}}
\end{eqnarray}
is found by virtue of the definition of the decay constant~(\ref{Amplitudes}), regardless of its trivial sign. Thus, we can easily check the GOR relation for pions in Eq.~(\ref{GORPi}), from which the pion mass and chiral condensate are denoted by $m_\pi^2=\sqrt{2}\bar{c}m_q/\sigma_0$ and Eq.~(\ref{CondensateLSM}), respectively.

As for the baryonic sector, again we take $\mu_q=0$ to achieve concise relations. In the LSM framework, even in the vacuum ${\cal P}^5$ and $\sigma$ mixes due to the baryon-number violation that was absent in the ChPT analysis, as explicitly shown in Eq.~(\ref{MixingPositiveP}). The mixing is solved by introducing mass eigenstates $\tilde{\cal P}^5$ and $\tilde{\sigma}_0$ via
\begin{eqnarray}
\left(
\begin{array}{c}
\tilde{\cal P}^5 \\
\tilde{\sigma} \\
\end{array}
\right)
=
\left(
\begin{array}{cc}
\cos\vartheta & -\sin\vartheta \\
\sin\vartheta & \cos\vartheta \\
\end{array}
\right)
\left(
\begin{array}{c}
{\cal P}^5 \\
{\sigma} \\
\end{array}
\right)\ , \label{MixingP5Sigma}
\end{eqnarray}
 where the mixing angle $\vartheta$ is determined to satisfy $\tan\vartheta=\Delta/\sigma_0$ from Eq.~(\ref{MixingPositiveP}). The corresponding mass eigenvalues read
\begin{eqnarray}
 m_{\tilde{\cal P}^5}^2 =m_\pi^2 \ \ , \ \ \ m_{\tilde{\sigma}}^2 = m_\pi^2 + 2\tilde{\lambda}(\sigma_0^2+\Delta^2)\ .
\end{eqnarray}
Inverting the mixing matrix~(\ref{MixingP5Sigma}), ${\cal P}^5$ and $\sigma$ are expressed as a function of $\tilde{\cal P}^5$ and the current $j_{X\mu}^5$ in the vacuum in Eq.~(\ref{JXLSM}) can be rewritten into
 \begin{eqnarray}
 j_{X\mu}^5 =  \frac{\sigma_0\cos\theta + \Delta\sin\theta}{\sqrt{2}}\partial_\mu \tilde{\cal P}^5 + \cdots = \sqrt{\frac{\sigma_0^2+\Delta^2}{2}}\partial_\mu\tilde{\cal P}^5 + \cdots\ ,
 \end{eqnarray}
 resulting in that
 \begin{eqnarray}
 f_5 = \sqrt{\frac{\sigma_0^2+\Delta^2}{2}}\ .
 \end{eqnarray}
Using $m_\pi^2=\sqrt{2}\bar{c}m_q/\sigma_0=\sqrt{2}\bar{c}j/\Delta$ at $\mu_q=0$ and Eq.~(\ref{CondensateLSM}), the GOR relation for the baryon in this limit, Eq.~(\ref{GORB5}), can be verified to hold.

Finally, from Eq.~(\ref{JBLSM}) the decay constant $f_B$ is evaluated to be
\begin{eqnarray}
f_B = 2\Delta
\end{eqnarray}
within the LSM. Meanwhile, $m_\pi^2 = \sqrt{2}j\bar{c}/\Delta$ at $\mu_q=0$. Hence, using these equations together with Eq.~(\ref{CondensateLSM}), one can confirm that the GOR relation associated with $U(1)_B$ symmetry~(\ref{GORBaryon}) is certainly satisfied.\footnote{At finite $\mu_q$, $\left(\frac{f_B}{2\sqrt{2}}\right)^2m_{B_4}^2 = -\frac{j\langle\psi\psi\rangle}{2}$ seems to hold, similarly to the ChPT framework.}

\subsection{Topological susceptibility }
\label{sec:Topological}

The hadron mass spectrum from the LSM has been presented in Sec.~\ref{sec:HadronMassLSM}, indicating that the $U(1)_A$ anomaly effects in the superfluid phase would be enhanced from the behavior of the lowest mode of $\eta$-$B'$-$\bar{B}'$ mixed state. One of the useful quantity to explore the $U(1)_A$ anomaly is the {\it topological susceptibility} which is defined by 
\begin{eqnarray}
\chi_{\rm top} \equiv -i\int d^4x\frac{\delta^2\Gamma_{\rm QC_2D}}{\delta\theta(x)\delta\theta(0)}\Bigg|_{\theta=0} = -i\int d^4x \langle0|{\rm T}^*Q(x)Q(0)|0\rangle \ ,
\end{eqnarray}
since $Q= g_s^2/(64\pi^2)\epsilon^{\mu\nu\rho\sigma}G^a_{\mu\nu}G^a_{\rho\sigma}$ ($G_{\mu\nu}^a= \partial_\mu A_\nu^a-\partial_\nu A_\mu^a + g_sA_\mu^a A_\nu^b$ is the gluon field strength) is nothing but the topological charge responsible for the anomaly. On the lattice, currently two groups, Japanese group and Russian group have simulated the topological susceptibility at finite $\mu_q$~\cite{Astrakhantsev:2020tdl,Iida:2024irv,Lombardo:2021jpn}. However, those results seem to be inconsistent even at qualitative level; The latter result indicates a suppression of $\chi_{\rm top}$ at large $\mu_q$ while the former one exhibits a constant behavior. Then, in this subsection we investigate the topological susceptibility at finite $\mu_q$ within the LSM to present useful information from a model study, and discuss the fate of the $U(1)_A$ anomaly effects in cold and dense QC$_2$D~\cite{Kawaguchi:2023olk}.

The QC$_2$D Lagrangian with the $\theta$ term is given by
\begin{eqnarray}
{\cal L}_{\rm QC_2D} = {\cal L}_{\rm QC_2D}^q -\frac{1}{4}G_{\mu\nu}^a G^{\mu\nu a} + \theta\frac{g_s^2}{64\pi^2}\epsilon^{\mu\nu\rho\sigma}G^a_{\mu\nu}G^a_{\rho\sigma}\ ,
\end{eqnarray}
where the quark part ${\cal L}_{\rm QC_2D}^q$ is defined by Eq.~(\ref{LQuarkQC2D}). After an $U(1)_A$ transformation of $\psi\to{\rm exp}[(i\theta/4)\gamma_5]\psi$, Fujikawa's method~\cite{Fujikawa:1979ay} yields a modified Lagrangian as
\begin{eqnarray}
{\cal L}_{\rm QC_2D}^{\theta} = \bar{\psi}i\Slash{D}\psi-m_q\bar{\psi}{\rm exp}[{(i\theta/2)\gamma_5]}\psi - \frac{1}{4}G_{\mu\nu}^aG^{\mu\nu a}\ , \label{ThetaLQC2D}
\end{eqnarray}
whose $\theta$ dependence is now absorbed into the fermion mass term. Therefore, the topological susceptibility is evaluated to be
\begin{eqnarray}
\chi_{\rm top} =  -i\int d^4x\frac{\delta^2\Gamma^\theta_{\rm QC_2D}}{\delta\theta(x)\delta\theta(0)}\Bigg|_{\theta=0} &=& -\frac{1}{4} \left[ m_q\langle\bar{\psi}\psi\rangle + im_q^2\chi_\eta\right] \nonumber\\
&=& \frac{im_q^2}{4}(\chi_\pi - \chi_\eta)\ , \label{ChiTopDef}
\end{eqnarray}
 with $\Gamma^\theta_{\rm QC_2D} = -i{\rm ln}Z_{\rm QC_2D}^\theta$ being the effective action generated by the rotated QC$_2$D Lagrangian~(\ref{ThetaLQC2D}). In this equation the meson susceptibilities are defined by
\begin{eqnarray}
\chi_\eta &=& \int d^4x\langle 0|{\rm T}{\cal O}_\eta(x){\cal O}_\eta(0)|0\rangle\ , \nonumber\\
\chi_\pi\delta^{ab} &=& \int d^4x\langle 0|{\rm T}{\cal O}_\pi^a(x){\cal O}_\pi^b(0)|0\rangle\ ,
\end{eqnarray}
where the composite operators have been defined in Eq.~(\ref{OperatorMeson}). Besides, in obtaining Eq.~(\ref{ChiTopDef}) we have made use of 
\begin{eqnarray}
\langle\bar{\psi}\psi\rangle = -im_q\chi_\pi \ ,
\end{eqnarray}
which is nothing but the first identity in Eq.~(\ref{ChiralWTI}).
Equation~(\ref{ChiTopDef}) indicates that the finite topological susceptibility is induced only when $\chi_\eta$ deviates from $\chi_\pi$. These susceptibility functions are two-point functions of the corresponding composite operators with vanishing momentum. Thus, unless state mixings occur, they would be essentially denoted by $\chi_\pi\propto -i/m_\pi^2$ and $\chi_\eta \propto -i/m_\eta^2$ where $m_\pi$ and $m_\eta$ are the pion and $\eta$ meson masses. The difference between $\eta$ mass and pion mass is generated by the $U(1)_A$ anomaly effect, so one can understand that the finite topological susceptibility is induced by the anomaly effect together with the current quark mass $m_q$~\cite{Kawaguchi:2023olk}.

The functions $\chi_\pi$ and $\chi_\eta$ are evaluated within the present LSM by virtue of the matching condition~(\ref{MatchingQCD}). That is, 
\begin{eqnarray}
\chi_\eta =  \frac{1}{i}\int d^4x\frac{\delta^2\Gamma_{\rm QC_2D}}{\delta p^0(x)\delta p^0(0)}\Bigg|_{\langle\zeta\rangle,\langle\zeta_\mu\rangle} =  \frac{1}{i}\int d^4x\frac{\delta^2\Gamma_{\rm LSM}}{\delta p^0(x)\delta p^0(0)}\Bigg|_{\langle\zeta\rangle,\langle\zeta_\mu\rangle}  = 2\bar{c}^2D_\eta(0) 
\ ,  \label{ChiEtaMatch}
\end{eqnarray}
and
\begin{eqnarray}
\chi_\pi \delta^{ab} = \frac{1}{i}\int d^4x\frac{\delta^2\Gamma_{\rm QC_2D}}{\delta p^a(x)\delta p^b(0)}\Bigg|_{\langle\zeta\rangle,\langle\zeta_\mu\rangle}  = \frac{1}{i}\int d^4x\frac{\delta^2\Gamma_{\rm LSM}}{\delta p^a(x)\delta p^b(0)}\Bigg|_{\langle\zeta\rangle,\langle\zeta_\mu\rangle}  = 2\delta^{ab}\bar{c}^2D_\pi(0) \nonumber\\ \label{ChiPiMatch}
\end{eqnarray}
($a,b=1-3$), respectively, where the spurious $p^a$ have been introduced in Sec.~\ref{sec:Spurion}. In these equation $D_\eta(p)$ and $D_\pi(p)$ are propagators of $\eta$ and pion, respectively. Using these effective-model expressions the topological susceptibility can be evaluated to be
\begin{eqnarray}
\chi_{\rm top} = \frac{i}{4}\big(m_\pi^{\rm (H)}\big)^2\big(\sigma_0^{\rm (H)}\big)^2\Big(D_\pi(0)-D_\eta(0)\Big)\ .
\end{eqnarray}

\begin{figure}[H]
  \begin{center}
    \begin{tabular}{cc}

      \begin{minipage}[c]{0.5\hsize}
       \centering
       \hspace*{-1cm} 
         \includegraphics*[scale=0.22]{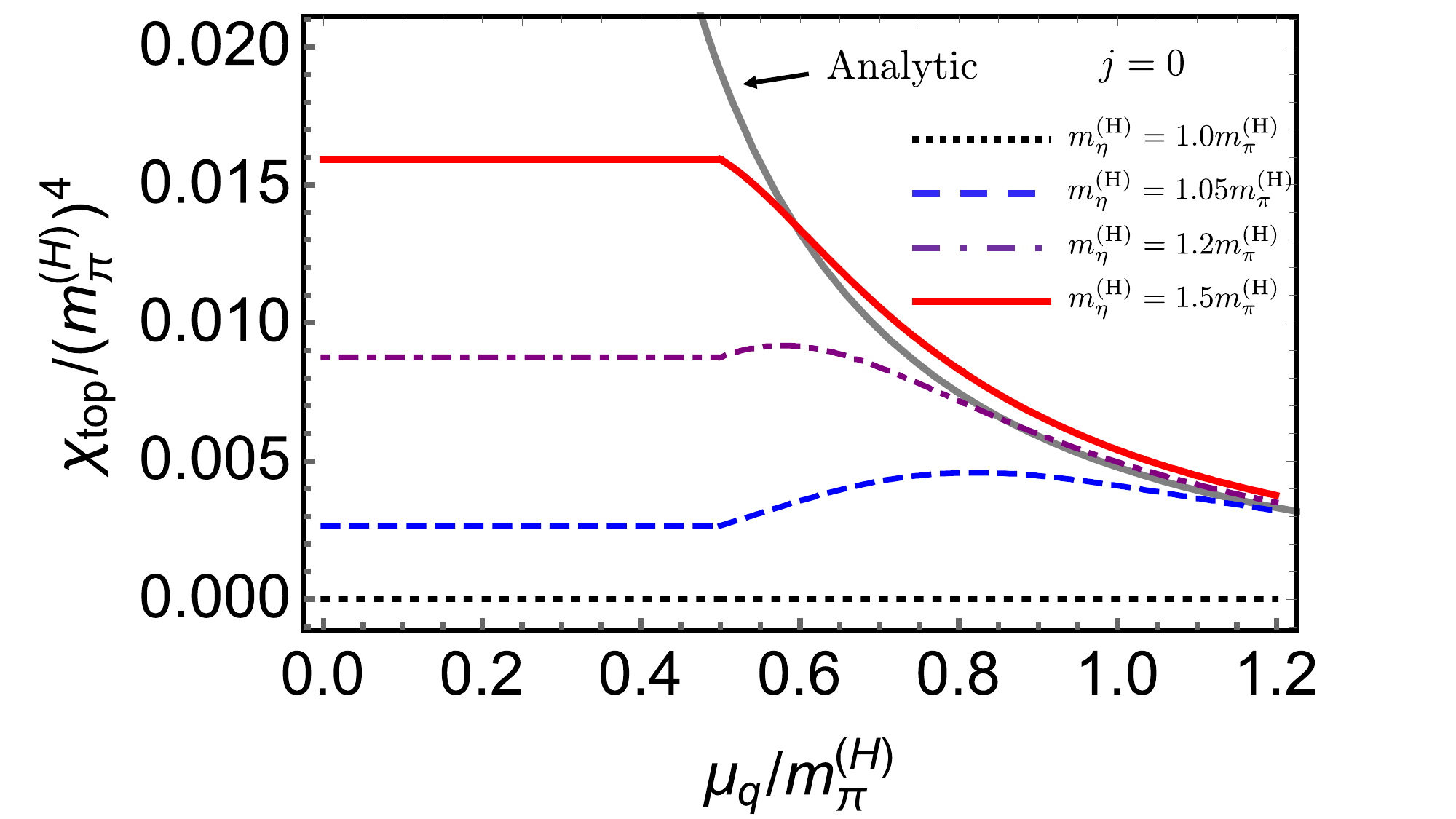}\\
         \end{minipage}

      \begin{minipage}[c]{0.4\hsize}
       \centering
        \hspace*{-1cm} 
          \includegraphics*[scale=0.22]{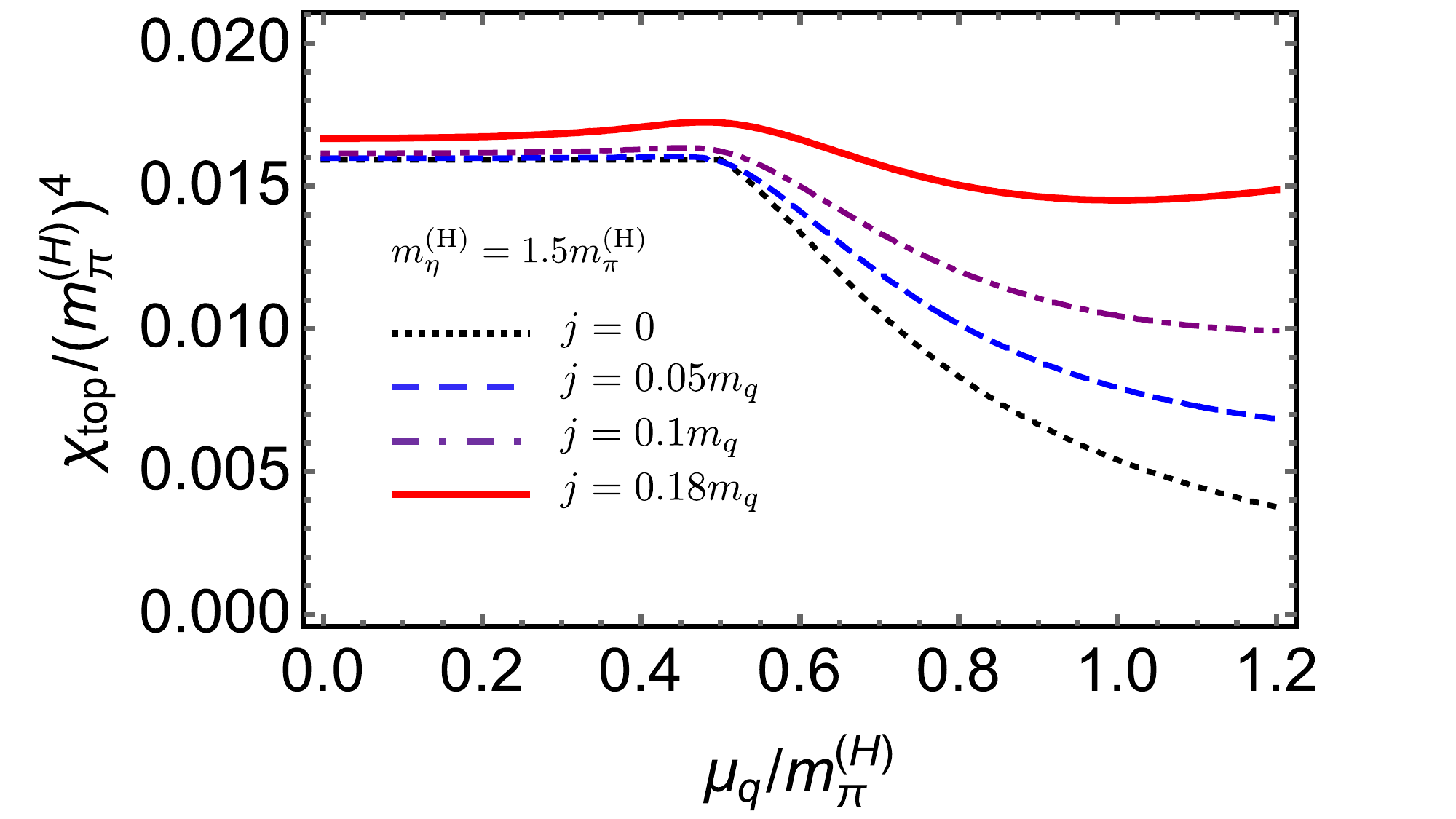}\\
      \end{minipage}

      \end{tabular}
 \caption{$\mu_q$ dependences of the topological susceptibility $\chi_{\rm top}$ normalized by $\big(m_\pi^{\rm (H)}\big)^4$. The left and right panels show $m_\eta^{\rm (H)}/m_\pi^{\rm (H)}$ dependences with $j=0$ and $j$ dependences with $m_\eta^{\rm (H)}/m_\pi^{\rm (H)}=1.5$, respectively.} 
\label{fig:ChiTop}
  \end{center}
\end{figure}

In the hadronic phase these propagators are simply given by
\begin{eqnarray}
D_\eta(p) = \frac{i}{p^2-\big(m_\eta^{\rm (H)}\big)^2}\  \ \ ,  \ \ \ \ \ D_\pi(p) = \frac{i}{p^2-\big(m_\pi^{\rm (H)}\big)^2}\ .
\end{eqnarray}
On the other hand, in the superfluid phase $D_\eta(p)$ is contaminated by mixings among $\eta$-$B'$-$\bar{B}$ (or  $\eta$-${\cal S}^4$-${\cal S}^5$) modes due to the $U(1)$ baryon-number violation, but is straightforwardly evaluated by picking up a $D_\eta$ component by inverting the $3\times3$ matrix~(\ref{MixingNegativeP}). The resultant $\mu_q$ dependences of the topological susceptibility with vanishing diquark source $j$ are depicted in the left panel of Fig.~\ref{fig:ChiTop}. In this figure we have chosen $m_\eta^{\rm (H)}/m_\pi^{\rm (H)}=1.0,1.05,1.2,1.5$ so as to take a closer look at the anomaly effect, where the anomaly effects are incorporated through the ${\rm det}\Sigma + {\rm det}\Sigma^\dagger$ term following Ref.~\cite{Kawaguchi:2023olk}. The figure implies that the topological susceptibility is always vanishing when the anomaly effect is absent. When the anomaly effect is switched on, in the hadronic phase a constant $\chi_{\rm top}$ is induced, the magnitude of which is enhanced as we impose the stronger effect.


The left panel of Fig.~\ref{fig:ChiTop} exhibits suppression of the topological susceptibility at large $\mu_q$, particularly for larger $m_\eta^{\rm (H)}/m_\pi^{\rm (H)}$. To explore this behavior in detail let us rewrite Eq.~(\ref{ChiTopDef}) in terms of the low-energy quantities. That is, with the help of the GOR relation~(\ref{GORPi}) we can express the topological susceptibility~(\ref{ChiTopDef}) as
\begin{eqnarray}
\chi_{\rm top} = \frac{f_\pi^2 m_\pi^2}{2}\delta_m \ \ \ \ \ \  \ {\rm with}\ \ \ \ \ \ \ \ \delta_m = 1-\frac{\chi_\eta}{\chi_\pi} \ . \label{ChiTopLEC}
\end{eqnarray}
In this equation $f_\pi$ and $m_\pi$ are the pion decay constant and pion mass in the superfluid phase, which reads $f_\pi = \sigma_0/\sqrt{2}$ and $m_\pi = 2\mu_q$ within the present LSM, respectively. Using the asymptotic value of $\chi_\eta/\chi_\pi \sim 1/3$~\cite{Suenaga:2022uqn} and an identity $f_\pi^2 m_\pi^4 = \big(f_\pi^{\rm (H)}\big)^2\big(m_\pi^{\rm (H)}\big)^4$ one can approximate $\chi_{\rm top}$ for sufficiently large $\mu_q$ as
\begin{eqnarray}
\chi_{\rm top} \sim\frac{\big(f_\pi^{\rm (H)}\big)^2\big(m_\pi^{\rm (H)}\big)^4}{3m_\pi^2} =  \frac{\big(f_\pi^{\rm (H)}\big)^2\big(m_\pi^{\rm (H)}\big)^4}{12}\mu_q^{-2}\ . \label{ChiTopAs}
\end{eqnarray}
In the left panel of Fig.~\ref{fig:ChiTop}, the black curve corresponds to this analytic solution which is in good agreement with the numerical behaviors. Therefore, we conclude that the asymptotic suppression of the topological susceptibility is accompanied by the increment of pion mass: $m_\pi=2\mu_q$ in the superfluid phase, i.e., the chiral restoration.

In the actual lattice simulation it would not be so easy to take a zero limit of the diquark source, so it is worth studying effects from the diquark source $j$ within the LSM analysis. These effects are incorporated by $\langle p^0\rangle= j$ for the spurion field, leading to
\begin{eqnarray}
\chi_{\rm top}^{\rm w/j} = \chi_{\rm top} + \delta\chi_{\rm top}\ ,
\end{eqnarray}
in which $\chi_{\rm top}$ is defined in Eq.~(\ref{ChiTopDef}) and the corrections driven by the diquark source read
\begin{eqnarray}
\delta\chi_{\rm top} = \frac{i}{2}m_qj\chi_{B_5'\eta} + \frac{i}{4}j^2(\chi_{B_4}-\chi_{B_5'})\ .
\end{eqnarray}
The first contribution represents a mixed effect from baryonic and mesonic sector proportional to $m_qj$ while the second one does a pure baryonic effect proportional to $j^2$, with the susceptibilities defined by
\begin{eqnarray}
\chi_{B_5'\eta} &=& \int d^4x\langle0|{\rm T}{\cal O}_\eta(x){\cal O}_{B_5'}(0)|0\rangle \ ,\nonumber\\
\chi_{B_4} &=& \int d^4x\langle0|{\rm T}{\cal O}_{B_4}(x){\cal O}_{B_4}(0)|0\rangle   \ ,\nonumber\\
\chi_{B_5'} &=& \int d^4x\langle0|{\rm T}{\cal O}_{B_5'}(x){\cal O}_{B_5'}(0)|0\rangle  \ . \label{SusceptibilityWithJ}
\end{eqnarray}
Here we have defined the following composite operator of the negative-parity diquark:
\begin{eqnarray}
{\cal O}_{B_5'} = -\frac{1}{2}\psi^TC\tau_c^2\tau_f^2\psi + {\rm H.c.}\ .
\end{eqnarray}
The susceptibility functions~(\ref{SusceptibilityWithJ}) can be evaluated within the LSM framework similarly to Eqs.~(\ref{ChiEtaMatch}) and~(\ref{ChiPiMatch}).

The resultant topological susceptibilities with $j/m_q=0,0.05,0.1,0.18$ and $m_\eta^{\rm (H)}/m_\pi^{\rm (H)}=1.5$ are exhibited in the right panel of Fig.~\ref{fig:ChiTop}. As $j$ is increased the suppression of $\chi_{\rm top}$ is diminished. In particular, when $j/m_q=0.18$ the topological susceptibility is approximately constant in a range of $0<\mu_q\lesssim 1.2m_\pi^{\rm (H)}$.

\subsection{Sound velocity}
\label{sec:SoundVelocity}

\begin{figure}[H]
\centering
\includegraphics[width=7.5cm]{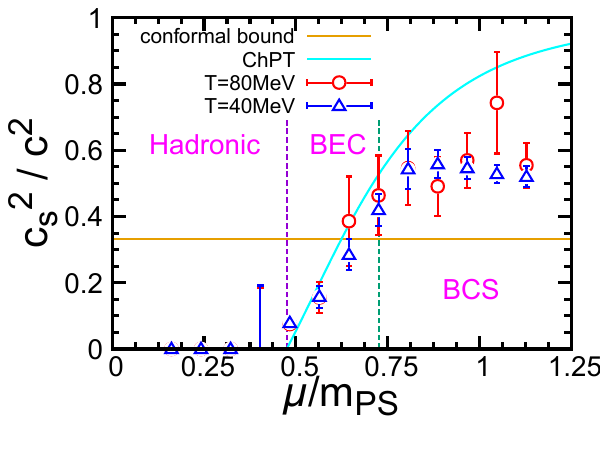}
\caption{The lattice result on the sound velocity at finite chemical potential. This figure is taken from Ref.~\cite{Iida:2024irv}.}
\label{fig:Cs2Lattice}
\end{figure}

Recently the sound velocity at low temperature was simulated on the lattice ~\cite{Iida:2022hyy,Iida:2024irv}, as exhibited in Fig.~\ref{fig:Cs2Lattice}, indicating that the sound velocity exceeds the conformal limit $\bar{c}_s^2=1/3$ for $\mu_q\gtrsim 0.7m_\pi^{\rm (H)}$. Meanwhile, we know that finally it must converge on the limiting value $\bar{c}_s^2$ from the following simple dimensional analysis. When the chemical potential is sufficiently large $\mu_q\gg \Lambda_{\rm QC_2D}$, the pressure $p$ takes the form of ($\alpha$ is some constant)
\begin{eqnarray}
p \sim \alpha \mu_q^4\ , \label{PressureAs}
\end{eqnarray}
since the system is dominated by only $\mu_q$. Hence, the number density and its susceptibility are derived to be $n=4\alpha\mu_q^3$ and $\chi= 12\alpha\mu_q^2$, resulting in
\begin{eqnarray}
c_s^2 \sim \frac{4\alpha\mu_q^3}{\mu_q\times 12\alpha\mu_q^2} = \frac{1}{3}\ ,
\end{eqnarray}
with the help of the formula~(\ref{SoundVelocityFormula}). Therefore, the lattice result implies the existence of peak structures of $c_s^2$ at some $\mu_q$.

As investigated in Sec.~\ref{sec:Thermodynamic}, the sound velocity evaluated within the ChPT exceeds the conformal value $\bar{c}_s^2 = 1/3$ but monotonically approaches $c_s^2 \sim 1$ at large $\mu_q$ without exhibiting any peaks~\cite{Hands:2006ve,Son:2000by}. This behavior contradicts the above simple dimensional analysis. Such contradiction emerges since the ChPT framework is constructed upon the manifold of $G/H$, which requires a definite energy scale to break certain symmetries. In fact, the pressure~(\ref{PressureChPT}) is always proportional to the decay constant $f_0$. On the other hand, the LSM is based on a linear representation of the Pauli-G\"ursey $SU(4)$ symmetry that naturally allows us to enter the symmetry restored phase. Thus, there is no intrinsic energy scale to characterize the symmetry breaking and the correct asymptotic behavior of the sound velocity is expected to be reproduced. Keeping this expectation in mind, here we examine the sound velocity within the LSM, particularly focusing on influence from the chiral-partner structure, as an extended model of the ChPT.

From the effective potential~(\ref{VEffLSM}), the appropriately subtracted pressure derived within the LSM is evaluated to be~\cite{Kawaguchi:2024iaw}
\begin{eqnarray}
p^{\rm sub}_{\rm LSM} = p^{\rm sub}_{\rm ChPT} + \delta p \ , \label{PsubLSM}
\end{eqnarray}
where $p^{\rm sub}_{\rm ChPT}$ is the subtracted pressure from the ChPT~(\ref{PressureChPT}). The additional contribution $\delta p $ is ($\bar{\mu} = \mu_q/\mu_{\rm cr} = 2\mu_q/m_\pi^{\rm (H)}$)
\begin{eqnarray}
\delta p = \big(f_\pi^{\rm (H)}\big)^2\big(m_\pi^{\rm (H)}\big)^2\left[\frac{4}{\delta\bar{m}_{\sigma-\pi}^2}(\bar{\mu}^2-1)^2\right]\ , \label{DeltaP}
\end{eqnarray}
with
\begin{eqnarray}
\delta\bar{m}_{\sigma-\pi}^2 = \frac{\big(m_\sigma^{\rm (H)}\big)^2-\big(m_\pi^{\rm (H)}\big)^2}{\mu_{\rm cr}^2}\ ,
\end{eqnarray}
and $f_\pi^{\rm (H)} = \sigma_0^{\rm (H)}/\sqrt{2}$. In this equation 
\begin{eqnarray}
\big(m_\pi^{\rm (H)}\big)^2 &=& m_0^2 + \tilde{\lambda}\big(\sigma_0^{\rm (H)}\big)^2 \ ,\nonumber\\
\big(m_\sigma^{\rm (H)}\big)^2 &=& m_0^2 + 3\tilde{\lambda}\big(\sigma_0^{\rm (H)}\big)^2 \ ,
\end{eqnarray}
are the masses of pion and sigma meson in the hadronic phase, so that 
\begin{eqnarray}
\delta\bar{m}_{\sigma-\pi}^2 = \frac{2\tilde{\lambda}\big(\sigma_0^{\rm (H)}\big)^2 }{\mu_{\rm cr}^2}\ .
\end{eqnarray}
Thus, in a limit of $\mu_q\to\infty$ we can see $\delta p \to \mu_q^4/\tilde{\lambda}$ which dominates over the ChPT result $p^{\rm sub}_{\rm ChPT}$ and 
\begin{eqnarray}
p_{\rm LSM}^{\rm sub} \to \frac{1}{\tilde{\lambda}}\mu_q^4\ .
\end{eqnarray}
This scaling is indeed consistent with the simple dimensional analysis~(\ref{PressureAs}). Notably the correction~(\ref{DeltaP}) is proportional to the inverse of the chiral-partner mass difference $\delta\bar{m}_{\sigma-\pi}^2$. In a limit of $m_\sigma^{\rm (H)}\to\infty$, $\delta p$ vanishes and the pressure is reduced to $p_{\rm ChPT}^{\rm sub}$, which is consistent with a fact that integrating out $\sigma$ meson from the LSM derives the ChPT.

From the pressure~(\ref{PsubLSM}), the energy density, the number density and its susceptibility are readily obtained in the following forms:
\begin{eqnarray}
\epsilon^{\rm sub}_{\rm LSM} &=& \epsilon^{\rm sub}_{\rm ChPT} + \delta \epsilon \ , \nonumber\\
n^{\rm sub}_{\rm LSM} &=& n^{\rm sub}_{\rm ChPT} + \delta n \ , \nonumber\\
\chi^{\rm sub}_{\rm LSM} &=& \chi^{\rm sub}_{\rm ChPT} + \delta \chi \ ,
\end{eqnarray}
with the corrections evaluated as
\begin{eqnarray}
\delta \epsilon &=& \big(f_\pi^{\rm (H)}\big)^2\big(m_\pi^{\rm (H)}\big)^2\left[\frac{4}{\delta\bar{m}_{\sigma-\pi}^2}(3\bar{\mu}^2+1)(\bar{\mu}^2-1)\right]\ , \nonumber\\
\delta n &=& \frac{2\big(f_\pi^{\rm (H)}\big)^2\big(m_\pi^{\rm (H)}\big)^2}{\mu_q}\left[\frac{8}{\delta\bar{m}_{\sigma-\pi}^2}(\bar{\mu}^4-\bar{\mu}^2)\right]\ , \nonumber\\
\delta \chi &=& 8\big(f_\pi^{\rm (H)}\big)^2\left[\frac{8}{\delta\bar{m}_{\sigma-\pi}^2}(3\bar{\mu}^2-1)\right]\ .
\end{eqnarray}
All these corrections vanish when taking $m_\sigma\to\infty$ so as to reproduce the corresponding ChPT results. The resultant sound velocity is given by
\begin{eqnarray}
\big(c_s^{\rm LSM}\big)^2 = \frac{n_{\rm ChPT} + \delta n}{\mu_q(\chi_{\rm ChPT} + \delta \chi)} = \frac{\left(1-1/\bar{\mu}^4\right) + 8\left(\bar{\mu}^2-1\right)/\delta\bar{m}^2_{\sigma-\pi}}{\left(1+3/\bar{\mu}^4\right) + 8\left(3\bar{\mu}^2-1\right)/\delta\bar{m}^2_{\sigma-\pi}}\ . \label{Cs2LSM}
\end{eqnarray}

\begin{figure}[H]
\centering
\includegraphics[width=8cm]{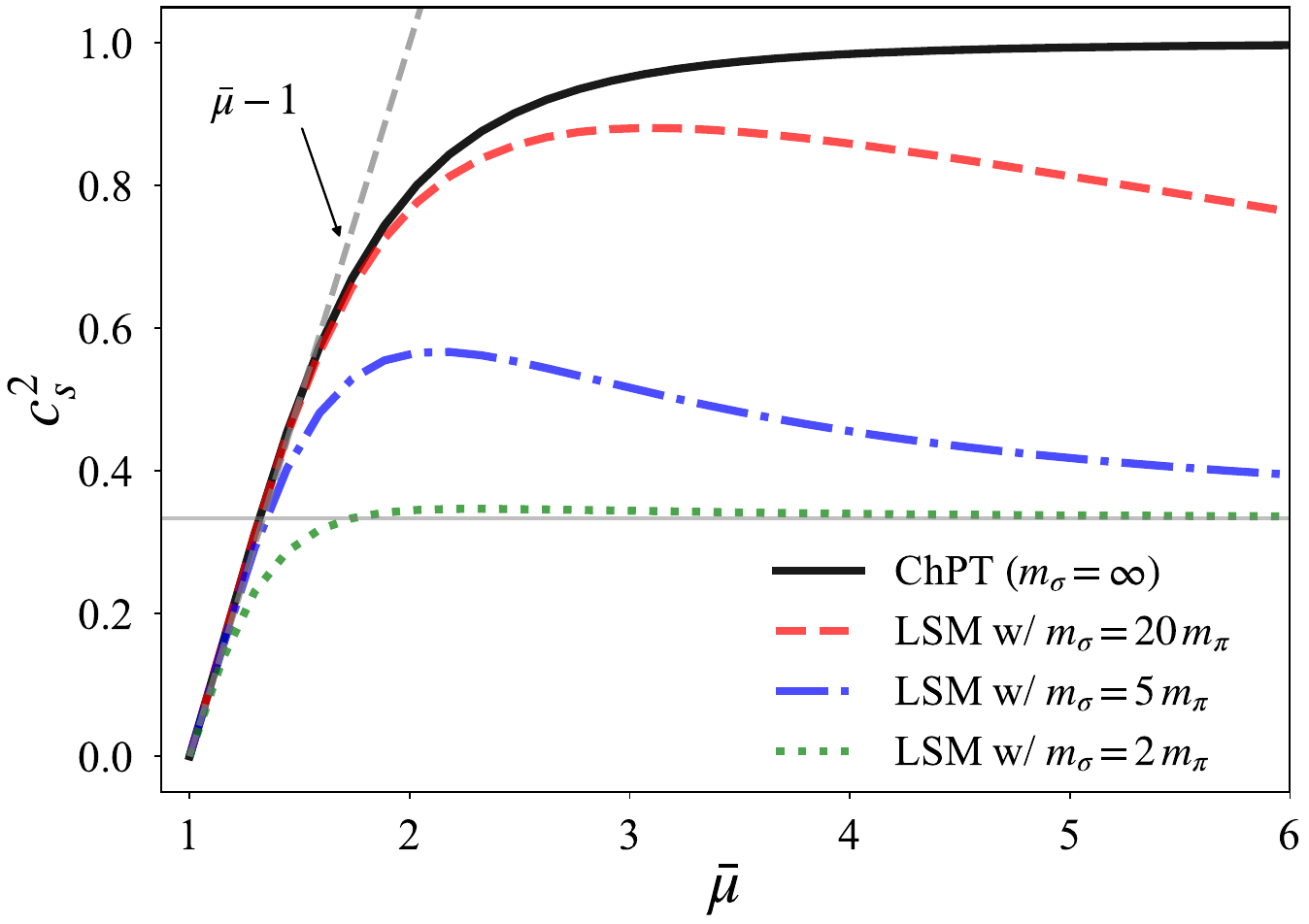}
\caption{$\bar{\mu}$ ($=\mu_q/\mu_{\rm cr} = 2\mu_q/m_\pi^{\rm (H)}$) dependences of the sound velocity $c_s^2$ with $m_\sigma^{\rm (H)}/m_\pi^{\rm (H)} = 2,5,20, \infty$. The dashed gray line denotes $c_s^2 = \bar{\mu}-1$ evaluated analytically. This figure is taken from Ref.~\cite{Kawaguchi:2024iaw}.}
\label{fig:SoundLSM}
\end{figure}   

Depicted in Fig.~\ref{fig:SoundLSM} is $\mu_q$ dependence of the sound velocity~(\ref{Cs2LSM}) with $m_\sigma^{\rm (H)}/m_\pi^{\rm (H)} = 2,5,20$ and $\infty$. The gray dashed line is an analytic solution expanded in the vicinity of $\mu_q\approx \mu_{\rm cr}$ in Eq.~(\ref{Cs2LSM}): $c_s^2\approx\bar{\mu}-1$, which is independent of $m_\sigma^{\rm (H)}$. This figure shows that the sound velocity peak is successfully reproduced within the present LSM where the chiral-partner contribution proportional to $1/\delta\bar{m}_{\sigma-\pi}^2$ is incorporated~\cite{Kawaguchi:2024iaw}. Thus, from this reproduction one can conclude that the LSM is capable of accessing more dense regime of QC$_2$D where the ChPT cannot apply. For quantitative comparisons, it is inevitable to include fluctuations and spin-$1$-hadron contributions.

\section{Extended linear sigma model (eLSM)}
\label{sec:ELSM}

\subsection{Model construction}
\label{sec:ELSMModel}

In Sec.~\ref{sec:HadronMassLSM} mass spectra of negative-parity as well as positive-parity spin-$0$ hadrons in cold and dense QC$_2$D has been explored within the LSM framework, and $\mu_q$ dependences of the hadron masses have been elucidated from symmetry viewpoints. Meanwhile, the lattice simulation indicates the flipping of the mass ordering where pion becomes heavier than $\rho$ meson in the superfluid phase~\cite{Hands:2007uc,Murakami:2022lmq}. This behavior implies that a model analysis including spin-$1$ hadrons is inevitable to further correctly explore low-energy physics of dense QC$_2$D. Thus, here, we invent the eLSM describing spin-$0$ and spin-$1$ hadrons in a unified way based on the linear representation of the Pauli-G\"ursey $SU(4)$ symmetry~\cite{Suenaga:2023xwa}.\footnote{The eLSM in three-color QCD was invented by Frankfurt group~\cite{Parganlija:2010fz,Parganlija:2012fy}.}

\begin{table}[htbp]
\caption{Quantum numbers of the spin-$1$ hadrons. }
\label{tab:Spin1}
\begin{center}
  \begin{tabular}{c||c|c|c} \hline \hline
\textbf{Hadrons} & \textbf{Spin and parity ($J^P$)} & \textbf{Quark number} & \textbf{Isospin}\\ \hline
$\omega$ & $1^-$ & $0$ & $0$ \\ 
$\rho$ & $1^-$ & $0$ & $1$ \\
$f_1$ & $1^+$ & $0$ & $0$ \\
$a_1$ & $1^+$ & $0$ & $1$ \\
$B_S$ ($\bar{B}_S$) & $1^+$ & $+2$ ($-2$) & $1$ \\
$B_{AS}$ ($\bar{B}_{AS}$) & $1^-$ & $+2$ ($-2$) & $0$ \\ \hline\hline
 \end{tabular}
\end{center}
\end{table}

Employing the linear representation, interpolating fields of low-lying spin-$1$ mesons and baryons are given by
\begin{eqnarray}
&& \omega^\mu \sim \bar{\psi}\gamma^\mu\psi \ , \ \ f_1^\mu \sim \bar{\psi}\gamma_5\gamma^\mu\psi \ , \ \  \rho^{0,\mu} \sim \bar{\psi}\tau_f^3\gamma^\mu\psi \ , \nonumber\\
&& \rho^{\pm,\mu} \sim \frac{1}{\sqrt{2}}\bar{\psi}\tau_f^\mp\gamma^\mu\psi  \ , \ \  a_1^{0,\mu} \sim \bar{\psi}\tau_f^3\gamma_5\gamma^\mu\psi \ , \ \ a_1^{\pm,\mu} \sim \frac{1}{\sqrt{2}}\bar{\psi}\tau_f^\mp\gamma_5\gamma^\mu\psi \ ,  \label{Spin1Meson}
\end{eqnarray}
and 
\begin{eqnarray}
&& B_{S}^{I_z=0,\mu} \sim-\frac{i}{\sqrt{2}}\psi^T{\cal C}\gamma^\mu\tau_c^2\tau_f^1\psi\ , \ \  B_{S}^{I_z=\pm1,\mu} \sim -\frac{i}{2}\psi^T{\cal C}\gamma^\mu\tau_c^2({\bm 1}_f\pm \tau_f^3)\psi \ , \nonumber\\
&& B_{AS}^\mu \sim -\frac{1}{\sqrt{2}}\psi^T{\cal C}\gamma_5\gamma^\mu\tau_c^2\tau_f^2\psi \ , \ \  \bar{B}_{S}^{I_z=0,\pm1,\mu} = (B_{S}^{I_z=0,\pm1,\mu})^\dagger \ , \ \  \bar{B}_{AS}^\mu = (B_{AS}^\mu)^\dagger \ , \label{Spin1Baryon}
\end{eqnarray}
respectively, the quantum numbers of which are summarized in Table~\ref{tab:Spin1}. Then, a useful $4\times4$ matrix describing quark bilinear fields of the spin-$1$ hadrons is introduced as
\begin{eqnarray}
\Phi^\mu_{ij} \sim \Psi_j^\dagger\sigma^\mu\Psi_i\ , \label{PhiMuInt}
\end{eqnarray}
as a sibling of $\Sigma$ in Eq.~(\ref{PhiPAssign}), for which the hadron fields can be embedded in the following manner:
\begin{eqnarray}
\Phi^\mu = \frac{1}{2} \left(
\begin{array}{cccc}
\frac{\omega+\rho^{0}-(f_1+a_1^0)}{\sqrt{2}}  & \rho^+-a_1^+& \sqrt{2}  B_S^{I_z=+1} & B_S^{I_z=0}-B_{AS} \\
\rho^--a_1^- & \frac{\omega-\rho^{0}-(f_1-a_1^0)}{\sqrt{2}}  & B_S^{I_z=0} + B_{AS} & \sqrt{2}B_S^{I_z=-1} \\
\sqrt{2}\bar{B}_S^{I_z=-1} & \bar{B}_S^{I_z=0}+ \bar{B}_{AS} & -\frac{\omega+\rho^{0}+f_1+a_1^0}{\sqrt{2}} & -(\rho^-+a_1^-) \\
\bar{B}_S^{I_z=0}-\bar{B}_{AS} & \sqrt{2}\bar{B}_S^{I_z=+1} & -(\rho^++a_1^+) & -\frac{\omega-\rho^{0}+f_1-a_1^0}{\sqrt{2}} \\
\end{array}
\right)^\mu \ ,\label{PhiDef}
\end{eqnarray}
from Eqs.~(\ref{Spin1Meson}) and~(\ref{Spin1Baryon}). This matrix is reduced to
\begin{eqnarray}
\Phi^\mu = \left(\sum_{i=1}^{10}V^i S^i -\sum_{a=0}^5V'^aX^a \right)^\mu \label{PhiHadron}
\end{eqnarray}
such that symmetry properties of the spin-$1$ hadrons become clear, when assigning the hadron fields as
\begin{eqnarray}
&& \omega = V^0\ , \ \  \rho^\pm = \frac{V^1\mp iV^2}{\sqrt{2}}\ , \ \ \rho^0 = V^3\ , \ \  f_1=V'^0 \ , \ \ a_1^\pm = \frac{V'^1\mp iV'^2}{\sqrt{2}}\ , \ \ a_1^0 = V'^3\ , \nonumber\\
&& B_S^{I_z=0} = \frac{V^9 + iV^{10}}{\sqrt{2}}\ , \ \  \bar{B}_S^{I_z=0} = \frac{V^9 - iV^{10}}{\sqrt{2}}  \ , \ \  B_S^{I_z=\pm1} = \frac{(V^5+iV^6) \pm (V^7+iV^8)}{2} \ , \nonumber\\
&& \bar{B}_S^{I_z=\pm1} = \frac{(V^5-iV^6) \mp (V^7-iV^8)}{2} \ , \ \ B_{AS} = \frac{V'^5-iV'^4}{\sqrt{2}}\ , \ \ \bar{B}_{AS} = \frac{V'^5+iV'^4}{\sqrt{2}}\ .
\end{eqnarray}
The Pauli-G\"ursey $SU(4)$ transformation law of $\Phi^\mu$ is
\begin{eqnarray}
\Phi^\mu \to g \Phi^\mu g^\dagger \ \ \ {\rm with}\ \ \  g\in SU(4)\ , \label{PhiSU4}
\end{eqnarray}
from Eq.~(\ref{PhiMuInt}).

From the transformation laws~(\ref{SigmaSU4}) and~(\ref{PhiSU4}), an effective Lagrangian describing the low-lying spin-$0$ and spin-$1$ hadrons of QC$_2$D comprehensively, i.e., eLSM Lagrangian, is readily obtained as~\cite{Suenaga:2023xwa}
\begin{eqnarray}
{\cal L}_{\rm eLSM} &=& {\rm tr}[{D}_\mu \Sigma^\dagger {D}^\mu\Sigma]-m_0^2{\rm tr}[\Sigma^\dagger\Sigma]-\lambda_1\big({\rm tr}[\Sigma^\dagger\Sigma]\big)^2-\lambda_2{\rm tr}[(\Sigma^\dagger\Sigma)^2] \nonumber\\
&& +\bar{c}{\rm tr}[\zeta^\dagger\Sigma+\Sigma^\dagger \zeta]  +{\cal L}_{\rm anom.}- \frac{1}{2}{\rm tr}[\Phi_{\mu\nu}\Phi^{\mu\nu}] + m_1^2{\rm tr}[\Phi_\mu\Phi^\mu] \nonumber\\
&& + ig_3{\rm tr}\big[\Phi_{\mu\nu}[\Phi^\mu,\Phi^\nu]\big] + h_1{\rm tr}[\Sigma^\dagger\Sigma]{\rm tr}[\Phi_\mu\Phi^\mu] + h_2{\rm tr}[\Sigma\Sigma^\dagger\Phi_\mu\Phi^\mu] \nonumber\\
&& + h_3{\rm tr}[\Phi_\mu^T\Sigma^\dagger\Phi^\mu\Sigma] + g_4{\rm tr}[\Phi_\mu\Phi_\nu\Phi^\mu\Phi^\nu] + g_5{\rm tr}[\Phi_\mu\Phi^\mu\Phi_\nu\Phi^\nu] \nonumber\\
&& + g_6{\rm tr}[\Phi_\mu\Phi^\mu]{\rm tr}[\Phi_\nu\Phi^\nu] + g_7{\rm tr}[\Phi_\mu\Phi_\nu]{\rm tr}[\Phi^\mu\Phi^\nu] \ .
\label{ELSMLag}
\end{eqnarray}
In this Lagrangian
\begin{eqnarray}
\Phi_{\mu\nu} \equiv D_\mu\Phi_\nu-D_\nu\Phi_\mu 
\end{eqnarray}
is a field strength for the spin-$1$ hadrons, and the covariant derivatives for $\Sigma$ and $\Phi$ take the forms of
\begin{eqnarray}
&& D_\mu\Sigma \equiv \partial_\mu\Sigma-i\zeta_\mu\Sigma-i\Sigma \zeta^T_\mu-ig_1\Phi_\mu\Sigma-ig_2\Sigma\Phi_\mu^T\ , \nonumber\\
&& D_\mu \Phi_\nu \equiv \partial_\mu\Phi_\nu-i[\zeta_\mu,\Phi_\nu]\ ,
\end{eqnarray}
respectively, with the spurion field $\zeta^\mu$. It should be noted that
\begin{eqnarray}
&& {\rm tr}[D_\mu\Sigma^\dagger D^\mu\Sigma] + h_2{\rm tr}[\Sigma\Sigma^\dagger\Phi_\mu\Phi^\mu] + h_3{\rm tr}[\Phi_\mu^T\Sigma^\dagger\Phi^\mu\Sigma] \nonumber\\
 &=& {\rm tr}[\partial_\mu\Sigma^\dagger\partial^\mu\Sigma] + (g_1+g_2){\rm tr}[\Sigma\Sigma^\dagger\{\Phi_\mu, \zeta^\mu\} ]+ 2(g_1+g_2){\rm tr}[\Phi_\mu^T\Sigma^\dagger \zeta^\mu\Sigma] \nonumber\\
&+& i(g_1+g_2){\rm tr}[\Phi_\mu(\partial^\mu\Sigma\Sigma^\dagger-\Sigma\partial^\mu\Sigma^\dagger)] +(g_1^2+g_2^2+h_2){\rm tr}[\Sigma\Sigma^\dagger\Phi_\mu\Phi^\mu] \nonumber\\
&+& (2g_1g_2+h_3) {\rm tr}[\Phi_\mu^T\Sigma^\dagger\Phi^\mu\Sigma]\label{DerivativeExp}
\end{eqnarray}
holds, from which $\Sigma^T=-\Sigma$, implying that the four parameters $g_1$, $g_2$, $h_2$ and $h_3$ can be rearranged into the following three ones:
\begin{eqnarray}
C_1 &\equiv& g_1+g_2\ , \nonumber\\
C_2 &\equiv& g_1^2+g_2^2 + h_2\ , \nonumber\\
C_3 &\equiv& 2g_1g_2+h_3\ . \label{C123Def}
\end{eqnarray}

The eLSM Lagrangian~(\ref{ELSMLag}) contains effectively $14$ parameters, regardless of the anomalous contributions, which are hard to be completely fixed owing to the current limited lattice results. Here, to pick up only leading contributions, first, we assume the large $N_c$ limit that would be also supported by the so-called Zweig rule for spin-$1$ sectors. Thus, $\lambda_1=h_1=g_6=g_7=0$. Also, we again ignore the anomaly effects. Next, we assume $C\equiv C_1=C_2$ since essentially those parameters play the same role that controls mixings between spin-$0$ and spin-$1$ hadrons. Then, as for the couplings among spin-$1$ hadrons, we employ the following relations:
\begin{eqnarray}
g_3 = g_\Phi\ , \ \ g_4=-g_5=g_\Phi^2\ , \label{GPhiDef}
\end{eqnarray}
which can be inferred by the ${\cal O}(p^2)$ contributions of the HLS formalism~\cite{Harada:2010vy}. After those reduction, seven free parameters are left.

\subsection{Hadron mass spectrum}
\label{sec:ELSMMass}

In this subsection, we investigate $\mu_q$ dependences of the spin-$1$ hadron masses predicted by our eLSM.

Toward delineating the hadron mass spectrum, we need to take into account mean field contributions appropriately. In the present analysis we assume the following four mean fields:
\begin{eqnarray}
 \sigma_0 = \langle \sigma\rangle \ , \ \ \Delta = \langle B^5\rangle\ , \ \  \bar{\omega} = \langle\omega_{\mu=0}\rangle\ , \ \  \bar{V} =\langle V'^4_{\mu=0}\rangle\ . \label{MeanFields}
\end{eqnarray}
The spin-$0$ mean fields $\sigma_0$ and $\Delta$ correspond to the chiral and diquark condensates, respectively, similarly to the analysis in Sec.~\ref{sec:HadronMassLSM}. The third one, $\bar{\omega}$, is a mean field of $\omega$ meson modifying the chemical potential effects. The last one, $\bar{V}$, is responsible for a mean field of the iso-singlet and vector diquark, which is allowed by the $U(1)_B$ violation in the baryon superfund phase. The configuration of those mean field is determined by solving each the stationary condition. The resultant gap equations are rather complicated, so we leave their concrete forms to Ref.~\cite{Suenaga:2023xwa}.

\begin{figure}[H]
\centering
\includegraphics[width=8cm]{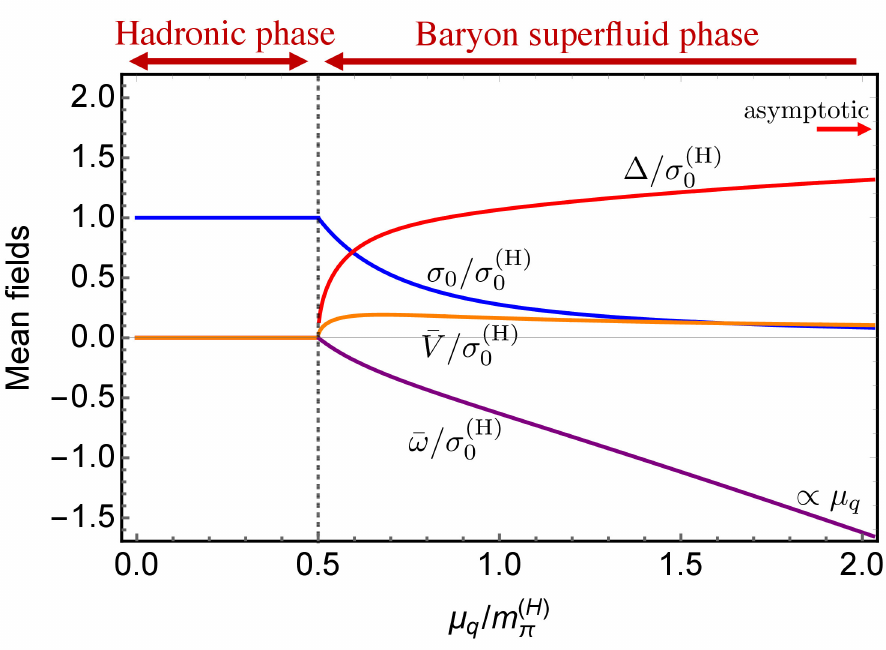}
\caption{$\mu_q$ dependences of the four mean fields: $\sigma_0$, $\Delta$, $\bar{\omega}$ and $\bar{V}$. The figure is taken from Ref.~\cite{Suenaga:2023xwa}.}
\label{fig:MeanFieldFour}
\end{figure}

In the following numerical analysis, we will adopt
\begin{eqnarray}
m^{(\rm H)}_\rho = 908\, {\rm MeV} \ , \ \ m_{a_1}^{\rm (H)} = 1614\, {\rm MeV} \ , \label{InputSpin1}
\end{eqnarray}
as inputs associated with the spin-$1$ hadron masses simulated on the lattice~\cite{Murakami:2022lmq,Murakami2022}, in addition to the inputs~(\ref{InputSpin0}) and~(\ref{TypicalSigma}). Hence, there remains only two free parameters, $C$ and $g_\Phi$. When choosing $C=12$, $\mu_q$ dependences of the mean fields~(\ref{MeanFields}) can be determined as exhibited in Fig.~\ref{fig:MeanFieldFour}, regardless of the value of $g_\Phi$.

Figure.~\ref{fig:MeanFieldFour} indicates that only $\sigma_0$ is finite in the hadronic phase whereas the remaining mean fields are always vanishing there. In the superfluid phase induced by a nonzero $\Delta$, the spin-$1$ mean fields $\bar{\omega}$ and $\bar{V}$ also acquire their non-vanishing values. In particular, $\bar{\omega}$ grows linearly with respect to $\mu_q$. Meanwhile, the gap $\Delta$ converges on a certain value at sufficiently large $\mu_q$, which is indicated by the arrow in this figure. The remaining mean fields $\sigma_0$ and $\bar{V}$ asymptotically vanish for $\mu_q\to\infty$. It should be noted that the critical chemical potential to enter the baryon superfluid phase is again given by $\mu_{\rm cr} = m_\pi^{\rm (H)}/2$ as the other chiral effective models predict, as long as we take into account the additional two spin-$1$ meson fields correctly. 


\begin{figure}[H]
  \begin{center}
    \begin{tabular}{cc}

      \begin{minipage}[c]{0.5\hsize}
       \centering
       \hspace*{-2.5cm} 
         \includegraphics*[scale=0.35]{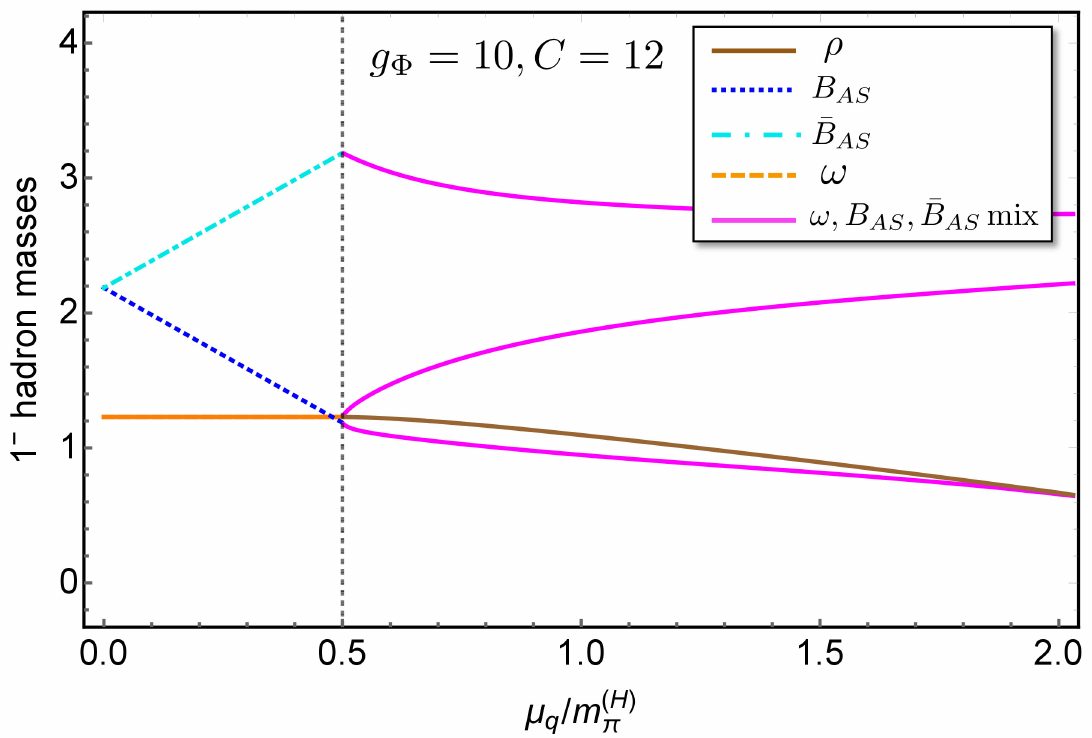}\\
         \end{minipage}

      \begin{minipage}[c]{0.4\hsize}
       \centering
        \hspace*{-1.1cm} 
          \includegraphics*[scale=0.35]{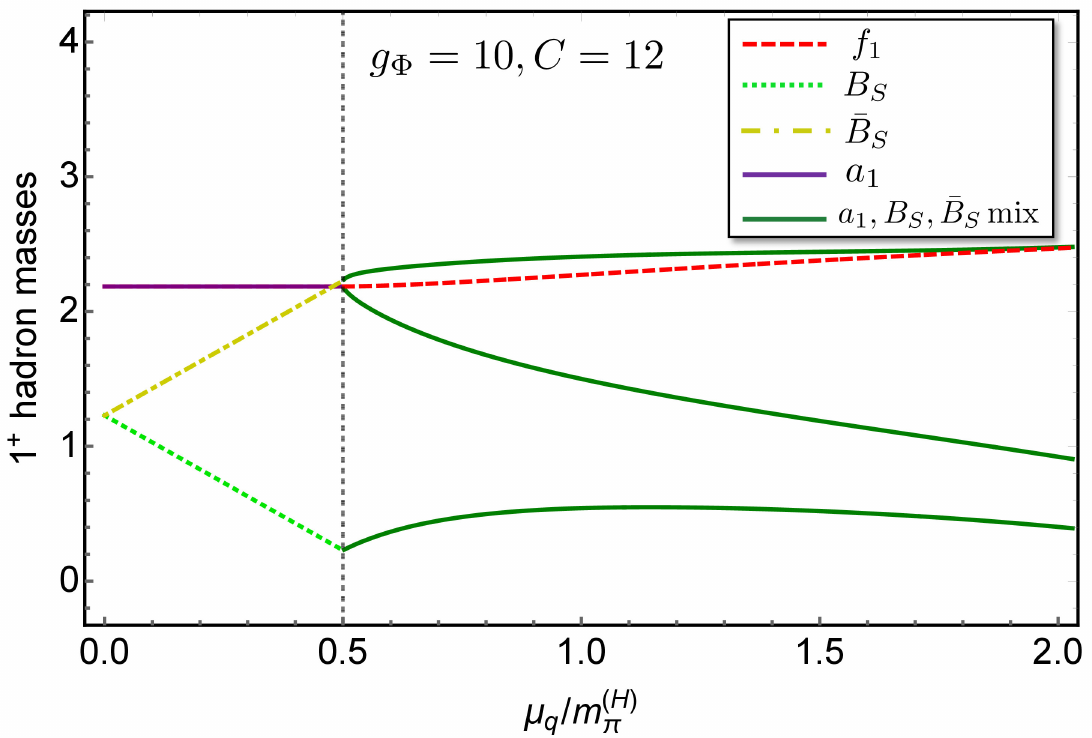}\\
      \end{minipage}

      \end{tabular}
 \caption{$\mu_q$ dependences of the $1^-$ (left) and $1^+$ (right) hadron masses with $g_\Phi=10$ and $C=12$. The figures are taken from Ref.~\cite{Suenaga:2023xwa}. } 
\label{fig:Spin1_G10_C12}
  \end{center}
\end{figure}
\begin{figure}[H]
  \begin{center}
    \begin{tabular}{cc}

      \begin{minipage}[c]{0.5\hsize}
       \centering
       \hspace*{-2.5cm} 
         \includegraphics*[scale=0.35]{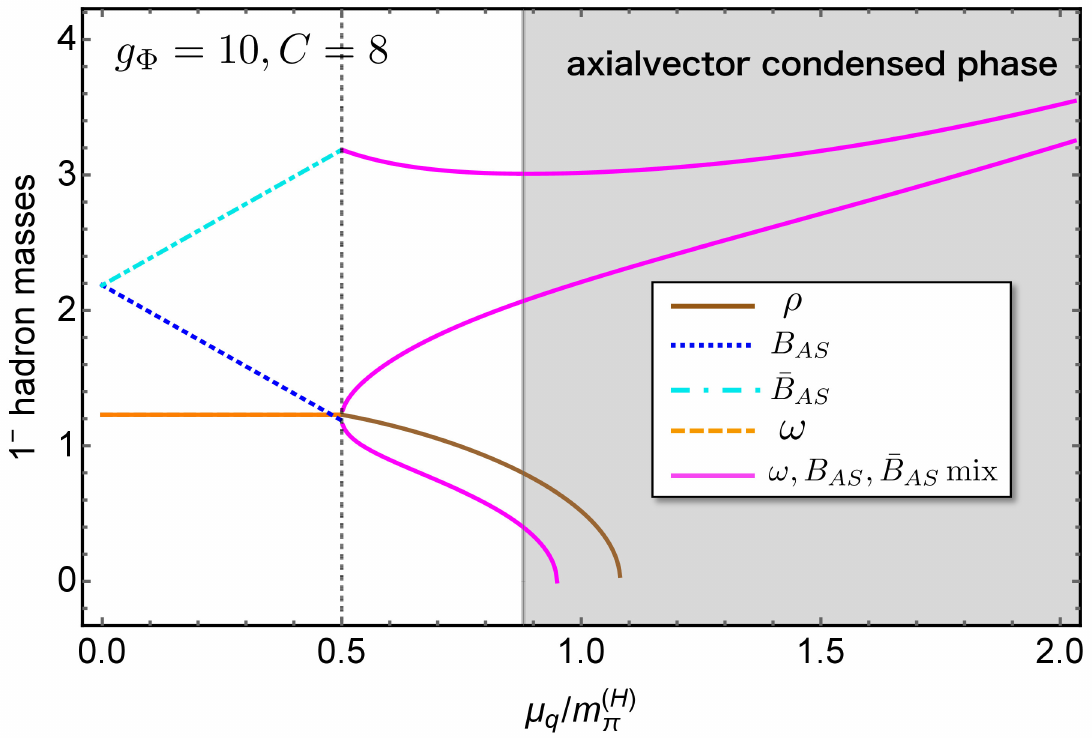}\\
         \end{minipage}

      \begin{minipage}[c]{0.4\hsize}
       \centering
        \hspace*{-1.1cm} 
          \includegraphics*[scale=0.35]{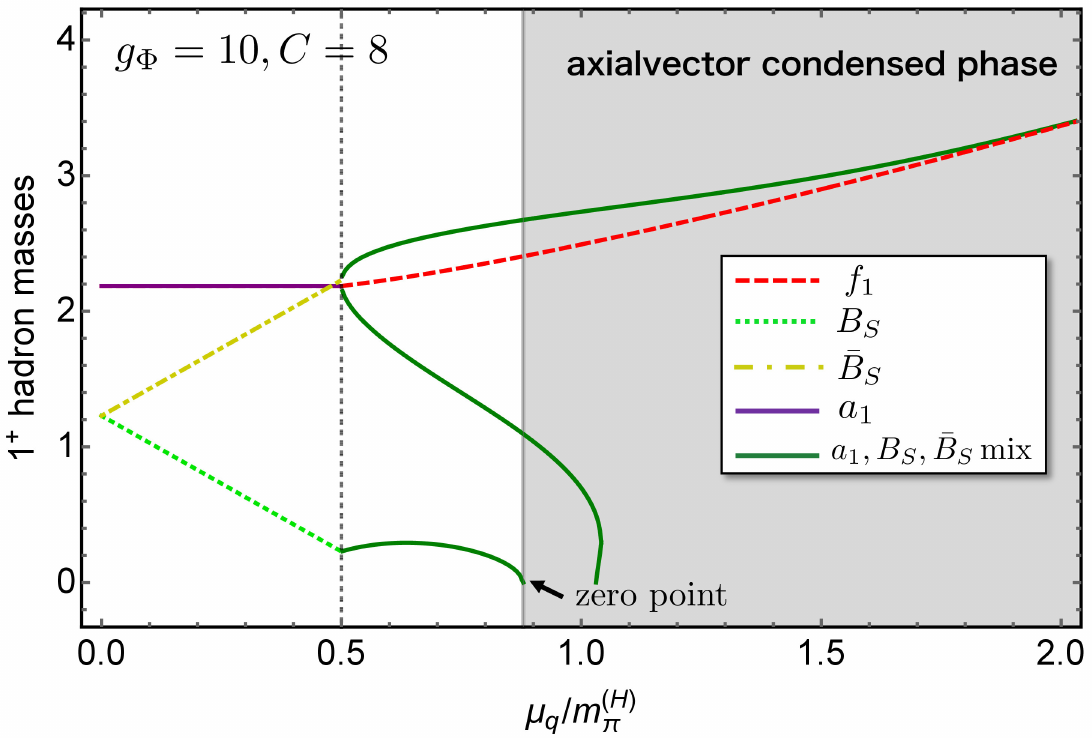}\\
      \end{minipage}

      \end{tabular}
 \caption{$\mu_q$ dependences of the $1^-$ (left) and $1^+$ (right) hadron masses with $g_\Phi=10$ and $C=8$. The figures are taken from Ref.~\cite{Suenaga:2023xwa}. } 
\label{fig:Spin1_G10_C8}
  \end{center}
\end{figure}

We are now ready to examine the hadron mass spectra of the spin-$1$ hadrons at finite $\mu_q$, since their mass formulas are straightforwardly obtained by reading off the quadratic terms from the reduced eLSM Lagrangian. The resulting formulas are complicated due to considerable mixings, so that we do not present here. (For the detail please see Appendixes of Ref.~\cite{Suenaga:2023xwa}.)

Depicted in Figs.~\ref{fig:Spin1_G10_C12} and~\ref{fig:Spin1_G10_C8} are the spin-$1$ hadron mass spectrum with $(g_\Phi,C)=(10,12)$ and $(g_\Phi,C)=(10,8)$, respectively, where the parameters are tuned to reproduce the mass reduction of $\rho$ meson in the superfluid phase measured on the lattice. The figures indicate that all the spin-$1$ hadron masses are constant or just linearly corrected with $\mu_q$ in the hadronic phase, similarly to the spin-$0$ meson masses. In the superfluid phase, meanwhile, several nonlinear behaviors are obtained owing to state mixings from the $U(1)_B$ violation; The three pink curves in the left panels denote the $\omega$-$B_{\rm AS}$-$\bar{B}_{\rm AS}$ mixed stats, while the green ones in the right panels denote the $a_1$-$B_{\rm S}$-$\bar{B}_{\rm S}$ mixed stats. In Fig.~\ref{fig:Spin1_G10_C8}, the colored area represents the axialvector condensed phase triggered by the mass of the lowest state of $a_1$-$B_{\rm S}$-$\bar{B}_{\rm S}$ mixed mode converges on zero. Possibility of the (axial)vector condensations were also predicted in Ref.~\cite{Lenaghan:2001sd}, although the gap equation to determine the ground-state configuration was not solved consistently. Thus, it would be challenging to seek for such $SO(3)$-violating phases in the future lattice simulations.

\begin{figure}[H]
\centering
\includegraphics[width=9cm]{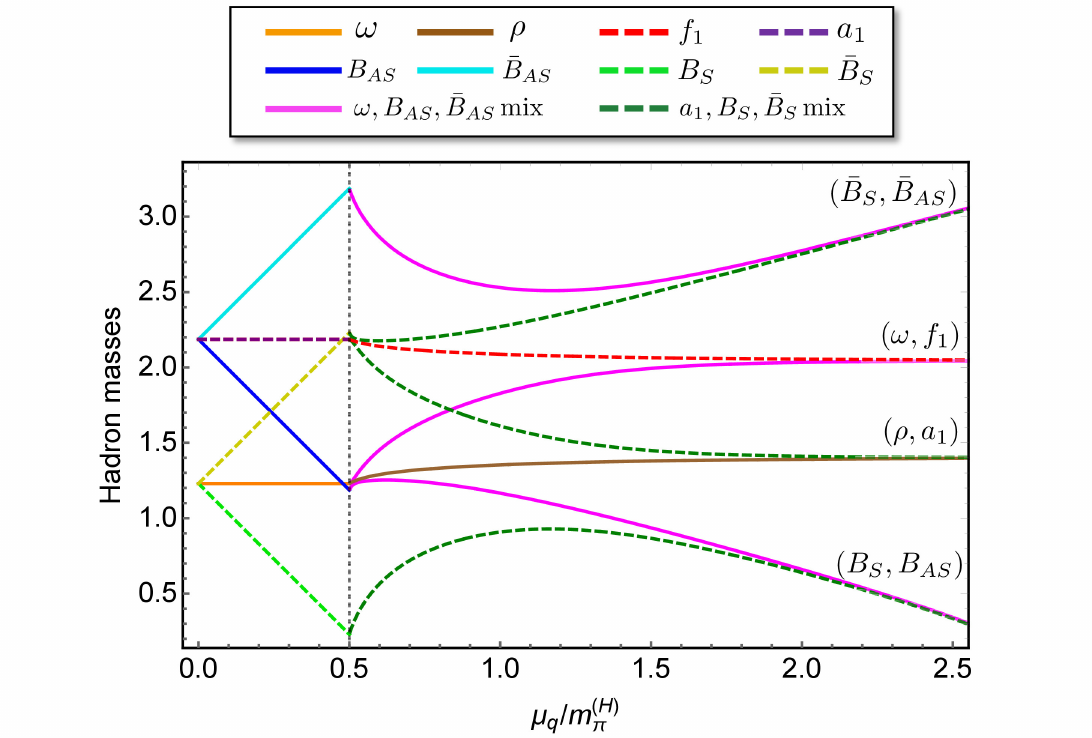}
\caption{$\mu_q$ dependences of all the spin-$1$ hadron masses. We have employed $(g_\Phi,C) = (10,16)$ to see the mass degeneracies of chiral partners clearly. This figure is taken from Ref.~\cite{Suenaga:2023xwa}.}
\label{fig:ChiralPartner1}
\end{figure}   

The mass degeneracies between the parity partners, i.e., chiral partners, are realized among the spin-$1$ hadrons similar to the spin-$0$ ones. To see this behavior we show $\mu_q$ dependences of the masses of all $1^\pm$ hadrons in Fig.~\ref{fig:ChiralPartner1}. In this figure we have adopted $(g_\Phi,C)=(10,16)$ so as to clearly confirm the mass degeneracies and plotted up to $\mu_q=2.5$. This figure indicates that the degeneracies hold for $(B_S,B_{AS})$, $(\rho,a_1)$, $(\omega,f_1)$ and $(\bar{B}_S,\bar{B}_{AS})$.

\section{Conclusions}
\label{sec:Conclusion}

In this review, I have summarized the main points of recent works on cold and dense QC$_2$D by means of the LSM, which is capable of describing the low-energy hadron spectrum in the baryon superfluid phase correctly~\cite{Suenaga:2022uqn,Kawaguchi:2023olk,Suenaga:2023xwa,Kawaguchi:2024iaw}, as an extension of the ChPT. 

As for the spin-$0$ hadron mass spectrum, in the baryon superfluid phase, the LSM has yielded a massless (the lowest) mode in the iso-singlet $0^+$ system which can be regarded as the NG boson of $U(1)_B$ symmetry breaking. Besides, a nonlinearly suppressed second-lowest mode has been found in the iso-singlet $0^-$ system. Those lowest-lying behaviors are qualitatively consistent with the lattice results. From a quantitative comparison of the latter nonlinear mass suppression, an enhancement of the $U(1)_A$ anomaly effects on the hadrons has been predicted. The mass spectrum of $0^\pm$ hadrons and some GOR relations in the presence of the diquark source have also been newly evaluated.

As for the spin-$1$ hadron mass spectrum, we could find several parameter sets for which the $\rho$ meson mass reduction in the superfluid phase observed by the lattice simulation is reproduced. Then, a possibility of the (axial)vector condensations violating the $SO(3)$ rotational symmetry has been discussed. For both the spin-$0$ and spin-$1$ hadrons, mass degeneracies between the parity partners, i.e., the chiral-partner structure, at higher density have been predicted.

Our WTI-based LSM analysis has implied that topological susceptibility in cold and dense QC$_2$D is suppressed followed by the chiral-symmetry restoration. If the $U(1)_A$ anomaly effect is assumed to be enhanced in such dense system, however, the suppression is weakened. Thus, the fate of the topological susceptibility largely depends on the behavior of $U(1)_A$ anomaly at hadronic level.

We have also seen that the peak structure of a (squared) sound velocity in the superfluid phase can be successfully reproduced within our LSM framework whereas the ChPT analysis cannot. This fact, in addition to the reproduction of the low-energy hadron spectrum in the superfluid phase, implies that the LSM constructed upon the linear representation of quark fields is applicable in the deeper regime of the crossover from hadronic to quark matter.

In what follows, I will present some topics related to the QC$_2$D study.
Similarly to QC$_2$D, the isospin QCD (QCD$_I$) where the isospin chemical potential is included, can also be regarded as one of the useful testing grounds toward elucidation of cold and dense QCD, thanks to disappearance of the sign problem in lattice simulations~\cite{Son:2000xc,Son:2000by,Splittorff:2000mm,Lu:2019diy,GomezNicola:2022asf,Brandt:2022hwy,Abbott:2023coj,Abbott:2024vhj}. The present LSM is easily translated into the QCD$_I$ language, and hence, dense QCD$_I$ would be another helpful field in order to check the results harvested from the LSM analysis in QC$_2$D.  Examinations in those systems are expected to provide useful informations on the equation of state of dense matter which are crucial to explain observation data of neutron stars~\cite{Komoltsev:2021jzg,Koehn:2024set}.

QC$_2$D is not only useful for delineating cold and dense QCD but also related to dark matter candidates, such as strongly interacting massive particles~\cite{Hochberg:2014kqa,Detmold:2014kba,Hochberg:2015vrg,Kamada:2022zwb,Kulkarni:2022bvh,Chu:2024rrv,Dengler:2024maq}. In this regard, it would be intriguing if the present LSM is capable of contributing to those beyond standard analyses.

QC$_2$D has an advantage that (anti)diquarks are counted as color-singlet hadrons, while in three-color QCD they cannot be a direct observable. In the latter real-life world, diquark properties play an important role in determining chiral dynamics of singly heavy baryons (SHBs) made of a heavy quark and a diquark, by virtue of the heavy-quark effective theory~\cite{manohar2000heavy}. Thus, examination of  diquarks in QC$_2$D from both theoretical and lattice studies is expected to provide useful information on the SHB spectroscopy e.g., the so-called ``inverse mass hierarchy'' induced by the $U(1)_A$ anomaly for the unobserved chiral partner SHBs~\cite{Harada:2019udr,Suenaga:2023tcy,Suenaga:2024vwr}. As long as we stick to zero chemical potential, lattice simulations with $2+1$ flavors are straightforward for any number of color.\footnote{In three-color QCD, lattice studies on the diquarks by means of, e.g., gauge-fixing treatment, potential problem, and static color-source method, are being conducted~\cite{Hess:1998sd,Alexandrou:2006cq,Babich:2007ah,Bi:2015ifa,Francis:2021vrr,Watanabe:2021nwe}.} In this regard, lattice simulations focused on diquarks in QC$_2$D with $N_f=2+1$ would be a challenging issue toward elucidation of the SHB properties in our world from chiral symmetry and the $U(1)_A$ anomaly.

Those applications imply that although QC$_2$D is a ``virtual'' theory played by QCD-like quarks and gluons, plenty of benefits are expected broadly, not to mention the numerical experiments in cold and dense medium.

\section*{Acknowledgments}
The author was supported by the JSPS KAKENHI Grant No. 23K03377 and No. 23H05439. The author thanks Drs.~Kei Iida, Etsuko Itou and Kotaro Murakami for fruitful discussions on lattice computations and their numerical results. The author also thanks Dr. Mamiya Kawaguchi for useful discussions on effective models.

\bibliography{reference}

\end{document}